\documentclass[aps,nofootinbib,amsmath,prc,showpacs,groupedaddress,12pt,superscriptaddress,notitlepage]{revtex4-1}

\usepackage[paperwidth=210mm,paperheight=297mm,centering,hmargin=2.4cm,tmargin=2.8cm,bmargin=3.5cm]{geometry}

\usepackage{amsfonts}
\usepackage{amsmath}
\usepackage{amssymb}
\usepackage{empheq}
\usepackage{epsfig}
\usepackage{latexsym}
\usepackage{paralist}
\usepackage{fancyhdr}
\usepackage{graphicx}

\numberwithin{equation}{section}

\usepackage[vcentermath]{youngtab}
\usepackage{young}
\usepackage{ytableau}
\usepackage{etex}
\usepackage{braket}
\usepackage{float}
\usepackage{slashed}
\usepackage{xcolor}
\usepackage{subcaption}
\usepackage{soul}

\usepackage{dcolumn}

\makeatletter
\@addtoreset{equation}{section}

\makeatother

\def\bea{\begin{eqnarray}} 
\def\eea{\end{eqnarray}}
\def\be{\begin{equation}} 
\def\ee{\end{equation}} 
\def\ba{\begin{array}}
\def\ea{\end{array}} 
\def\nn{\nonumber}

\def\be{\begin{equation}}
\def\ee{\end{equation}}
\def\bea{\begin{eqnarray}}
\def\eea{\end{eqnarray}}

\def\nn{\nonumber}

\def\cS{{\cal S}}

\renewcommand{\thefootnote}{\fnsymbol{footnote}}

\let\oldtitle\title
\renewcommand{\title}[1]{\oldtitle{\color{blue}{#1}}}

\let\oldeqref\eqref
\let\oldcite\cite
\renewcommand{\eqref}[1]{{\color{blue}\oldeqref{#1}}}
\renewcommand{\cite}[1]{{\color{blue}\oldcite{#1}}}

\makeatletter
\let\reftagform@=\tagform@
\def\tagform@#1{\maketag@@@{\ignorespaces\textcolor{blue}{(\ignorespaces #1 \unskip\@@italiccorr \ignorespaces)\ignorespaces}}}
\makeatother

\makeatletter
\renewcommand{\p@subsection}{}
\renewcommand{\p@subsubsection}{}
\makeatother

\usepackage{mathpazo}
\usepackage{amsmath}

\begin{document}

\title{Leading order CFT analysis of multi-scalar theories in $d>2$}

\author{A.\ Codello}
\email{a.codello@gmail.com}
\affiliation{Department of Physics, Southern University of Science and Technology, Shenzhen 518055, China}

\author{M.\ Safari}
\email{safari@bo.infn.it}
\affiliation{INFN - Sezione di Bologna, via Irnerio 46, 40126 Bologna, Italy}
\affiliation{
Dipartimento di Fisica e Astronomia,
via Irnerio 46, 40126 Bologna, Italy}

\author{G.\ P.\ Vacca}
\email{vacca@bo.infn.it}
\affiliation{INFN - Sezione di Bologna, via Irnerio 46, 40126 Bologna, Italy}

\author{O.\ Zanusso}
\email{omar.zanusso@uni-jena.de}
\affiliation{
Theoretisch-Physikalisches Institut, Friedrich-Schiller-Universit\"{a}t Jena,
Max-Wien-Platz 1, 07743 Jena, Germany}
\affiliation{INFN - Sezione di Bologna, via Irnerio 46, 40126 Bologna, Italy}

\begin{abstract}
\vspace{3mm}
We investigate multi-field multicritical scalar theories
using CFT constraints on two- and three-point functions combined with the Schwinger-Dyson equation. 
This is done in general and without assuming any symmetry for the models,
which we just define to admit a Landau-Ginzburg description
that includes the most general critical interactions built from monomials of the form $\phi_{i_1} \cdots \phi_{i_m}$.
For all such models we analyze to the leading order of the $\epsilon$-expansion the anomalous dimensions of the fields and those of the composite quadratic operators.
For models with even $m$ we extend the analysis to an infinite tower of composite operators of arbitrary order.
The results are supplemented by the computation of some families of structure constants.
We also find the equations which constrain the nontrivial critical theories at leading order and
show that they coincide with the ones obtained with functional perturbative RG methods.
This is done for the case $m=3$ as well as for all the even models. We ultimately specialize to $S_q$ symmetric models,
which are related to the $q$-{\tt state Potts} universality class, and focus on three realizations appearing
below the upper critical dimensions $6$, $4$ and $\frac{10}{3}$, which can thus be nontrivial CFTs in three dimensions.
\end{abstract}

\maketitle

\renewcommand{\thefootnote}{\arabic{footnote}}
\setcounter{footnote}{0}

\newpage
\section{Introduction}\label{sect:introduction}

The concept of universal behavior in physical systems is very fruitful and has been successfully spread to other quantitative sciences.
At the theoretical level quantum and statistical field theories are important tools to study the approach to criticality of physical systems,
where most of the details of the microscopic interactions are washed off by the presence of a second order phase transition and universal features emerge. 
In fact, only the degrees of freedom involved, the symmetries and the dimensionality of the system play a crucial role. 
In the so-called theory space of all quantum field theories (QFTs) these universal features appear as special points corresponding to critical theories in which the correlation length diverges and physics is nontrivial at all scales. These special QFTs are often lifted to become conformal field theories (CFTs) i.e.\ scale invariance together with Poincare symmetry are promoted to full conformal invariance~\cite{Luty:2012ww}.

Below their upper critical dimension, which is defined as the dimensionality above which a QFT exhibits Gaussian exponents,
these QFTs are generally interacting.
In order to investigate these systems nonperturbatively one can resort to numerical techniques, when such an approach is feasible.
These investigations might, for example, take the form of Monte-Carlo simulations which are based on the application of the Metropolis algorithm on a lattice~(see for example \cite{lattice} and references therein).
From our -- admittedly theoreticians' -- point of view simulations are useful to benchmark the results obtained with alternative analytical, sometimes approximate, methods.

Historically the most recognized analytical method to investigate theories which exhibit a second order phase transition is the renormalization group (RG) approach~\cite{Cardy:1996xt}, especially after the impetus given by the pioneering work of Wilson~\cite{Wilson:1971bg}. 
Along this line both perturbative and nonperturbative RG approaches can be employed.
The latter has acquired the name of exact RG, which is typically formulated at functional level in terms of flow equations for the Wilsonian action~\cite{Wegner:1972ih,Polchinski:1983gv}
or for the 1PI effective average action~\cite{Wetterich:1992yh,Morris:1994ie}.
Perturbative investigations have been around since the early days~\cite{Wilson:1971dc,Wilson:1973jj} 
and have mainly lead to results expressed in the $\epsilon$-expansion
and its resummation below the upper critical dimension of a QFT~\cite{epsilon}.

In general RG methods, especially if nonperturbative, can give access to global flow, i.e.\ flows which cover the full theory space.
Consequently such flows have an enormous amount of information but require an equal amount of difficult computations.
One usually needs to resort to some approximations in order to obtain tractable RG equations in the full theory space,
but reasonably precise results can be nevertheless obtained, as shown for example 
in the computation of the critical exponents of the {\tt Ising} universality class and its $O(N)$-symmetric extensions~\cite{Guida:1998bx}.
The investigations of several different universality classes -- new and old -- continues to this day
and certainly will not lose momentum any time soon~\cite{Pelissetto:2000ek,Gracey:2015tta, Gracey:2016mio,Gracey:2017oqf}. 

Soon after the early developments of the Wilsonian action,
it has been observed that the perturbative RG too can be conveniently formulated at functional level~\cite{Jack:1982sr,ODwyer:2007brp}.
In this approach, later referred to as the functional perturbative RG, one constructs beta functions which encode
the scale dependence of several couplings at the same time
and obtains results for several quantities in a more efficient way.
This method has also another important advantage which is relevant for this paper:
it can be used with a lot of generality in that very little of the system under investigation must be specified a priori.
For example, it can be used for theories with rather arbitrary interacting potential (as we will do in this paper),
as well as for families of scalar theories with both unitary and nonunitary interactions~\cite{Codello:2017hhh,Codello:2017epp},
and even for higher derivative theories in which the derivative interactions are present at criticality~\cite{Safari:2017irw,Safari:2017tgs}. 
Since it may also happen that internal symmetries emerge at critical points~\cite{Brezin:1973jt, Zia:1974nv,Michel:1983in, TMTB,Osborn:2017ucf}, 
this method can be taken as a starting point to investigate theories not constrained a priori by any symmetry, even including supersymmetry \cite{Wallace:1975ez,Gies:2017tod}. 

In the past few years alternative methods based on conformal invariance have gained considerable popularity
and shown increasing success.
The general strategy of these methods is to focus on the critical points in the space of all theories,
assume that scale invariance is promoted to be local,
and consequently exploit the enhanced conformal symmetry of the system.
It is not an understatement to say that if one assumes that critical points are CFTs, even in $d>2$,
there is a significant advantage when computing close-to-criticality quantities
because of the constraints on correlators imposed by conformal symmetry.
These ideas are at the base of the conformal bootstrap approach~\cite{Rattazzi:2008pe,ElShowk:2012ht}, which follows an early suggestion by A.~Polyakov and is based on the consistency conditions that are obtained by rewriting
the conformal partial wave expansion in two of the $s$, $t$ and $u$ channels thanks to operator product expansion (OPE) associativity.
This method was employed in the analysis of some critical theories and is currently giving
the most precise evaluation of the critical exponents of the universality class of the Ising model~\cite{Simmons-Duffin:2015qma},
and is able to deal with various symmetry groups (see for example \cite{Stergiou:2018gjj}).
Nevertheless, in order to push the analysis to the best accuracy, a good amount of computing power is required
even for the conformal bootstrap.
In this light, CFT methods have taken the stage as consistent and numerically effective substitutes of both lattice and RG methods at criticality.

Besides the numerical achievements of the conformal bootstrap,
several analytic realizations of the underlying idea have been developed, including some
which involve perturbative expansions in small parameters such as $\epsilon$.
Among these methods we mention those based on the singularity structure of conformal blocks~\cite{Gliozzi:2016ysv} and their Mellin representation~\cite{Gopakumar:2016wkt}, and on the large spin expansions~\cite{Alday:2016njk,Alday:2017zzv,Henriksson:2018myn}.

In this work we concentrate on a CFT-based method which determines the conformal data
of a theory in the $\epsilon$-expansion by requiring consistency between
the Schwinger-Dyson equation (SDE), related to a general action at criticality,
and conformal symmetry in the Gaussian limit $\epsilon\to0$ \cite{Rychkov:2015naa}. 
We refer to this method as SDE+CFT for brevity
and very briefly discuss the some properties of both SDE and CFT in the following paragraphs.
The interplay of these properties is the essential bulding block of this paper's analysis.

The SDE, which at the leading order are nothing but the generalizations of the classical equations of motion,
can constrain at the operatorial (or functional) level the correlators of a given QFT.
Contact terms at separate points are absent and any insertion of the equations of motion in a correlator
constructed with a string of operators provides a nontrivial relations among correlators.
In particular this is also true if the QFT is a CFT,
so that for any state of the CFT and for any list of operators $O_i$
resulting from the representation of the conformal group
one has the relation
\begin{align}
\left\langle\frac{\delta S}{\delta\phi}(x) \, O_1(y) \, O_2 (z) \dots\right\rangle  = 0,
\label{prop}
\end{align}
where $S$ is the conformally invariant action. Furthermore, conformal symmetry greatly constrains the correlators appearing in the above equation, even in $d>2$.
Adopting a basis $O_a$ of normalized scalar primary operators with scaling dimensions $\Delta_a$,
the two point correlators have the following form:
\begin{align}\label{cft-2pf}
\braket{O_a(x)  O_b(y)} =\frac{\delta_{ab}} {|x-y|^{2 \Delta_a}} .
\end{align}
The three-point correlator for scalar primary operators is also strongly constrained by conformal symmetry and reads
\begin{align}\label{cft-3pf}
\braket{O_a(x)  O_b(y) O_c(z)} =\frac{C_{abc} }{|x-y|^{\Delta_{a}+\Delta_{b}-\Delta_{c}} 
|y-z|^{\Delta_{b}+\Delta_{c}-\Delta_{a}} |z-x|^{\Delta_{c}+\Delta_{a}-\Delta_{b}}} \,,
\end{align}
where $C_{abc}$ are the structure constants of the CFT.

Thanks to the power of conformal symmetry, a
CFT is completely and uniquely determined by providing a basis of primary operators $O_a(x)$,
the scaling dimensions $\Delta_a$ and the structure constants $C_{abc}$, which together are known as \emph{CFT data}.
The idea of the SDE+CFT approach is to move below the upper critical dimension $d_c$, above which the theory is Gaussian,
and interpolate the nontrivial correlators shown above with those of the trivial Gaussian theory as a function of the critical coupling.
The consistency of this interpolation determines the leading order corrections in $\epsilon$ of some conformal data
when one exploits further relations between operators that are primary only in the Gaussian limit.
In this work we restrict ourselves to the information we can extract from the analysis of two- and three-point functions,
which is the current state-of-the-art of the approach.
Investigations based on this approach
have been applied up to now to scalar theories with and without $O(N)$ symmerty~
\cite{Rychkov:2015naa,Basu:2015gpa,Nakayama:2016cim,Nii:2016lpa,Hasegawa:2016piv},
and extended to unitary and nonunitary families of multicritical single-scalar theories \cite{Codello:2017qek,Codello:2017epp,Safari:2017irw}. 

Here we shall make a step further and apply this method to the study of  multicritical scalar theories with multiple (say $N$ different) fields $\phi_i$,
and with a generic interaction encoded in the potential
\be
V=\frac{1}{m!} V_{i_1 \cdots i_m}\, \phi_{i_1} \cdots \phi_{i_m},
\label{mcpotential}
\ee
where a sum over repeated indices is understood and no symmetry properties are considered. From the above form of the potential the upper critical dimension is determined to be $d_c=2m/(m-2)$. 
In general the number of independent monomials is $\binom{m+N-1}{m}$ and for each one of them one can introduce a coupling.
Moreover the fact that the quadratic part of the action (standard kinetic term in this case) is invariant under linear $SO(N)$ field transformation $\phi\to R\phi$,
which lead to equivalent theories describing the same physics, imposes further constraints on the couplings associated to inequivalent theories.
This can be analyzed with group theoretical methods~\cite{Zia:1974nv,Michel:1983in, TMTB}, 
also introducing invariants on the space of couplings under such field redefinitions~\cite{Osborn:2017ucf} in terms of which any universal quantity is expected to be expressed.
Discrete symmetries such as permutations can also be taken into account.

In the main text we are able to write explicit eigenvalue equations that depend functionally on the potential
and from which many universal conformal data at the leading nontrivial perturbative order can be extracted.
In particular we derive expressions for the anomalous dimensions of the fields, the anomalous dimensions of the quadratic composite operators, and for several classes of structure constants.
When treating even unitary models with $m=2n$ we are able to extend the procedure
to an infinite tower of higher order composite operators besides the quadratic ones.
For all even models
we write the equations which fix, at the leading order,
possible critical potentials in terms of the parameter $\epsilon$.
It turns out that these equations coincide with the fixed point equations
obtained for a generic potential using the functional perturbative RG approach.
This provides further insight on how some information of one approach (RG) is encoded in the other (CFT) and viceversa.

Likewise, for the cubic nonunitary model with $m=3$ we obtain, in complete analogy to the single field case,
results for anomalous dimensions of the fields, for the quadratic composite operators, and for some structure constants.
We are also able to fix as a function of $\epsilon$ the critical potentials. Similar to the even case,
the critical conditions coincide wih those of the functional perturbative RG approach.
For higher order nonunitary models with $m=2n-1$ and $n>2$ we are not able
to find enough constraints on the critical potential to set it in terms of $\epsilon$,
which is again a situation in complete analogy to the single-field case.

We then specialize our very general results by giving a more explicit form to the potential that is constrained by symmetry.  
As an interesting example, we choose the symmetry to be the permutation group $S_q$ acting on the fields with $q=N+1$
and we study it in detail. This symmetry group corresponds to Potts-like field theories, which include as special cases the standard
field-theoretical cubic realization of the Potts universality class, the reduced Potts model, and -- in principle -- infinitely many
generalizations.
Despite being much less constraining than $O(N)$ symmetry
(which nevertheless emerges as an effective symmetry for some fixed points),
the group tensor structures appearing in the potential can be naturally factorized
thus reducing strongly the number of independent parameters.
The Potts models~\cite{Potts:1951rk,Baxter:2000ez,Zia:1975ha,Nienhuis:1979mb,Wu:1982ra,Fortuin:1971dw,Zinati:2017hdy,Delfino:2017biz} are quite ubiquitous in statistical mechanics: Several interesting models can be obtained
if one takes analytic continuations of $q$. The most relevant continuations for this paper are to the value $q=1$,
which is related to models of percolation, and to $q=0$,
which is related to the random cluster model known as spanning forest \cite{Deng:2006ur}.\footnote{In absence of better nomenclature,
we address all continuum field-theoretical models with $S_q$ symmetry as ``Potts model''.
As it will be shown in the paper, these include the standard universality class that describes the standard lattice Potts model at criticality,
as well as some of its multicritical generalizations which we classify by the order of the critical interaction in the corresponding potentials.
}
The easiest way to construct an $S_q$-invariant potential interaction is to follow a standard vector representation of the $S_q$ group.
We concentrate on the Landau-Ginzburg description of Potts models which have upper critical dimensions $d_c>3$,
and therefore can be nontrivial in $d=3$.
This restricts our specific investigation to the cubic \cite{Amit:1976pz}, quartic \cite{Rong:2017cow}, and quintic potentials for which we obtain some universal conformal data, recovering as usual several RG results.

"Multi-field generality" and "functional description" are ingredients that
bring this work close, in spirit, to that of Osborn \& Stergiou \cite{Osborn:2017ucf}
in which several similar questions are addressed using multiloop perturbative RG methods instead of the CFT+SDE technique.
Our work should also be regarded as a companion to a forthcoming paper
devoted to the study of multi-field multicritical Potts models with functional perturbatve RG techniques \cite{CSVZ4}.
While the CFT+SDE methods used in this work are still limited to the leading order of the $\epsilon$-expansion,
their value lies in the fact that they outline the importance of conformal invariance at criticality
and they facilitate the computation of conformal data and, more generally, of the OPE.

In the next section, which is the most important of the paper,
we apply the CFT+SDE technique to a multi-field multicritical model with a general potential,
providing in several subsections general expressions for the conformal data in terms of the potential.
We shall then introduce in Section~\ref{sect:potts_models} the Potts model and discuss its field representations
and group invariants, along with some useful relations,
and introduce the relevant Landau-Ginzburg representation we shall later use. Having imposed the $S_q$ symmetry
we also introduce operators on the space of quadratic fields that project them into irreducible representations with definite anomalous dimensions, which we also give.
In Section~\ref{sect:potts-cft} we present the analysis and the results for the specialized cubic, (restricted) quartic,
and quintic Potts universality classes. We then present our conclusions.
The paper ends with two appendices. The first contains some useful relations for free theory correlators
which are used extensively in the text.
The second includes three parts reporting in order: the reduction relations for $S_q$ symmetric tensors,
some computational details for the quintic model, and few useful RG results~\cite{CSVZ4} needed for the quintic model.

\section{CFT data from classical equations of motion: general results}\label{sect:sde-cft-analysis}
The first part of this paper is devoted to an analysis of general scalar theories with a number of $N$ fields, where no symmetry is imposed on the model. 
In the Landau-Ginzburg description, these models are expressed by the following action
\be
S=\int \! d^dx \left[{\textstyle{\frac{1}{2}}}\,\partial \phi_i \cdot \partial \phi_i + V(\phi) \right] \,,
\label{general_model}
\ee
and therefore characterized by a standard kinetic term and interactions induced by a generic local potential with a critical dimension $d_c$.
Considering multicritical potentials such as in Eq.~\eqref{mcpotential} that depend on the product of $m$ fields, one has the relation $d_c=2m/(m-2)$.
We shall in general adopt the perturbative $\epsilon$-expansion technique below the critical dimension $d=d_c-\epsilon$. All the fields have the same canonical dimension $\delta=d/2-1=2/(m-2)-\epsilon/2$. 
The method we employ is based on the use of the Schwinger-Dyson equations (SDE) combined with the assumption of conformal symmetry of the critical model. Critical information is extracted from the study of two and three point correlators, whose functional form is completely fixed in terms the conformal data parameters. Our analysis gives access to some of these conformal data at leading order in the $\epsilon$-expansion (different quantities can have a different power in $\epsilon$ at leading order).

We devote separate subsections to the computation of the field anomalous dimension, the critical exponents of the mass operators, critical exponents of all higher order operators 
for even models, and some structure constants (or OPE coefficients). Finally we shall show for the case of $m=3$ corresponding to $d_c=6$, the case $m=4$ corresponding to $d_c=4$, and then for general models with even $m>4$, how the CFT constraints together with the Schwinger-Dyson equations can be used to fix the critical theory, i.e. the $\epsilon$ dependence of the couplings present in the potential. In particular we show that these constraints are exactly the same as the fixed point conditions of beta functions which appear in the functional perturbative RG approach~\cite{ODwyer:2007brp, Codello:2017hhh,Codello:2017epp, Osborn:2017ucf}.

We stress that the results given in this section are general and, as such, depend on the generic potential $V$ which defines a multicritical model. We shall then restrict ourselves in the next Sections to specific models having in particular the $S_q$ symmetry.

\subsection{Field anomalous dimension}\label{sect:eta-cft}
The SDE-based computation of the anomalous dimension for multi-field scalar theories follows closely that of the single-field case \cite{Codello:2017qek}. Let us consider the general multi-field action \eqref{general_model} which leads to the equation of motion 
\be \label{eom}
\Box \phi_i = V_i,
\ee
where lower indices on the potential refer to its field derivatives, as in the special case \eqref{mcpotential}. We shall use the parameter $n=m/2$ to label the families of multicritical models
where $n$ is the multicriticality label corresponding to the power of the classically marginal potential $\phi^{2n}$ in the theory \cite{ODwyer:2007brp,Codello:2017hhh,Codello:2017epp}. Notice that for the cubic and quintic models which we shall study in detail later in Section~\ref{sect:potts-cft} $n$ is a half odd number. 

In general the fields $ \phi_i $ are not necessarily scaling operators and there can be $N$ different anomalous dimensions associated to the true scaling fields, which correspond to the defining primaries of the CFT. 
These are related to the scaling fields $ \tilde\phi_i $ through a linear transformation which leaves the kinetic term invariant 
\be 
\tilde\phi_i = R_{il}\phi_l, \qquad R^TR=1,
\ee
Then, having a definite scaling, the two-point function of the fields $\tilde\phi_i$ takes the following form
\be 
\langle \tilde\phi_i(x) \tilde\phi_j(y)\rangle = \frac{\tilde c\delta_{ij}}{|x-y|^{2\Delta_i}} ,
\label{2Pc}
\ee
where $\tilde c$ is a constant and $\Delta_i$ is the dimension of the field $\tilde \phi_i$.
Notice that the matrix $R$ can always be chosen such as to diagonalize also the space of fields with the same dimension. In terms of the scaling fields one may also define $V(\phi_i)=\tilde V(\tilde \phi_i)$. 

The scaling dimension $\Delta_i$ is the sum of the field canonical dimension $\delta$ and anomalous dimension $\gamma_i$     
\be 
\Delta_i = \delta + \gamma_i, \qquad \delta = \frac{1}{n-1} - \frac{\epsilon}{2}.
\ee
One can also notice that the composite operator $V_i(\phi)$ has scaling dimension $\Delta_i+2$, when interactions are turned on below the upper critical dimension, which means that a recombination of conformal multiplets takes place.

One can find the anomalous dimension $\gamma_i$ solving a simple  equation obtained, as in the single field case, 
applying $\Box_x\Box_y$ to Eq.~\eqref{2Pc}. One gets
\be \label{2}
\langle \tilde V_a(x) \tilde V_b(y) \rangle = \Box^2_x\frac{\tilde c}{|x-y|^{2\Delta}}\;\delta_{ab}.
\ee
where the equations of motion have been used on the l.h.s. The r.h.s of this equation can be straightforwardly calculated and at leading order gives
\be
\Box^2_x\frac{\tilde c \delta_{ab}}{|x-y|^{2\Delta}} = \frac{16\Delta(\Delta-\delta)(\Delta+1)(\Delta+1-\delta)}{|x-y|^{2\Delta+4}} \tilde c \delta_{ab}\;\stackrel{\mathrm{LO}}{=}\; \frac{16\delta_c(\delta_c+1)\gamma}{|x-y|^{2\delta_c+4}} c \delta_{ab},
\ee
where $\delta_c$ is the critical ($\epsilon=0$) value of $\delta$ defined in Appendix \ref{free} and $c$ is the free theory value of $\tilde c$ given by Eq.~\eqref{c}. 
For the calculation of the l.h.s one can Taylor expand the potential and use \eqref{2pf-free} to get at leading order
\be \label{vavb}
\langle \tilde V_a(x) \tilde V_b(y) \rangle \;\stackrel{\mathrm{LO}}{=}\; \sum_\ell \frac{1}{\ell!} \frac{c^\ell}{|x-y|^{2\ell\delta_c}}\;
\tilde V_{ai_1\cdots i_\ell}\, \tilde V_{bi_1\cdots i_\ell}\Big|_{\phi=0}\,,
\ee
For the multicritical model $m=2n$ (with $n$ integer or half integer) this expression picks only the $\ell =2n-1$ term in the sum.
In this case, noticing that $2\delta_c+4=2(2n-1)\delta_c$, Eq.~\eqref{2} at leading order gives the anomalous dimension of the field $\tilde\phi_a$ 
\be  \label{eta-cft}
\gamma_a \,\delta_{ab} \stackrel{\mathrm{LO}}{=} \frac{(n-1)^2}{8(2n)!}\,c^{2(n-1)}\;\tilde V_{ai_1\cdots i_{2n-1}}\, \tilde V_{bi_1\cdots i_{2n-1}}\Big|_{\phi=0}.
\ee 
Notice that in the Potts models which we shall consider later on, symmetry properties enforce the r.h.s to be proportional to $\delta_{ab}$. Written in terms of the original field the above equation becomes
\be 
\gamma_a \delta_{ab} = \frac{(n-1)^2}{8(2n)!}\,c^{2(n-1)} R^T_{ac} V_{ci_1i_2\cdots i_{2n-1}} V_{di_1i_2\cdots i_{2n-1}} R_{db},
\ee
which means that the matrix of anomalous dimensions is
\be  \label{adm} 
\boxed{\gamma_{ab} = \frac{(n-1)^2}{8(2n)!}\,c^{2(n-1)} V_{ai_1i_2\cdots i_{2n-1}} V_{bi_1i_2\cdots i_{2n-1}},}
\ee
and that $R$ is the matrix that diagonalizes it. In the rest of the paper we drop the tilde on the fields and the potential and always assume, unless otherwise stated, to work in the diagonal basis.

Let us make an aside comment here. In the RG analysis of physical systems close to criticality  
the approach to criticality is controlled by parameters such as the temperature, the simplest example being the Ising model with quartic interaction. In this model using for example dimensional regularization one has to tune to zero the mass operator, which is relevant at criticality. In the multi-field case in order to reach such a condition for all fields, insisting to tune only one parameter, one is forced to require that all the "bare masses" coincide. This requirement is equivalent to the so called zero trace property on generic quartic interactions $v_{ijkk}=v \delta_{ij}$~\cite{Brezin:1973jt}, which implies that at the fixed point 
all the anomalous dimensions are equal. In single-field multicritical models where there as more than one relevant operator, the approach to criticality can be controlled by introducing other tuning parameters. Requiring the same number of parameters, in the multi-field case, to control the approach to criticality, one is forced to introduce other conditions on the critical potential to have the same bare mass and bare couplings of all the relevant operators which we would like to tune to zero. 
In our general CFT approach of this paper we are not concerned with these extra requirements and keep the arguments as general as possible, imposing no symmetry on the model.

\subsection{Quadratic operators}\label{ss:qo}

We will now move on to the computation of the critical exponents corresponding to the mass operators $\phi_i\phi_j$. It is useful for this purpose to review first how the critical exponent $\gamma_2$ of the operator $\phi^2$ is obtained in single-field scalar theories \cite{Codello:2017qek}, where the critical potential includes only the marginal interaction $V(\phi)=\frac{g}{(2n)!}\phi^{2n}$. Here $n$ is either an integer or a half odd number. For $n\neq 2$ this is done by applying the operator $\Box_x\Box_y$ to the three-point function $\langle\phi(x) \,\phi(y)\, \phi^{2}(z)  \rangle$ and calculating it at leading order in two ways: first by using the SDE
\be \label{eqn-sf-1}
\langle \square_x\phi(x) \,\square_y\phi(y)\, \phi^2(z)  \rangle 
\,\stackrel{\mathrm{LO}}{=}\, \frac{g^{2}}{(2n-1)!^2} \frac{C^{\mathrm{free}}_{2n-1,2n-1,2}}{|x-y|^4|x-z|^{2\delta_c}|y-z|^{2\delta_c}}\,,
\ee
and second by direct application of the operator $\Box_x\Box_y$ to the expression for the three-point function 
\be \label{eqn-sf-2}
\square_x\square_y \langle \phi(x) \,\phi(y)\, \phi^2(z)  \rangle
\,\stackrel{\mathrm{LO}}{=}\, C^{\mathrm{free}}_{1,1,2} \,\frac{8(n-2)(\gamma_2 - 2\gamma)}{(n-1)^2}  \frac{1}{|x-y|^4|x-z|^{2\delta_c}|y-z|^{2\delta_c}}\,.
\ee
The value of $\gamma_2$ is obtained by equating these two. The two structure constants in the free theory given above can be found from the general formula \eqref{c3_free}.  In the multi-field case the situation is slightly different as there is more than one mass operators $\phi_i\phi_j$. So, although these do not mix with derivative operators or operators with more fields, they can mix together. One can assign a (anomalous) scaling dimension only to particular combinations of them that make a scaling operator. Suppose for instance that the combination
\be 
S_{pq}\; \phi_p\phi_q 
\ee
makes a scaling operator, where $S_{pq}$ is a tensor symmetric in its pair of indices. Let us denote the anomalous dimension of this operator by $\gamma^S_2$, where the label $S$ refers to the particular choice of $S_{pq}$. Then the two equations for the single-field case \eqref{eqn-sf-1} and \eqref{eqn-sf-2} are generalized respectively  to 
\be \label{bibjpq}
\langle \square_x\phi_i(x) \,\square_y\phi_j(y)\; [S_{pq}\, \phi_p\phi_q](z)  \rangle 
\,\stackrel{\mathrm{LO}}{=}\, \frac{V_{i\,i_1\cdots i_{2n-1}}V_{j\,j_1\cdots j_{2n-1}}}{(2n-1)!^2} \frac{C^{\mathrm{free}}_{i_1\cdots i_{2n-1},\, j_1\cdots j_{2n-1},\, pq}\,S_{pq}}{|x-y|^4|x-z|^{2\delta_c}|y-z|^{2\delta_c}},
\ee
\be  \label{bbijpq}
\square_x\square_y \langle \phi_i(x) \,\phi_j(y)\; [S_{pq}\,\phi_p\phi_q](z)  \rangle
\,\stackrel{\mathrm{LO}}{=}\,\frac{8(n-2)(\gamma^S_2 - \gamma^i-\gamma^j)}{(n-1)^2}  \frac{C^{\mathrm{free}}_{i,j,pq}\,S_{pq} }{|x-y|^4|x-z|^{2\delta_c}|y-z|^{2\delta_c}},
\ee
where the coupling $g$ has been replaced by the $2n$th derivative of the potential, which is evaluated at $\phi=0$, as will be understood in the rest of the paper. The two free structure constants in \eqref{bibjpq} and \eqref{bbijpq} are defined respectively as the coefficients in the correlation functions 
\be 
\left\langle [\phi_{i_1}\cdots \phi_{i_{2n-1}}](x) [\phi_{j_1}\cdots \phi_{j_{2n-1}}](y) [\phi_p\phi_q](z) \right\rangle, \qquad 
\left\langle \phi_i(x) \phi_j(y) [\phi_p\phi_q](z) \right\rangle,
\ee
in the free theory. These structure constants in the free theory are related to their single-field analogues as
\be 
C^{\mathrm{free}}_{i,j,pq} =  C^{\mathrm{free}}_{1,1,2} \; \delta^i_{(p}\delta^j_{q)}, \qquad
C^{\mathrm{free}}_{i_1\cdots i_{2n-1},\, j_1\cdots j_{2n-1},\, pq} = C^{\mathrm{free}}_{2n-1,2n-1,2} \; (\delta^{i_1}_{p}\delta^{j_1}_{q}\, \delta^{i_2}_{j_2}\cdots \delta^{i_{2n-1}}_{j_{2n-1}})\,,
\ee
where in the last expression the Kronecker delta functions are enclosed in parenthesis indicating that the $i_l$, $j_l$ and the $p,q$ indices are separately symmetrized. These notations are introduced in Appendix~\ref{free} in the more general sense. Equating the two equations \eqref{bibjpq} and \eqref{bbijpq} and simplifying a bit leads to 
\be \label{crit-cft}
(\gamma^S_2 - \gamma^i-\gamma^j) \; S_{ij} = \frac{(n-1)^2c^{2(n-1)}}{8(n-2)(2n-2)!}\, V_{i\,p\,i_2\cdots i_{2n-1}}V_{j\,q\,i_2\cdots i_{2n-1}} \,S_{pq}.
\ee
In models where all anomalous dimensions $\gamma^i=\gamma$ are equal, it is clear from \eqref{eta-cft} that $S_{ij} = \delta_{ij}$ is always an eigenvector with eigenvalue $\gamma_2^1$ given by
\be \label{gamma20-eta}
\gamma_2^1 = 2\frac{n^2-1}{n-2}\eta,
\ee
where $\eta=2\gamma$. Eq.~\eqref{crit-cft} may be written entirely in terms of the potential using \eqref{eta-cft}. The result is
\begin{empheq}[box=\fbox]{align}
\gamma^S_2\, S_{ij} &= \frac{(n-1)^2c^{2(n-1)}}{8(n-2)(2n-2)!}\, V_{i\,p\,i_2\cdots i_{2n-1}}V_{j\,q\,i_2\cdots i_{2n-1}} \,S_{pq} \,\,
\nn\\
&\,+ \frac{(n-1)^2c^{2(n-1)}}{8(2n)!}\, V_{i\,i_1\cdots i_{2n-1}}V_{p\,i_1\cdots i_{2n-1}} \,S_{pj}\,\, \label{gammaS2}
\\
&\,+ \frac{(n-1)^2c^{2(n-1)}}{8(2n)!}\, V_{j\,i_1\cdots i_{2n-1}}V_{p\,i_1\cdots i_{2n-1}} \,S_{pi} \nn
\end{empheq}
Solving this eigenvalue equation one can find the scaling operators and their anomalous dimensions. There is one important exception to the above equation, and that is the operator $V_{ijk}\phi^j\phi^k$ for the case $n=3/2$. This is because this operator which has a definite scaling property is not a primary but a descendant operator. In this case the anomalous dimension, which we can call $\gamma^i_2$, does not satisfy \eqref{gammaS2} and is obtain instead from the exact relation $\gamma^i_2 = \gamma_i + \epsilon/2$. 

For the special case of $n=2$ the situation is different. In this case, for the single-field theory information on the anomalous dimension is obtained by comparing the two expressions \cite{Codello:2017qek}
\be 
\langle \square_x\phi(x) \,\phi(y)\, \phi^2(z)  \rangle 
\,\stackrel{\mathrm{LO}}{=}\, \frac{g}{3!} \frac{C^{\mathrm{free}}_{3,1,2}}{|x-y|^2|x-z|^{2\delta_c+2}|y-z|^{2\delta_c-2}}, 
\ee
\be
\square_x \langle \phi(x) \,\phi(y)\, \phi^2(z)  \rangle
\,\stackrel{\mathrm{LO}}{=}\,\frac{2C^{\mathrm{free}}_{1,1,2} \,(\gamma_2 - 2\gamma) }{|x-y|^2|x-z|^{2\delta_c+2}|y-z|^{2\delta_c-2}}.
\ee
The multifield generalization of these two equations is 
\be 
\langle \square_x\phi_i(x) \,\phi_j(y)\, [S_{ab}\,\phi_{a}\phi_{b}](z)  \rangle 
\,\stackrel{\mathrm{LO}}{=}\, \frac{V_{ipql}}{3!} \frac{C^{\mathrm{free}}_{pql,j,ab}\,S_{ab}}{|x-y|^2|x-z|^{2\delta_c+2}|y-z|^{2\delta_c-2}}, 
\ee
\be
\square_x \langle \phi_i(x) \,\phi_j(y)\; [S_{ab}\,\phi_{a}\phi_{b}](z) \rangle
\,\stackrel{\mathrm{LO}}{=}\,\frac{2C^{\mathrm{free}}_{i,j,ab} \,S_{ab}\,(\gamma^S_2 - \gamma_i-\gamma_j)}{|x-y|^2|x-z|^{2\delta_c+2}|y-z|^{2\delta_c-2}}.
\ee
Equating the two gives
\be 
2C^{\mathrm{free}}_{1,1,2}\,\delta^{i}_{(a}\delta^j_{b)} \,S_{ab}\,(\gamma^S_2 - \gamma_i-\gamma_j) = \frac{1}{3!}V_{ipql}\, C^{\mathrm{free}}_{3,1,2}\,\delta^{(p}_j\delta^q_{(a}\delta^{l)}_{b)}\,S_{ab}\,.
\ee
After some simplification, and noticing that $\gamma_i$ are of second order in the potential, this equation reduces to 
\be  \label{crit-cft-2}
\boxed{
\gamma^S_2  \, S_{ij} = \frac{c}{4} V_{ijab}\,S_{ab},
}
\ee
indicating that $S_{ab}$ is an eigenvector of $V_{ijab}$ in this case. 
The analyses leading to equations \eqref{gammaS2} and \eqref{crit-cft-2} cannot be pushed further unless we specify the model, or at least the symmetry. These two equations are studied further and solved in Section \ref{sect:potts_models} for Potts field theories in particular dimensions. Another possible path to reduce the complexity in a general analysis would be to restrict the number of fields in order to deal with a finite (possibly small) number of couplings even in absence of a symmetry and consequently on the number of possible critical theories, satisfying set of equations which we shall derive in Section~\ref{sec:FPcond}.

\subsection{Higher order composite operators: recurrence relation and its solution} \label{ss:rr} 
So far we have discussed the anomalous dimensions of the fields $\phi_i$ and those of the quadratic operators for a general multicritical scalar model with integer or half-odd $n$. For the unitary "even" critical theories with 
integer $n$ we move on and seek scaling operators of arbitrary order $k$ and their anomalous dimensions. For simplicity of notation we sometimes use the following abbreviation
\be
\cS_{k}=S_{i_1 \cdots i_k} \phi_{i_1} \cdots \phi_{i_k},
\label{scaling_op}
\ee
the anomalous dimensions of which we denote by $\gamma^S_k$. Let us also note that at the leading perturbative order we are considering here, \eqref{scaling_op} is the most general form the scaling operators can take, while from the next-to-leading order derivative operators start mixing as well. In the RG language this translates to the fact that in the beta function of the potential the leading order contribution comes only from the potential itself, and the contribution from the wave-function renormalization  appears only at next-to-leading order, and so on for higher derivative operators.

In analogy with the analysis for single-field models detailed in \cite{Codello:2017qek} we shall study the three point correlation functions
\be \label{box-1kk1}
\langle \square_x\phi_i(x) \, \cS_{k}(y) \cS_{k+1}(z) \rangle\,.
\ee
This quantity can be evaluated at leading order by using the equations of motion, which gives
{\setlength\arraycolsep{2pt}
\bea 
&& \langle \square_x\phi_i(x) \,[S_{j_i\cdots j_k}\phi_{j_1}\cdots\phi_{ j_k}](y)\,[S_{l_1\cdots l_{k+1}}\phi_{l_1}\cdots \phi_{{k+1}}](z) \rangle
\nn\\
&\stackrel{\mathrm{LO}}{=}& \frac{1}{(2n-1)!}  \langle [V_{ii_1\cdots i_{2n-1}} \phi_{i_1}\cdots \phi_{i_{2n-1}}](x) \,[S_{j_i\cdots j_k}\phi_{j_1}\cdots\phi_{ j_k}](y)\, [S_{l_1\cdots l_{k+1}}\phi_{l_1}\cdots \phi_{{k+1}}](z) \rangle \nn\\
&\stackrel{\mathrm{LO}}{=}& \frac{C^{\mathrm{free}}_{2n-1,k,k+1}}{(2n-1)!} \,  \frac{V_{ii_1\cdots i_rj_1\cdots j_s}S_{j_1\cdots j_s l_1\cdots l_t}S_{l_1\cdots l_ti_1\cdots i_r}}{|x-y|^2|y-z|^{2k\delta_c-2}|z-x|^{2\delta_c+2}}.
\eea}%
In the last step, for the leading order contribution, we have used Eq.~\eqref{3pfree} where we rename $l_{12}=r$, $l_{23}=s$ and $l_{31}=t$ which are defined as
\be \label{rst-cond}
\left\lbrace
\ba{l}
r+s=2n-1 \\
s+t = k \\
t+r = k+1 \\
\ea 
\right.
\qquad \Rightarrow \qquad
\left\lbrace
\ba{l}
r= n \\
s= n-1 \\
t= k+1-n \\
\ea 
\right.
\ee
On the other hand, applying the box directly we get, again at leading order
\be 
\langle \square_x\phi_i(x) \, \cS_{k}(y) \cS_{k+1}(z) \rangle\stackrel{\mathrm{LO}}{=} C^{\mathrm{free}}_{1,k,k+1} \,  \frac{2(\gamma^S_{k+1}-\gamma^S_k-\gamma_i)}{n-1} \frac{S_{ii_1\cdots i_k}S_{i_1\cdots i_k}}{|x-y|^2|y-z|^{2k\delta_c-2}|z-x|^{2\delta_c+2}}.
\ee
Comparing these two and using 
\be 
C^{\mathrm{free}}_{2n-1,k,k+1} = \frac{(2n-1)!k!(k+1)!}{(n-1)n!(k-n+1)!}c^{n+k}, \qquad
C^{\mathrm{free}}_{1,k,k+1} = \frac{k!(k+1)!}{k!} c^{k+1} = (k+1)! c^{k+1},
\ee
we obtain a recurrence relation for the critical exponents
\be \label{rr} 
(\gamma^S_{k+1}-\gamma^S_k-\gamma_i) S_{ii_1\cdots i_k}S_{i_1\cdots i_k} = c_{n,k} V_{ii_1\cdots i_rj_1\cdots j_s}S_{j_1\cdots j_s l_1\cdots l_t}S_{l_1\cdots l_ti_1\cdots i_r}\,,
\ee
where for convenience we have introduced the quantity
\be 
c_{n,k} = \frac{n-1}{2(2n-1)!}\frac{C^{\mathrm{free}}_{2n-1,k,k+1}}{C^{\mathrm{free}}_{1,k,k+1}} = 
\frac{c^{n-1}}{2(n-2)!n!} \frac{k!}{(k-n+1)!}.
\label{cnk}
\ee
From the constraints \eqref{rst-cond} it is clear that the smallest $k$ for which the r.h.s of \eqref{rr} does not vanish is $k=n-1$, which means that the anomalous dimensions are linear in the couplings starting from $\gamma^S_n$, while all the lower ones, $\gamma^S_k$ with $k<n$, are at least quadratic. So we start from Eq.~\eqref{rr} with $k=n-1$. The anomalous dimensions $\gamma^S_{n-1}$ and $\gamma^i$ are of higher order and can be omitted, giving
\be \label{gamma-n} 
\gamma^S_n S_{ii_1\cdots i_{n-1}}S_{i_1\cdots i_{n-1}} \stackrel{\mathrm{LO}}{=} c_{n,n-1} V_{ii_1\cdots i_nj_1\cdots j_{n-1}}S_{j_1\cdots j_{n-1}}S_{i_1\cdots i_n}\,,
\ee
where we have used the fact that $r=n$, $s=n-1$ and $t=0$ in this case. On the r.h.s. of this equation we rename the indices as $j_1\cdots j_{n-1} \to  i_1\cdots i_{n-1}$ and $i_1\cdots i_{n} \to  j_1\cdots j_{n}$ and then on both sides of the equation we rename $i \to i_n$. We observe then that the coefficients of $S_{i_1\cdots i_{n-1}}$ are independent of it and since such tensors form a complete basis for symmetric $(n-1)$-index tensors one can drop $S_{i_1\cdots i_{n-1}}$ on both sides. This leads, after permuting some indices thanks to symmetry, to
\be \label{gamma-S-n} 
\gamma^S_n S_{i_1\cdots i_n} = c_{n,n-1} V_{i_1\cdots i_nj_1\cdots j_n}S_{j_1\cdots j_n},
\ee
which is an eigenvalue equation for operators with $n$ fields. Let us now move to the next case $k=n$. This time Eq.~\eqref{rr} gives
\be 
(\gamma^S_{n+1}-\gamma^S_n) S_{ii_1\cdots i_n}S_{i_1\cdots i_n} \stackrel{\mathrm{LO}}{=}
 c_{n,n} V_{ii_1\cdots i_nj_1\cdots j_{n-1}}S_{j_1\cdots j_{n-1}l}S_{li_1\cdots i_n}\,.
\ee
Here the coefficients are not independent of $S_{i_1\cdots i_n}$ because of $\gamma^S_n$, but one can eliminate it using Eq.~\eqref{gamma-S-n}. This gives 
{\setlength\arraycolsep{2pt}
\bea
\hspace{-0.7cm}
\gamma^S_{n+1} S_{ii_1\cdots i_n}S_{i_1\cdots i_n} &=& \gamma^S_n S_{ii_1\cdots i_n}S_{i_1\cdots i_n} + c_{n,n} V_{ii_1\cdots i_nj_1\cdots j_{n-1}}S_{j_1\cdots j_{n-1}l}S_{li_1\cdots i_n} 
\nn\\
&=&  c_{n,n-1} V_{i_1\cdots i_nj_1\cdots j_n} S_{ii_1\cdots i_n} S_{j_1\cdots j_n} + c_{n,n} V_{ii_1\cdots i_nj_1\cdots j_{n-1}}S_{li_1\cdots i_n}S_{j_1\cdots j_{n-1}l} 
\nn\\
&=&  c_{n,n-1} V_{j_1\cdots j_ni_1\cdots i_n} S_{ij_1\cdots j_n} S_{i_1\cdots i_n} + c_{n,n} V_{ij_1\cdots j_ni_1\cdots i_{n-1}}S_{i_nj_1\cdots j_n}S_{i_1\cdots i_n},
\eea}%
where in the last step several indices have been renamed conveniently to make $S_{i_1\cdots i_n}$ appear in all terms as on the l.h.s. 
Since the coefficients are now independent of $S_{i_1\cdots i_n}$ one can drop this tensor provided that we symmetrize the indices $i_1\cdots i_n$ (when needed we denote such operation by enclosing the indices in round brackets) obtaining
\be 
\gamma^S_{n+1} S_{i_1\cdots i_{n+1}} =  c_{n,n-1} V_{j_1\cdots j_ni_1\cdots i_n} S_{i_{n+1}j_1\cdots j_n} + c_{n,n} V_{j_1\cdots j_ni_{n+1}(i_1\cdots i_{n-1}}S_{i_n)j_1\cdots j_n} \,.
\ee
This equation is therefore manifestly symmetric in the indices $i_1\cdots i_n$, but one may wonder if the r.h.s is symmetric in all $i_1\cdots i_{n+1}$ indices as expected from the l.h.s. At first sight this does not seem clear, but it turns out that the two coefficients $c_{n,n-1}$ and $c_{n,n}$ have just the right ratio to make the r.h.s symmetric. In fact from Eq.~\eqref{cnk} one has $c_{n,n}=nc_{n,n-1}$ and we can write
{\setlength\arraycolsep{2pt}
\bea 
\gamma^S_{n+1} S_{i_1\cdots i_{n+1}} &=&  c_{n,n-1}\left(V_{j_1\cdots j_ni_1\cdots i_n} S_{i_{n+1}j_1\cdots j_n} + n V_{j_1\cdots j_ni_{n+1}(i_1\cdots i_{n-1}}S_{i_n)j_1\cdots j_n}\right) 
\nn\\
&=&  (n+1)c_{n,n-1} V_{j_1\cdots j_n(i_1\cdots i_n}S_{i_{n+1})j_1\cdots j_n}
\nn\\
&=&  (c_{n,n-1}+c_{n,n}) V_{j_1\cdots j_n(i_1\cdots i_n}S_{i_{n+1})j_1\cdots j_n}
\nn\\
&=&  \frac{2}{n} c_{n,n+1} V_{j_1\cdots j_n(i_1\cdots i_n}S_{i_{n+1})j_1\cdots j_n} \,. \label{gamma-n+1}
\eea}%
This gives an eigenvalue equation for scaling operators of the next level, i.e. operators with $n+1$ fields. We would like to generalize this result to operators with an arbitrary number of fields. Indeed it is natural to guess that for $l\geq 0$ the following relation holds
{\setlength\arraycolsep{2pt}
\bea 
\gamma^S_{n+l} S_{i_1\cdots i_{n+l}} &=& (c_{n,n-1}+c_{n,n}+\cdots+c_{n,n-1+l}) V_{j_1\cdots j_n(i_1\cdots i_n}S_{i_{n+1}\cdots i_{n+l})j_1\cdots j_n}
\nn\\
&=& \frac{l+1}{n} c_{n,n+l} V_{j_1\cdots j_n(i_1\cdots i_n}S_{i_{n+1}\cdots i_{n+l})j_1\cdots j_n} \label{gamma-kc}\,,
\label{gamma_gen}
\eea}%
where in the second line we have used 
\be 
\sum_{i=0}^l c_{n,n-1+i} = \frac{l+1}{n} c_{n,n+l},
\ee
which, using the definition \eqref{cnk}, is equivalent to the following identity which can be easily checked
\be 
\sum_{i=0}^l \frac{(n-1+i)!}{i!} = \frac{(l+n)!}{n\,l!}  = \frac{l+1}{n} \frac{(l+n)!}{(l+1)!}\,.
\ee
The relation~\eqref{gamma_gen} can be proved by induction. Having shown the $l=0$ case \eqref{gamma-n} and assuming the above formula is true, Eq.~\eqref{rr} for $k=n+l$ gives
{\setlength\arraycolsep{2pt}
\bea
\gamma^S_{n+l+1} S_{ii_1\cdots i_{n+l}}S_{i_1\cdots i_{n+l}} &=& \gamma^S_{n+l} S_{ii_1\cdots i_{n+l}}S_{i_1\cdots i_{n+l}} + c_{n,n+l} V_{ii_1\cdots i_nj_1\cdots j_{n-1}}S_{j_1\cdots j_{n-1}l_1\cdots l_{l+1}}S_{l_1\cdots l_{l+1}i_1\cdots i_n} 
\nn\\[2mm]
&=& c_{n,n+l} \frac{l+1}{n} V_{j_1\cdots j_ni_1\cdots i_n}S_{i_{n+1}\cdots i_{n+l}j_1\cdots j_n}S_{ii_1\cdots i_{n+l}} 
\nn\\
&+& c_{n,n+l} V_{ii_1\cdots i_nj_1\cdots j_{n-1}}S_{j_1\cdots j_{n-1}l_1\cdots l_{l+1}}S_{l_1\cdots l_{l+1}i_1\cdots i_n} 
\nn\\[2mm]
&=& c_{n,n+l} \frac{l+1}{n} V_{i_1\cdots i_nj_1\cdots j_n}S_{i_1\cdots i_{n+l}}S_{ij_1\cdots j_n i_{n+1}\cdots i_{n+l}} 
\nn\\
&+& c_{n,n+l} V_{il_1\cdots l_ni_1\cdots i_{n-1}}S_{i_1\cdots i_{n+l}}S_{i_n\cdots i_{n+l}l_1\cdots l_n}\,,
\eea}%
where again in the last step indices have been conveniently renamed in order to make $S_{i_1\cdots i_{n+l}}$ appear in all terms as that on the l.h.s. Dropping this tensor and setting $i=i_{n+l+1}$ we get
{\setlength\arraycolsep{2pt}
\bea
\gamma^S_{n+l+1} S_{i_1\cdots i_{n+l+1}}
&=& c_{n,n+l} \frac{l+1}{n} V_{j_1\cdots j_n(i_1\cdots i_n}S_{i_{n+1}\cdots i_{n+l}) i_{n+l+1}j_1\cdots j_n} 
\nn\\
&+& c_{n,n+l} V_{i_{n+l+1}l_1\cdots l_n(i_1\cdots i_{n-1}}S_{i_n\cdots i_{n+l})l_1\cdots l_n}\,,
\eea}%
where some necessary symmetrizations have been introduced.
We can now manipulate this expression further to show that the r.h.s is symmetric in $i_1\cdots i_{n+l+1}$
{\setlength\arraycolsep{2pt}
\bea
\gamma^S_{n+l+1} S_{i_1\cdots i_{n+l+1}}\hspace{-1.5cm}&{}&\nn\\
&=& \frac{1}{n} c_{n,n+l} \left((l+1) V_{j_1\cdots j_n(i_1\cdots i_n}S_{i_{n+1}\cdots i_{n+l}) i_{n+l+1}j_1\cdots j_n} 
+n V_{i_{n+l+1}l_1\cdots l_n(i_1\cdots i_{n-1}}S_{i_n\cdots i_{n+l})l_1\cdots l_n}\right)
\nn\\
&=& \frac{1}{n} c_{n,n+l} \frac{(n+l+1)!}{(n+l)!}  V_{j_1\cdots j_n(i_1\cdots i_n}S_{i_{n+1}\cdots i_{n+l} i_{n+l+1})j_1\cdots j_n} 
\nn\\
&=& \frac{n+l+1}{n} c_{n,n+l}  V_{j_1\cdots j_n(i_1\cdots i_n}S_{i_{n+1}\cdots i_{n+l} i_{n+l+1})j_1\cdots j_n} 
\nn\\
&=& \frac{l+2}{n} c_{n,n+l+1}  V_{j_1\cdots j_n(i_1\cdots i_n}S_{i_{n+1}\cdots i_{n+l} i_{n+l+1})j_1\cdots j_n} \,.
\eea}%
This completes our induction reasoning and proves \eqref{gamma-kc}, which can be written more explicitly also as
\be  
\boxed{\gamma^S_{n+l} S_{i_1\cdots i_{n+l}} = \frac{(n-1)c^{n-1}}{2n!^2} \frac{(n+l)!}{l!}\, V_{j_1\cdots j_n(i_1\cdots i_n}S_{i_{n+1}\cdots i_{n+l})j_1\cdots j_n} \label{gamma-k}\,.}
\ee
Solving this eigenvalue equation one can find both the scaling operators and their anomalous dimensions at leading order, $O(\epsilon)$ in this case, for any $l\ge 0$.

A comment is in order here. The starting point of the derivation of \eqref{gamma-k} was to use the structure of the three-point function 
\be
\langle \phi_i(x) \, \cS_{k}(y) \cS_{k+1}(z) \rangle\,
\ee
and apply one box operator $\Box_x$ to it. In fact for $k=2n-2$ and $k=2n-1$ the structure that we have used for this three-point function may not be valid because in each case one of the operators $\cS_{k}$ or $\cS_{k+1}$ present in the three-point function can be a descendant and not a primary operator. This means that the recurrence relation \eqref{rr} might be not necessarily valid for $k=2n-2$ and $k=2n-1$, so that Eq.~\eqref{gamma-k} is valid at least up to $l=n-2$ from which one can obtain the anomalous dimension
$\gamma^S_{2n-2}$. However, as we will see later the validity of the recurrence relation \eqref{rr} for $k=2n-2$ and $k=2n-1$ can be shown in a different way, implying that \eqref{gamma-k} is true for all $l\geq 0$.  

In fact, in the two cases $k=2n-2$ and $k=2n-1$, when the operator $\cS_{2n-1}$ is a descendant operator, one can instead use the e.o.m to write the quantity \eqref{box-1kk1} as
{\setlength\arraycolsep{2pt}
\bea  
&& \Box_x \Box_y \langle \phi^i(x)\phi^j(y)\cS_{2n-2}(z) \rangle, \qquad k=2n-2  \label{bbijS1} \\[1mm]
&& \Box_x \Box_y \langle \phi^i(x)\phi^j(y)\cS_{2n}(z) \rangle, \hspace{12.9mm} k=2n-1 \label{bbijS2}\,.
\eea}%
The first one \eqref{bbijS1} gives another eigenvalue equation for $\gamma^S_{2n-2}$ which we already have from \eqref{gamma-k}. Comparing the two gives the functional fixed point equation for all multi-field multicritical even models. The details of this is given in Section~\ref{ss:gem}. 

The anomalous dimensions $\gamma^S_{2n-1}$ corresponding to descendant operators are obtained from 
an identity relating the scaling dimensions of the descendant operators to those of the fields $\phi^i$, and can be shown to satisfy \eqref{gamma-k} for $k=2n-2$. Finally, the second equation \eqref{bbijS2} gives the missing eigenvalue equation for $\gamma^S_{2n}$.   
In other words we will obtain a relation for $\gamma^S_{2n}$ where instead of the tensor $S_{i_1\cdots i_{2n-1}}$  the descendent structure $V_{i i_1\cdots i_{2n-1}}$ is present with the corresponding anomalous dimension. 
This is exactly what is needed to complete the space of composite operators of order $2n-1$ in the recurrence relation~\eqref{rr}.
These will be discussed in detail in Section~\ref{ss:mprr}.

\subsection{Structure constants} \label{ssec:ope}

Apart from the anomalous dimensions of the quadratic and higher order operators that we have discussed so far, conformal symmetry along with the equations of motion provide information on several classes of leading order structure constants. This has been shown in the single-field case in \cite{Codello:2017qek}.  It is straightforward to extend the computation of structure constants of single-field theories 
to the multi-field case. In this section we make such a generalization and provide compact formulas for some structure constants in multicritical and multi-field even or odd models. 

\subsubsection{Generalization of $C_{1,2p,2q-1}$} \label{ssec:c12p2q-1}

Consider for instance general even models for which the multicriticality label $n$ is an integer. 
Several set of structure constants had been computed for the single-field case in \cite{Codello:2017qek}, for example 
\be \label{c12p2q-single}
C_{1,2p,2q-1} = \frac{g}{(2n-1)!} \frac{(n-1)^2}{4(p-q)(p-q+1)}C^{\mathrm{free}}_{2n-1,2p,2q-1},
\ee
which is valid in the range $q+p \geq n$, $q-p \geq 1-n$ and $q-p \neq 0,1$. 
By arguments similar to those of the previous section this can be straightforwardly generalized to the multi-field case as 
{\setlength\arraycolsep{2pt}
\bea \label{c12p2q-1}
C_{i,j_1\cdots j_{2p},k_1\cdots k_{2q-1}} &=& \frac{V_{ii_1\cdots i_{2n-1}}}{(2n-1)!} \frac{(n-1)^2}{4(p-q)(p-q+1)}C^{\mathrm{free}}_{i_1\cdots i_{2n-1},j_1\cdots j_{2p},k_1\cdots k_{2q-1}} \nn \\
&=&  \frac{V_{ii_1\cdots i_{2n-1}}}{(2n-1)!} \frac{(n-1)^2}{4(p-q)(p-q+1)} C^{\mathrm{free}}_{2n-1,2p,2q-1} \nn\\
&\times & (\delta^{i_1}_{j_{s+1}} \cdots \delta^{i_r}_{j_{2p}} \, \delta^{j_1}_{k_{t+1}}  \cdots \delta^{j_s}_{k_{2q-1}} \, \delta^{k_1}_{i_{r+1}} \cdots \delta^{k_t}_{i_{2n-1}}) \,,
\eea}%
where $q,p$ are constrained as in the single-field case, and the integers $r,s,t$ satisfy the relation
\be 
2n-1= r+t, \quad 2p = r+s, \quad 2q-1 = s+t.
\ee
The parentheses in the third line enclosing the deltas indicate that the $i_l$s, the $j_l$s and the $k_l$s are separately symmetrized.  
To obtain the structure constant that is defined as the coefficient appearing in the three-point function
\be \label{3pf-even}
\left\langle \phi_i(x) \cS_{2p}(y) \tilde{\cS}_{2q-1}(z) \right\rangle 
\ee
one needs to contract the symmetric tensors $S,\tilde{S}$ with \eqref{c12p2q-1} and find
\be  \label{ciuv-even}
\boxed{
C_{\phi_i \cS_{2p} \tilde{\cS}_{2q-1}} = \frac{V_{il_1\cdots l_r k_1\cdots k_t}\,S_{j_1\cdots j_s l_1 \cdots l_r} \, \tilde{S}_{k_1\cdots k_t j_1 \cdots j_s}}{(2n-1)!} \frac{(n-1)^2}{4(p-q)(p-q+1)}C^{\mathrm{free}}_{2n-1,2p,2q-1}.
}
\ee  
Notice that the tensors $S,\tilde{S}$ in \eqref{3pf-even} must be chosen such that the corresponding operators have a definite scaling, i.e. they satisfy \eqref{gamma-k}. They also must not be descendant operators, which can only occur for $\tilde{S}_{2n-1}$, that is when $q=n$.

The free theory structure constant in the above equation is proportional to $c^{n+p+q-1}$ according to the general formula (A.8) of \cite{Codello:2017qek}. The CFT normalization requires rescaling the fields $\phi = \sqrt{c}\hat\phi$ such that the two point function of $\phi$ are normalized to unity. With this normalization the $c$ factors in the free structure constants are removed and, defining $\hat V(\hat\phi)=V(\phi)$, in the new equation of motion a factor of $c^{-1}$ will appear on the r.h.s as if $V\rightarrow V/c$ in \eqref{eom}. Also, in terms of the rescaled field the $2n$th field derivative of the potentials will be $c^n$ times the original one. More explicitly
\be 
\Box \hat \phi_i = \frac{1}{c} \hat V_i, \qquad
\hat V_{i_1\cdots i_{2n}} = c^n V_{i_1\cdots i_{2n}}.
\ee
Combining these two we find that in the CFT normalization, dropping the hat on the new fields and potential, one must remove the $c$ factors in the free structure constants and make the replacement $V_{i_1\cdots i_{2n}}\rightarrow c^{n-1} V_{i_1\cdots i_{2n}}$. We shall make such a choice of normalization in Section~\ref{sect:potts-cft}, when studying some Potts models,
but only when we give explicit $\epsilon$ dependent expressions for the structure constants. 

\subsubsection{Generalization of $C_{1,2p,2q}$ and $C_{1,2p-1,2q-1}$} \label{ssec:c12p2q}

Consider now general odd models which include the cubic and quintic models that we are especially interested in and write the half-odd multicriticality label as $n=\ell+1/2$ where $\ell$ is an integer. 
For scalar theories with a single degree of freedom the following structure constant was computed in \cite{Codello:2017qek}
\be \label{c12p2q-single-2}
C_{1,2p,2q} = \frac{g}{(2\ell)!} \frac{(2\ell-1)^2}{4(4(p-q)^2-1)}C^{\mathrm{free}}_{2\ell,2p,2q},
\ee
which is valid only in the range $q+p\geq \ell$ and $|q-p|\leq \ell$. In the multi-field case this generalizes to 
{\setlength\arraycolsep{2pt}
\bea \label{c12p2q}
C_{i,j_1\cdots j_{2p},k_1\cdots k_{2q}} &=& \frac{V_{ii_1\cdots i_{2\ell}}}{(2\ell)!} \frac{(2\ell-1)^2}{4(4(p-q)^2-1)}C^{\mathrm{free}}_{i_1\cdots i_{2\ell},j_1\cdots j_{2p},k_1\cdots k_{2q}} \\
&=& \frac{V_{ii_1\cdots i_{2\ell}}}{(2\ell)!} \frac{(2\ell-1)^2}{4(4(p-q)^2-1)}C^{\mathrm{free}}_{2\ell,2p,2q}(\delta^{i_1}_{j_{s+1}} \cdots \delta^{i_r}_{j_{2p}} \, \delta^{j_1}_{k_{t+1}}  \cdots \delta^{j_s}_{k_{2q}} \, \delta^{k_1}_{i_{r+1}} \cdots \delta^{k_t}_{i_{2\ell}}) \nn
\eea}%
with $q,p$ constrained as in the single-field case and where the integers $r,s,t$ satisfy the relation
\be 
2\ell= r+t, \quad 2p = r+s, \quad 2q = s+t.
\ee
The structure constant defined as the coefficient of the three point function
\be \label{3pf}
\left\langle \phi_i(x) \cS_{2p}(y) \tilde{\cS}_{2q}(z) \right\rangle
\ee
is obtained by contracting \eqref{c12p2q} with $S,\tilde S$, which are symmetric tensors that satisfy \eqref{gamma-k} and therefore give rise to scaling operators. This leads to
\be  \label{ci2p2q}
\boxed{
C_{\phi_i \cS_{2p} \tilde{\cS}_{2q}}  = \frac{V_{il_1\cdots l_r k_1\cdots k_t}\,S_{j_1\cdots j_s l_1 \cdots l_r} \, \tilde{S}_{k_1\cdots k_t j_1 \cdots j_s}}{(2\ell)!} \frac{(2\ell-1)^2}{4(4(p-q)^2-1)}C^{\mathrm{free}}_{2\ell,2p,2q}.
}
\ee 
Notice also that these operators must not be descendants. This can only occur for $\cS_{2\ell}$ and $\tilde\cS_{2\ell}$, that is when $q=\ell$. 
Finally, the structure constants $C_{1,2p-1,2q-1}$ and consequently its multi-field generalization are given respectively by \eqref{c12p2q-single} and \eqref{ci2p2q} after making the shift $p\rightarrow p-\frac{1}{2}$ and $q\rightarrow q-\frac{1}{2}$, 
that is, one can immediately write
\be  \label{ci2p-12q-1}
\boxed{
C_{\phi_i \cS_{2p-1} \tilde{\cS}_{2q-1}} = \frac{V_{il_1\cdots l_r k_1\cdots k_t}\,S_{j_1\cdots j_s l_1 \cdots l_r} \, \tilde{S}_{k_1\cdots k_t j_1 \cdots j_s}}{(2\ell)!} \frac{(2\ell-1)^2}{4(4(p-q)^2-1)}C^{\mathrm{free}}_{2\ell,2p-1,2q-1},
}
\ee 
where now $q,p$ fall in the range $q+p\geq \ell+1$ and $|q-p|\leq \ell$, and the integers $r,s,t$ satisfy the relation
\be 
2\ell= r+t, \quad 2p-1 = r+s, \quad 2q-1 = s+t.
\ee

\subsubsection{Generalization of $C_{1,1,1}$} \label{ssec:c111}

Again, for odd models where $n=\ell+1/2$ it is straightforward to find the generalization of the OPE coefficient $C_{1,1,1}$ to the multi-field case. Excluding the case $\ell=1$ which requires a separate treatment and can be extracted from the computation of the previous subsection setting $p=q=1$ inside Eq.~\eqref{ci2p-12q-1}, for all other values of $\ell$ this is obtained by evaluating the following expressions at leading order
\be 
\square_x\square_y\square_z\langle\phi_i(x)\phi_j(y)\phi_k(z)\rangle \,\stackrel{\mathrm{LO}}{=}\, \frac{2^8\ell(\ell-1)}{(2\ell-1)^6} \frac{C_{\phi_i \phi_j \phi_k}}{|x-y|^{\delta_c+2}|x-z|^{\delta_c+2}|y-z|^{\delta_c+2}},
\ee
\bea
\langle\square_x\phi_i(x)\square_y\phi_j(y)\square_z\phi_k(z)\rangle &\,\stackrel{\mathrm{LO}}{=}&\,\frac{V_{i\,a_1\cdots a_\ell\,b_1\cdots b_\ell}V_{j\,b_1\cdots b_\ell\,c_1\cdots c_\ell}V_{k\,c_1\cdots c_\ell\,a_1\cdots a_\ell}}{(2\ell)!^3}\times \nonumber\\
&{}&\frac{C^\mathrm{free}_{2\ell,\,2\ell,\,2\ell}}{|x\!-\!y|^{2\ell\delta_c}|x\!-\!z|^{2\ell\delta_c}|y\!-\!z|^{2\ell\delta_c}},
\eea
and equating the two, recalling that $2\ell\delta_c = \delta_c + 2$. This gives
\be \label{cijk}
\boxed{
C_{\phi_i \phi_j \phi_k} = \frac{(2\ell-1)^6c^{3\ell}}{2^8\ell(\ell-1)\ell!^3}\,V_{i\,a_1\cdots a_\ell\,b_1\cdots b_\ell}V_{j\,b_1\cdots b_\ell\,c_1\cdots c_\ell}V_{k\,c_1\cdots c_\ell\,a_1\cdots a_\ell},
}
\ee
where we have directly used $C^\mathrm{free}_{2\ell,\,2\ell,\,2\ell} =(2\ell)!^3c^{3\ell}/\ell!^3$.
%

\subsubsection{Generalization of $C_{1,1,2k}$} \label{ssec:c112k}

The generalization of the structure constant $C_{1,1,2k}$ to the multi-field case is defined as the coefficient appearing in the three-point correlation function
\be 
\left\langle \phi_i(x) \phi_j(y) \cS_{2k}(z) \right\rangle
\ee
where the operator $\cS_{2k}$ with $2k$ fields is 
a scaling operator satisfying \eqref{gamma-k}. Using the result of \cite{Codello:2017qek} and following the arguments of the previous sections this is straightforwardly calculated
\begin{empheq}[box=\fbox]{align}
C_{\phi_i \phi_j \cS_{2k}} &= \frac{(n-1)^4c^{2n+k-1}}{16k(k-1)(k-n)(k-n+1)} \frac{(2k)!}{k!^2(2n-k-1)!^2} \,\,\nn\\
&\times  V_{ii_1\cdots i_{2n-k-1}a_1\cdots a_k}V_{ji_1\cdots i_{2n-k-1}b_1\cdots b_k}S_{a_1\cdots a_k \,b_1\cdots b_k}. \label{ciju}
\end{empheq}
As in the single-field case, in this equation $n$ is either an integer or a half-odd number and $k$ is constrained to the range $2\leq k \leq 2n-1$ and $k\neq n,n-1$.

\subsection{"Fixed point" equation from CFT}\label{sec:FPcond}
We conclude this section showing in general how the constraints imposed by conformal symmetry on two and three point functions together with the use of the Schwinger-Dyson equations can fix the possible critical theories at leading order in $\epsilon$.
We shall follow a path which is slightly different from the one employed in~\cite{Nii:2016lpa, Codello:2017qek}, and do not directly rely on the conditions on the scaling dimensions of descendant operators from the equation of motion (when the interactions are turned on below the critical dimension).
Interestingly enough we find conditions which can be simplified to match exactly the fixed point condition of the RG approach in its functional form~\cite{ODwyer:2007brp, Codello:2017hhh,Codello:2017epp, Osborn:2017ucf, CSVZ4} which we dubbed functional perturbative RG approach. It is well known that in general, fixed point equations admit solutions which are characterized by some internal symmetries not necessarily realized away from criticality, giving a scenario where critical theories can have a higher level of symmetry, or an emergent symmetry. Therefore all the discussions in the literature with RG techniques regarding possible symmetry enhancements at criticality~\cite{Zia:1974nv,Michel:1983in, TMTB,Osborn:2017ucf} are directly applicable also in this CFT perturbative framework, at least in the cases shown below, i.e. all unitary multicritical models and the one with a cubic potential.

\subsubsection{The $d_c=6$ case} \label{ss:fp-cubic}

The only odd model that we are able to analyze in this respect is the one corresponding to $n=3/2$. Let us consider the three-point function of the scaling fields $\phi_i$, which takes the following form
\be 
\langle \phi_i(x) \phi_j(y) \phi_k(z)\rangle = \frac{C_{\phi_i\phi_j\phi_k}}{|x-y|^{\Delta_i+\Delta_j-\Delta_k}|y-z|^{\Delta_j+\Delta_k-\Delta_i}|x-z|^{\Delta_i+\Delta_k-\Delta_j}}  , \quad C_{\phi_i\phi_j\phi_k} = -\frac{c^2}{4} V_{ijk},
\ee
where the structure constant has been computed by setting $m=p=q=1$ in \eqref{ci2p-12q-1}. 
Acting with three Laplacians on the general scaling form one obtains at leading order
\be 
\Box_x\Box_y\Box_z\langle \phi_i(x)\phi_j(y)\phi_k(z)\rangle \stackrel{\mathrm{LO}}{=} \frac{32(\epsilon\!-\!2(\gamma_i\!+\!\gamma_j\!+\!\gamma_k))}{|x\!-\!y|^4|y\!-\!z|^4|x\!-\!z|^4}C_{\phi_i\phi_j\phi_k} = \frac{8c^2(2(\gamma_i\!+\!\gamma_j\!+\!\gamma_k)\!-\!\epsilon)}{|x\!-\!y|^4|y\!-\!z|^4|x\!-\!z|^4} V_{ijk},
\label{eqcub3box2}
\ee
while using the SDE one gets
{\setlength\arraycolsep{2pt}
\bea
\langle \Box_x\phi_i(x) \Box_y\phi_j(y) \Box_z\phi_k(z)\rangle &\stackrel{\mathrm{LO}}{=}&
\frac{1}{2!^3} V_{iab}V_{jcd}V_{kef} \langle [\phi_a\phi_b](x) [\phi_c\phi_d](y)[\phi_e\phi_f](z)\rangle 
\nn\\ 
&=&\frac{c^3\, V_{iab} V_{jbc}V_{kca}}{|x-y|^4|y-z|^4|x-z|^4}.
\label{eqcub3box1}
\eea}%
Equating the expressions on the right hand side of Eqs.~\eqref{eqcub3box2} and~\eqref{eqcub3box1} one finds
\be \label{fp-6}
\boxed{
8(2(\gamma_i+\gamma_j+\gamma_k)-\epsilon) V_{ijk} = c\, V_{iab} V_{jbc} V_{kca}.
}
\ee
Making the replacement $V\rightarrow 8 V/\sqrt{c}$ to accord with our RG conventions this becomes
\be 
(2(\gamma_i+\gamma_j+\gamma_k)-\epsilon)V_{ijk} = 8\, V_{iab} V_{jbc} V_{kca}.
\label{CFTd6cond}
\ee
One can easily verify that this is nothing but the functional fixed point equation obtained from RG, written in the diagonal basis. To do this let us give a look at the leading order, i.e. cubic, beta function 
which in terms of dimensionless variables becomes
\be 
\beta_v = -d v +\frac{d-2}{2}\phi_i v_i + \phi_i \gamma_{ij} v_j-\frac{2}{3} v_{ij} v_{jl} v_{li}\,.
\ee
Taking the third field-derivative and setting the result to zero we get the fixed point equation
\be 
0 = -d v_{ijk} +3\frac{d-2}{2} v_{ijk} + \gamma_{ia} v_{ajk}+\gamma_{ja} v_{aik}+\gamma_{ka} v_{aij}-4 v_{iab} v_{jbc} v_{kca}\,.
\ee
Setting $d=6-\epsilon$  one has
\be 
0 = -\frac{\epsilon}{2} v_{ijk} + \gamma_{ia} v_{ajk}+\gamma_{ja} v_{aik}+\gamma_{ka} v_{aij}-4 v_{iab} v_{jbc} v_{kca}\,,
\ee
which, in the diagonal basis where $\gamma_{ij}\rightarrow \gamma_i\delta_{ij}$, matches the condition found in Eq.~\eqref{CFTd6cond}.

\subsubsection{The $d_c=4$ case} 

Suppose that $S_{ij}\phi_i\phi_j$ is a scaling operator with anomalous dimension $\gamma^S_2$. We have seen that in $d_c=4$ quite generally the matrix $S_{ij}$ satisfies the eigenvalue equation \eqref{crit-cft-2}. We can write on the one hand
\be  
\square_x\square_y \langle \phi_i(x) \,\phi_j(y)\, [S_{kl}\phi_k\phi_l](z) \rangle
\stackrel{\mathrm{LO}}{=} C^{\mathrm{free}}_{112} \,  \frac{4\gamma^S_2(\epsilon-\gamma^S_2)}{|x-y|^{4}|x-z|^{2}|y-z|^{2}} S_{ij},
\ee
while on the other hand, using the SDE, we obtain
{\setlength\arraycolsep{2pt}
\bea 
\langle \square_x\phi_i(x) \,\square_y\phi_j(y)\, [S_{kl}\phi_k\phi_l](z) \rangle
&=& \frac{1}{3!^2}V_{iklm} V_{jpqr}\langle [\phi_k\phi_l\phi_m](x) \,[\phi_p\phi_q\phi_r](y)\, [S_{kl}\phi_k\phi_l](z) \rangle \nn\\
&\stackrel{\mathrm{LO}}{=}& \frac{C^{\mathrm{free}}_{332}}{3!^2} \,  \frac{V_{pqik} V_{pqjl} S_{kl}}{|x-y|^{4}|x-z|^{2}|y-z|^{2}}.
\eea}%
Taking into account that $C^{\mathrm{free}}_{332} = 3!^2c^4$ and $C^{\mathrm{free}}_{112} = 2c^2$, 
one then finds
\be 
c^4 V_{pqik} V_{pqjl} S_{kl} = 8c^2\gamma^S_2(\epsilon-\gamma^S_2) S_{ij}.
\ee
In order to rewrite such condition in a simpler form, we can perform some manipulations, eliminating the explicit dependence on the anomalous dimension of the quadratic operators using \eqref{crit-cft-2}
{\setlength\arraycolsep{2pt}
\bea
c^4 V_{pqik} V_{pqjl} S_{kl} &=& 8c^2\left[\epsilon\frac{c}{4} V_{ijkl}-\gamma^S_2\frac{c}{4} V_{ijkl}\right] S_{kl} \nn\\
&=& 8c^2\left[\epsilon\frac{c}{4} V_{ijkl}-\left(\frac{c}{4}\right)^2 V_{ijpq} V_{pqkl}\right] S_{kl} \,.
\eea}%
The r.h.s is now independent of $S_{kl}$, and the equation is valid for all $S_{kl}$, so it is valid for an arbitrary symmetric matrix. One can therefore drop $S_{kl}$ and symmetrize the factor in $kl$. Simplifying the result one obtains
\be \label{fp-4}
\boxed{
V_{pqik}V_{ljpq} + V_{pqil}V_{kjpq} + V_{ijpq} V_{pqkl} -\epsilon\,\frac{4}{c}\, V_{ijkl} =0\,.}
\ee
Making the rescaling $V\rightarrow 4V/c$ to match the RG normalization removes the $c/4$ factors and we finally get
\be \label{fp-cft-4}
V_{pqik} V_{ljpq}+ V_{pqil} V_{kjpq} + V_{ijpq} V_{pqkl} -\epsilon V_{ijkl} =0\,,
\ee
which is nothing but the functional fixed point equation from RG. 
Indeed recalling the functional perturbative RG beta function for the potential at leading order, 
written in dimensionaless variables
\be 
\beta_v = -d v +\frac{d-2}{2}\phi_i v_i 
+\frac{1}{2} v_{ij} v_{ij}\,
\ee
and taking four field derivatives, one obtains
\be  
0 = -d v_{ijkl} +2(d-2) v_{ijkl} + 
v_{pqij} v_{pqkl}+v_{pqik} v_{pqjl}+v_{pqil} v_{pqjk}\,.
\ee
Setting $d=4-\epsilon$ then
\be \label{fp-rg-4} 
v_{pqij} v_{pqkl}+v_{pqik} v_{pqjl}+v_{pqil} v_{pqjk} 
-\epsilon v_{ijkl}=0\,.
\ee
in agreement with the result \eqref{fp-cft-4} from conformal symmetry.

\subsubsection{General even models} \label{ss:gem}

We pointed out in Section~\ref{ss:rr} that the anomalous dimension $\gamma^S_{2n-2}$ which is given by the solution \eqref{gamma-k} to the recurrence relations \eqref{rr} can be obtained also in a different way, and that the consistency of the two results gives rise to the fixed point conditions for all multicritical even models. In this section we show this in detail. The first step is to consider the multi-field generalization of the structure constant $C_{1,2p,2q-1}$ given in \eqref{ciuv-even}. For $p=n-1$ and $q=1$ this is obtained by applying $\Box_x$ to the following three-point function 
\be 
\left\langle \phi_i(x)\phi_j(y) \cS_{2n-2}(z)\right\rangle, 
\ee
and evaluating it once by direct application of $\Box_x$ and once by using the e.o.m. For this case, in the notation of Eq.~\eqref{ciuv-even} we have
\be 
\left\lbrace
\ba{l}
t+r=2n-1 \\
r+s = 2n-2 \\
s+t = 1 \\
\ea 
\right.
\qquad \Rightarrow \qquad
\left\lbrace
\ba{l}
r= 2n-2 \\
s= 0 \\
t= 1 \\
\ea 
\right.
\ee
which leads to the following expression 
{\setlength\arraycolsep{2pt}
\bea  
C_{\phi_i\phi_j\mathcal{S}_{2n-2}} &=& \frac{(n-1)^2}{(2n-1)!} \frac{C^{\mathrm{free}}_{2n-1,2n-2,1}}{4(n-2)(n-1)} V_{il_1\cdots l_{2n-2}j} S_{l_1\cdots l_{2n-2}}
\nn\\
&=& \frac{(n-1)c^{2n-1}}{4(n-2)} V_{ijl_1\cdots l_{2n-2}} S_{l_1\cdots l_{2n-2}} \,,
\eea}%
where we have used $C^{\mathrm{free}}_{2n-1,2n-2,1} = (2n-1)! c^{2n-1}$.
Let us now apply two boxes to the above three-point function as suggested in Section \ref{ss:rr}. The result coming from the use of SDE is
{\setlength\arraycolsep{2pt}
\bea  
&& \left\langle \Box_x \phi_i(x)\Box_y\phi_j(y) [S_{l_1\cdots l_{2n-2}} \phi_{l_1}\!\cdots \phi_{l_{2n-2}}](z)\right\rangle
\nn\\
&=& \frac{1}{(2n-1)!^2} V_{ii_1\cdots i_{2n-1}}V_{jj_1\cdots j_{2n-1}} \left\langle [\phi_{i_1} \!\cdots \phi_{i_{2n-1}}](x)[\phi_{j_1} \!\cdots \phi_{j_{2n-1}}](y) [S_{l_1\cdots l_{2n-2}} \phi_{l_1}\!\cdots \phi_{l_{2n-2}}](z)\right\rangle 
\nn\\
&=& \frac{1}{(2n-1)!^2} V_{ii_1\cdots i_{n-1}j_1\cdots j_n}V_{jj_1\cdots j_nl_1\cdots l_{n-1}}S_{l_1\cdots l_{n-1}i_1\cdots i_{n-1}} \frac{C^{\mathrm{free}}_{2n-1,2n-1,2n-2}}{|x-y|^{2\delta_c+2}|y-z|^2|z-x|^2}
\nn\\
&=& \frac{(2n-2)!}{(n-1)!^2n!} V_{ii_1\cdots i_{n-1}j_1\cdots j_n}V_{jj_1\cdots j_nl_1\cdots l_{n-1}}S_{l_1\cdots l_{n-1}i_1\cdots i_{n-1}} \frac{c^{3n-2}}{|x-y|^{2\delta_c+2}|y-z|^2|z-x|^2},
\eea}%
where we have used
\be 
C^{\mathrm{free}}_{2n-1,2n-1,2n-2} = \frac{(2n-1)!^2(2n-2)!}{(n-1)!^2n!} c^{3n-2}
\ee
and the counting of the indices in the third line comes from
\be 
\left\lbrace
\ba{l}
r+s = 2n-1 \\
s+t = 2n-1 \\
t+r = 2n-2 \\
\ea 
\right.
\qquad \Rightarrow \qquad
\left\lbrace
\ba{l}
r = n-1 \\
s = n \\
t = n-1 \\
\ea 
\right.
\ee
On the other hand, direct application of the boxes gives
{\setlength\arraycolsep{2pt}
\bea  
\Box_x \Box_y \left\langle \phi_i(x)\phi_j(y) \cS_{2n-2}(z)\right\rangle \hspace{-2cm} && \nn\\
&=& 
\Box_x \Box_y \frac{C_{\phi_i\phi_j\mathcal{S}_{2n-2}}}{|x-y|^{(4-2n)\delta-\gamma^S_{2n-2}}|y-z|^{(2n-2)\delta+\gamma^S_{2n-2}}|z-x|^{(2n-2)\delta+\gamma^S_{2n-2}}} 
\nn\\
&=& \frac{8(n-2)}{(n-1)^2} \left((n-1)\epsilon-\gamma^S_{2n-2}\right)\frac{C_{\phi_i\phi_j\mathcal{S}_{2n-2}}}{|x-y|^{2\delta_c+2}|y-z|^2|z-x|^2} \,.
\eea}%
Equating the two results and simplifying a bit leads to 
{\setlength\arraycolsep{2pt}
\bea
&& \frac{2}{n-1}\left((n-1)\epsilon-\gamma_{2n-2}^S\right) V_{ijl_1\cdots l_{2n-2}}S_{l_1\cdots l_{2n-2}} 
\nn\\
&=& \frac{(2n-2)!c^{n-1}}{n!(n-1)!^2} V_{ii_1\cdots i_{n-1}j_1\cdots j_n} V_{j j_1\cdots j_n l_1\cdots l_{n-1}} S_{l_1\cdots l_{n-1}i_1\cdots i_{n-1}}\,.
\eea}%
One can now eliminate $\gamma_{2n-2}^S$ using the formula \eqref{gamma-k} for $l=n-2$. This gives
{\setlength\arraycolsep{2pt}
\bea
(n-1)\epsilon\, V_{ijl_1\cdots l_{2n-2}}S_{l_1\cdots l_{2n-2}} &=&  V_{ijl_1\cdots l_{2n-2}} \frac{n-1}{n} c_{n,2n-2} V_{j_1\cdots j_n(l_1\cdots l_n} S_{l_{n+1}\cdots l_{2n-2})j_1\cdots j_n} 
\nn\\
&+& \frac{(2n-3)!c^{n-1}}{n!(n-2)!^2} V_{ii_1\cdots i_{n-1}j_1\cdots j_n} V_{j j_1\cdots j_n l_1\cdots l_{n-1}} S_{l_1\cdots l_{n-1}i_1\cdots i_{n-1}}
\nn\\[2mm]
&=& c_{n,2n-2}\frac{n-1}{n} V_{ijl_1\cdots l_{2n-2}} V_{j_1\cdots j_nl_1\cdots l_n} S_{l_{n+1}\cdots l_{2n-2}j_1\cdots j_n} 
\nn\\
&+& c_{n,2n-2} V_{ii_1\cdots i_{n-1}j_1\cdots j_n} V_{j j_1\cdots j_n l_1\cdots l_{n-1}} S_{l_1\cdots l_{n-1}i_1\cdots i_{n-1}}
\nn\\[2mm]
&=& c_{n,2n-2}\frac{n-1}{n} V_{ijj_1\cdots j_nl_{n+1}\cdots l_{2n-2}} V_{j_1\cdots j_nl_1\cdots l_n} S_{l_1\cdots l_{2n-2}} 
\nn\\
&+& c_{n,2n-2} V_{il_n\cdots l_{2n-2}j_1\cdots j_n} V_{j j_1\cdots j_n l_1\cdots l_{n-1}} S_{l_1\cdots l_{2n-2}}\,,
\eea}%
where in the second equation we have used the relation
\be 
c_{n,2n-2} = \frac{c^{n-1}}{2(n-2)!n!} \frac{(2n-2)!}{(n-1)!} = \frac{(2n-3)!c^{n-1}}{n!(n-2)!^2}
\ee
and in the third equality we have renamed the dummy indices to make the indices on $S_{l_1\cdots l_{2n-2}}$ the same as those on the l.h.s.  Now we can drop $S_{l_1\cdots l_{2n-2}}$ from both sides and symmetrize the remaining tensors in $l_1\cdots l_{2n-2}$
{\setlength\arraycolsep{2pt}
\bea
(n-1)\epsilon\, V_{ijl_1\cdots l_{2n-2}} &=& c_{n,2n-2} \frac{n-1}{n} V_{j_1\cdots j_nij(l_{n+1}\cdots l_{2n-2}} V_{l_1\cdots l_n)j_1\cdots j_n} 
\nn\\
&+& c_{n,2n-2} V_{j_1\cdots j_ni(l_n\cdots l_{2n-2}} V_{l_1\cdots l_{n-1})j j_1\cdots j_n} 
\nn\\[3mm]
&=& c_{n,2n-2}\frac{1}{2n^2}\, 2n(n-1) V_{j_1\cdots j_nij(l_{n+1}\cdots l_{2n-2}} V_{l_1\cdots l_n)j_1\cdots j_n} 
\nn\\
&+& c_{n,2n-2}\frac{1}{2n^2}\, 2n^2 V_{j_1\cdots j_ni(l_n\cdots l_{2n-2}} V_{l_1\cdots l_{n-1})j j_1\cdots j_n} \nn\\[3mm]
&=& c_{n,2n-2}\frac{1}{2n^2}\, \frac{(2n)!}{(2n-2)!} V_{j_1\cdots j_n(ijl_{n+1}\cdots l_{2n-2}} V_{l_1\cdots l_n)j_1\cdots j_n} 
\nn\\[3mm]
&=& \frac{(n-1)(2n)!c^{n-1}}{4n!^3} V_{j_1\cdots j_n(ijl_{n+1}\cdots l_{2n-2}} V_{l_1\cdots l_n)j_1\cdots j_n} \,.
\eea}%
With a simple manipulation we have seen that the tensor on the r.h.s is not only symmetric in $l_1\cdots l_{2n-2}$ but also in $ijl_1\cdots l_{2n-2}$ as expected from the l.h.s. 
This leads to the following conditions on the couplings of the potential
\be \label{fp-even}
\boxed{
0 = (1-n)\epsilon\, V_{i_1\cdots i_{2n}} + \frac{(n-1)(2n)!}{4n!^3}c^{n-1} V_{j_1\cdots j_n(i_1\cdots i_n} V_{i_{n+1}\cdots i_{2n})j_1\cdots j_n}.} 
\ee
It is interesting to point out that comparing this fixed point equation with the recurrence relation \eqref{gamma-k} for $l=n$ one immediately notices that $S_{i_1\cdots i_{2n}}=V_{i_1\cdots i_{2n}}$, corresponding to the classically marginal operator 
\be \label{V}
V_{i_1\cdots i_{2n}} \phi_{i_1} \cdots \phi_{i_{2n}},
\ee
is always an eigenvector with eigenvalue
\be  \label{gamma-V-2n}
\gamma^V_{2n} = 2(n-1)\epsilon.
\ee
In RG terms, the dimension of the coupling corresponding to the operator \eqref{V} is therefore given at leading order by $\theta^V_{2n} = d-2n \delta - \gamma^V_{2n} = -(n-1)\epsilon$ which is a negative number, hence indicating that the fixed point is infrared stable along the direction of the operator \eqref{V}. The complete stability analysis of the solutions to \eqref{fp-even} requires solving the eigenvalue equation \eqref{gamma-k} for $l=n$.

In Eq.~\eqref{fp-even} one can also make the rescaling $V\rightarrow 4V/(n-1)c^{n-1}$ as done in RG
\be
0 = (1-n)\epsilon\, V_{i_1\cdots i_{2n}} + \frac{(2n)!}{n!^3} V_{j_1\cdots j_n(i_1\cdots i_n} V_{i_{n+1}\cdots i_{2n})j_1\cdots j_n}\,.
\ee
This is indeed nothing but the functional fixed point equation for a general even model with multicriticality label $n$ derived from RG. It is obtained by taking the $2n$th field derivative of the leading order beta functional 
\be 
\beta_v = -d v +\frac{d-2}{2}\phi_i v_i +\frac{1}{n!} v_{j_i\cdots j_n} v_{j_i\cdots j_n}\,,
\ee
and setting $d=2n/(n-1)-\epsilon$. It might be worth mentioning that the anomalous dimensions of higher order operators obtained in Section~\ref{ss:rr} can be extracted from the above beta function. One simply needs to deform the potential as $v\rightarrow v+ \delta v$ and linearize the r.h.s in the deformation $\delta v$, take $n+l$ field derivatives (where $l\geq 0$), and evaluate the result at the fixed point where only the $2n$th derivative of the potential persists. The tensor coefficient of $\delta v_{i_1\cdots i_{n+l}}$ then gives the stability matrix that coincide with \eqref{gamma-k}, of course after suitable rescaling of the potential. It might be worth mentioning that since $V_{i_1 \cdots i_m}$ constitute the most general set of couplings at criticality the equations \eqref{fp-6}, \eqref{fp-4} and \eqref{fp-even} are completely general at leading order and admit all possible fixed points, while at higher orders in perturbation theory these equations are corrected by higher powers of the interactions $V_{i_1 \cdots i_m}$.

\subsection{The missing pieces in the recurrence relation} \label{ss:mprr}

\subsubsection{The case $k=2n-2$}

As discussed towards the end of Section~\ref{ss:rr}, the reasonings in that section do not justify the validity of \eqref{rr} for $k=2n-2$ when the operator $\cS_{2n-1}$ is a descendant operator. Indeed for the multi-field models, not all operators $\cS_{2n-1}$ are primaries. The descendant ones take the form of the r.h.s of the equation of motion at the critical point
\be 
\Box_x \phi_i = \frac{1}{(2n-1)!} V_{ii_1\cdots i_{2n-1}} \phi_{i_1}\cdots \phi_{i_{2n-1}}\,.
\ee
This means that the descendant operators are a set of $2n-1$ index tensors labeled by $i$
\be \label{do}
\mathcal{V}_i = V_{ii_1\cdots i_{2n-1}}\phi_{i_1}\cdots \phi_{i_{2n-1}}, \quad i=1,2,\cdots , N\,.
\ee 
We therefore label the corresponding anomalous dimensions by the index $i$, and denote them from now on as $\gamma^i_{2n-1}$. Let us now insert this into Eq.~\eqref{gamma-k} (setting also $l=n-1$) to see what we get
{\setlength\arraycolsep{2pt}
\bea
\gamma^i_{2n-1} V_{ii_1\cdots i_{2n-1}} 
&=& \frac{(n-1)(2n)!c^{n-1}}{4n!^3}\, V_{j_1\cdots j_n(i_1\cdots i_n}V_{i_{n+1}\cdots i_{2n-1})ij_1\cdots j_n} 
\nn\\
&=& \frac{(n-1)(2n)!c^{n-1}}{4n!^3}V_{j_1\cdots j_n(i_1\cdots i_n}V_{i_{n+1}\cdots i_{2n-1}i)j_1\cdots j_n},
\eea}%
where in the second line the $i$ index is taken into the symmetrizing parenthesis. This can be done because
{\setlength\arraycolsep{2pt}
\bea
V_{j_1\cdots j_n(i_1\cdots i_n}V_{i_{n+1}\cdots i_{2n-1})ij_1\cdots j_n}  &=& \frac{1}{2}\left[V_{j_1\cdots j_n(i_1\cdots i_n}V_{i_{n+1}\cdots i_{2n-1})ij_1\cdots j_n} + V_{j_1\cdots j_ni(i_1\cdots i_{n-1}}V_{i_n\cdots i_{2n-1})j_1\cdots j_n}  \right]
\nn\\
&=& V_{j_1\cdots j_n(i_1\cdots i_n}V_{i_{n+1}\cdots i_{2n-1}i)j_1\cdots j_n}.
\eea}%
Now on the r.h.s of the above equation one can use the "fixed point" equation \eqref{fp-even} derived in the previous section to obtain   
\be
\gamma^i_{2n-1} V_{ii_1\cdots i_{2n-1}} = \frac{(n-1)(2n)!c^{n-1}}{4n!^3}V_{j_1\cdots j_n(i_1\cdots i_n}V_{i_{n+1}\cdots i_{2n-1}i)j_1\cdots j_n} = (n-1)\epsilon V_{ii_1\cdots i_{2n-1}}.
\ee
We therefore have 
\be  \label{gamma-2n-1}
\gamma^i_{2n-1} = (n-1)\epsilon,
\ee
consistent with $\Delta^\mathcal{V}_i=\Delta_i+2$, where $\Delta^\mathcal{V}_i$ is the scaling dimension of $\mathcal{V}_i$ defined in \eqref{do}. So the anomalous dimensions $\gamma^i_{2n-1}$ corresponding to descendant operators also satisfy \eqref{gamma-k}. This means that \eqref{rr} is valid for $k=2n-2$ for any $\cS_{2n-1}$.

\subsubsection{The case $k=2n-1$}

The final missing piece in the recurrence relations \eqref{rr} is the case $k=2n-1$. We briefly pointed out earlier in Section~\ref{ss:rr} that this missing information comes from an analysis of the three-point function 
\be 
\left\langle \phi_i(x)\phi_j(y) \cS_{2n}(z)\right\rangle, 
\ee
when two box operators $\Box_x \Box_y$ are applied to it. But before that, we need to do the same analysis when only one operator $\Box_x$ is applied. This gives the multi-field generalization of the structure constant $C_{1,1,2n}$ which is also obtained by evaluating Eq.~\eqref{ciuv-even} for $p=n$ and $q=1$.
This gives the structure constant
\be
C_{\phi_i\phi_j\mathcal{S}_{2n}} = \frac{n-1}{4n} \frac{C^{\mathrm{free}}_{2n-1,1,2n}}{(2n-1)!} V_{il_1\cdots l_{2n-1}} S_{jl_1\cdots l_{2n-1}} 
= \frac{(n-1)c^{2n}}{2} V_{il_1\cdots l_{2n-1}} S_{jl_1\cdots l_{2n-1}}\,,
\ee
where we have used $C^{\mathrm{free}}_{2n-1,1,2n} =(2n)! c^{2n}$. 
Let us now apply two boxes to the above three-point function. As usual we evaluate this once using the SDE and once by applying the box operators directly. The SDE method gives  
{\setlength\arraycolsep{2pt}
\bea  
&& \left\langle \Box_x \phi_i(x)\Box_y\phi_j(y) \mathcal{S}_{2n}(z)\right\rangle
\\
&=& \frac{1}{(2n-1)!^2} V_{ii_1\cdots i_{2n-1}}V_{jj_1\cdots j_{2n-1}} \left\langle [\phi_{i_1}\cdots \phi_{i_{2n-1}}](x)[\phi_{j_1}\cdots \phi_{j_{2n-1}}](y) \mathcal{S}_{2n}(z)\right\rangle 
\nn\\
&=& \frac{1}{(2n-1)!^2} V_{ii_1\cdots i_nj_1\cdots j_{n-1}}V_{jj_1\cdots j_{n-1}l_1\cdots l_n}S_{l_1\cdots l_ni_1\cdots i_n} \frac{C^{\mathrm{free}}_{2n-1,2n-1,2n}}{|x-y|^2|y-z|^{2\delta_c+2}|z-x|^{2\delta_c+2}}
\nn\\
&=& \frac{(2n)!}{(n-1)!n!^2}V_{ii_1\cdots i_nj_1\cdots j_{n-1}}V_{jj_1\cdots j_{n-1}l_1\cdots l_n}S_{l_1\cdots l_ni_1\cdots i_n} \frac{c^{3n-1}}{|x-y|^2|y-z|^{2\delta_c+2}|z-x|^{2\delta_c+2}},\nn
\eea}%
where we have used 
\be 
C^{\mathrm{free}}_{2n-1,2n-1,2n} = \frac{(2n-1)!^2(2n)!}{(n-1)!n!^2} c^{3n-1}\,.
\ee
Direct application of the boxes instead leads to
{\setlength\arraycolsep{2pt}
\bea  
\Box_x \Box_y \left\langle \phi_i(x)\phi_j(y) \mathcal{S}_{2n}(z)\right\rangle &=& \Box_x \Box_y \frac{C_{\phi_i\phi_j\mathcal{S}_{2n}}}{|x-y|^{-2-\gamma^S_{2n}}|y-z|^{2+2\delta+\gamma^S_{2n}}|z-x|^{2+2\delta+\gamma^S_{2n}}} 
\nn\\
&{}&\hspace{-2cm}= \frac{8n(\gamma^S_{2n}-(n-1)\epsilon)}{(n-1)^2} \frac{C_{\phi_i\phi_j\mathcal{S}_{2n}}}{|x-y|^2|y-z|^{2\delta_c+2}|z-x|^{2\delta_c+2}} .
\eea}%
Comparing the two results gives rise to the following eigenvalue equation
{\setlength\arraycolsep{2pt}
\bea  
&& \frac{8n(\gamma^S_{2n}-(n-1)\epsilon)}{(n-1)^2} \frac{(n-1)c^{2n}}{2} V_{il_1\cdots l_{2n-1}} S_{jl_1\cdots l_{2n-1}} 
\nn\\
&=& \frac{(2n)!c^{3n-1}}{(n-1)!n!^2}V_{ii_1\cdots i_nj_1\cdots j_{n-1}}V_{jj_1\cdots j_{n-1}l_1\cdots l_n}S_{l_1\cdots l_ni_1\cdots i_n}, 
\eea}%
which can be further simplified using Eq.~\eqref{gamma-2n-1} to replace the term $(n-1)\epsilon$ on the l.h.s with the anomalous dimension of the descendant operators, and also using the definition \eqref{cnk}
\be 
(\gamma^S_{2n}-\gamma^i_{2n-1}) V_{il_1\cdots l_{2n-1}} S_{jl_1\cdots l_{2n-1}} = c_{n,2n-1} 
V_{ii_1\cdots i_nj_1\cdots j_{n-1}}V_{jj_1\cdots j_{n-1}l_1\cdots l_n}S_{l_1\cdots l_ni_1\cdots i_n}.
\ee
This is precisely the missing piece in our recurrence relation \eqref{rr}, that is, the case $k=2n-1$ when the operator of order $2n-1$ is the descendant of $\phi_i$, obtained from the equation of motion. This implies that Eq.~\eqref{gamma-k} is valid for all $l\geq 0$.

\section{Potts models}\label{sect:potts_models}
We now consider a particular family of theories characterized by the $S_q$ symmetry, 
the Potts model \cite{Potts:1951rk},
which has been introduced as a spin-lattice model that generalizes the Ising model. 
Let $\left\{\sigma_l\right\}$ be a spin configuration
labeled by a regular lattice ${\cal L}$ where location $l \in {\cal L}$ and in which each single spin can take up to $q$ different values $\sigma_l=1,\dots,q$.
The model can be characterized by the microscopic Hamiltonian
\begin{eqnarray}\label{eq:hamiltonian}
 {\cal H} &=& -J \sum_{\left<lr\right>} \delta_{\sigma_l,\sigma_r}
\end{eqnarray}
in which the summation extends only to nearest-neighbor spins in the lattice.
The Kronecker delta
ensures that only nearby spins of the same value
change the energetic balance of the model. The net effect is that the model is ferromagnetic if $J>0$
and anti-ferromagnetic if $J<0$. The Hamiltonian \eqref{eq:hamiltonian} is invariant under
the action of the group $S_q$ of permutations of $q$ objects which acts globally on the set of $q$ spin states.
The model is a fundamental actor in the theory of phase transitions because for $J>0$ it can exhibit
either first or second order phase transition according to both the value of $q$ and the dimensionality $d$ of the lattice.

There is an alternative formulation on the lattice based on random clusters~\cite{Fortuin:1971dw}, equivalent for $q\ge 2$, which has the advantage of allowing for an analytic continuation in $q$.
A straightforward expectation is that the critical physics of the $q$-states Potts model can be captured
by an opportune field theoretic realization of an $S_q$-invariant model,
and that the renormalization group flow of such model admits either a Gaussian fixed point if the phase transition is first order (for values of $q$ above a certain dimensional dependent threshold $q_c(d)$), or a non-Gaussian fixed point if the phase transition is second order. For this latter case one expects the universal features of the model also for $d>2$ to be described by a CFT, if scale invariance is lifted to conformal invariance.~\footnote{Very recently some arguments linking 2d complex CFTs to weakly (small latent heat) first order phase transitions have been presented~\cite{Gorbenko:2018ncu,Gorbenko:2018dtm}.}

Several RG analysis of the Potts model, also for the specific analytic continuations to $q=1$ (percolation) or $q=0$ (spanning forests), are available in the literature. The analytic continuation can be performed within a chosen representation of the $S_q$ discrete symmetry group. Perturbatively the standard approach is based on the $\epsilon$-expansion below the upper critical dimension\cite{Zia:1975ha}. A first attempt to study within wilsonian exact RG was made in~\cite{Zinati:2017hdy}. In $d=2$ several exact results are available~\cite{Baxter:2000ez,Nienhuis:1979mb,Delfino:2017biz}. See also~\cite{Wu:1982ra} for a review of the Potts models.

\subsection{Zoology of $S_q$-invariant interactions}\label{sect:invariant_interactions}

For the purpose of constructing QFTs of Potts models, with $S_q$-invariant interactions,
let us first describe a useful representation introducing a set of $q$ vectors $e^\sigma$ which point in the directions
of the vertices of a $N$-simplex, i.e. a simplex in ${\mathbb R}^N$, for $N=q-1$.
The set of vectors satisfies the following properties
\begin{eqnarray}
 && e^\sigma\cdot e^{\sigma'} =
 \sum_{i=1}^{N} e_i^\sigma  e_i^{\sigma'} = (N+1) \delta_{\sigma,\sigma'}-1 \\
 &&\sum_{\sigma=1}^{N+1} e_i^\sigma =0\,, 
 \qquad
  \sum_{\sigma=1}^{N+1} e_i^\sigma  e_j^{\sigma} = (N+1) \delta_{ij}\,.
\end{eqnarray}
These relations also determine the vectors $e^\sigma$ uniquely, up to rotations and $S_q$ transformations. We can use the vectors $e^\sigma$ to find a representation of the Kronecker delta
\begin{eqnarray}
 \delta_{\sigma,\sigma'} &=&
 \frac{1+e^\sigma\cdot e^{\sigma'}}{q}
\end{eqnarray}
which reflects a manifest invariance under the action of the group in the $N$-dimensional space ${\mathbb R}^N$.

This approach is also the key to the construction of all possible $S_q$-invariant field theoretic interactions.
Indeed one can straightforwardly construct manifestly them by considering any number of copies of 
the field $\psi^\sigma = e^\sigma_i \phi_i$ and summing over the index $\sigma$ (from now on, repeated Latin indices will be summed over).
We shall restrict our attention to local nonderivative interactions, which for this type are
\begin{eqnarray}
 \sum_{\sigma=1}^q (\psi^\sigma)^l
\end{eqnarray}
given $l\in {\mathbb N}$. Clearly, any product of any number of these interactions is also an invariant. In general we have
\begin{eqnarray}
 \prod_{a=1}^p \left(\sum_{\sigma_a=1}^q (\psi^{\sigma_a})^{l_a}\right)
\end{eqnarray}
for $p\in {\mathbb N}$ and $l_a\in{\mathbb N}$ for each $a$. Expressing these invariants in terms of the basic fields $\phi_i$
allows one to write down the most general $S_q$-invariant actions.

Before showing how to use the field $\psi^\sigma$ to construct all possible $S_q$-invariant interactions let us start from the simplest nontrivial interacting action which is cubic
\begin{eqnarray} 
 S[\phi] &=& \int {\rm d}^d x \Bigl\{
 \frac{1}{2} \sum_i \partial_\mu \phi_i \partial^\mu \phi_i + \lambda \sum_\sigma (\psi^\sigma)^3
 \Bigr\}\,,
\end{eqnarray}
in which we also introduce a coupling to weight the interaction and a kinetic term for the field $\phi_i$.
Expanding the field $\psi^\sigma$ in its ``components'' we obtain a manifestly symmetric action
\begin{eqnarray}\label{eq:action}
 S[\phi]
 &=& \int {\rm d}^d x \Bigl\{
 \frac{1}{2} \sum_i \partial_\mu \phi_i \partial^\mu \phi_i + \lambda \sum_\sigma \sum_{i,j,k} e_i^\sigma e_j^\sigma e_k^\sigma \phi_i \phi_j \phi_k
 \Bigr\} \,.
\end{eqnarray}
The critical points of \eqref{eq:hamiltonian} and \eqref{eq:action} are achieved by tuning a single interaction to criticality in both cases. 
In particular in the latter QFT continuous description the relevant $S_q$ symmetric mass term is tuned to zero.
We can imagine that a more complete classification of all possible $S_q$-invariant actions, which goes beyond \eqref{eq:action}
and includes in general more interactions, might serve as a tool to uncover multicritical phases that generalize \eqref{eq:hamiltonian}
through the inclusion of more order parameters.

Assuming that close to the Gaussian point the multiplet of scalar fields has canonical dimension $(d-2)/2$,
for increasing number of derivatives and powers of the fields $\phi_i$ the local interactions have increasing mass dimension.
We are interested in writing down all possible local nonderivative interactions which can be marginal in any dimension $d>3$.
It is convenient to introduce the following tensors
\be
q^{(p)}_{i_1,\dots, i_p} = \frac{1}{N\!+\!1}Q^{(p)}_{i_1,\dots, i_p}  \quad,\quad  Q^{(p)}_{i_1,\dots, i_p} =\sum_\alpha e_{i_1}^\alpha \dots e_{i_p}^\alpha 
\ee
Notice that by construction the first two tensors can generally be simplified
\begin{eqnarray}
 q^{(1)}_{i_1} &=& \sum_\alpha e_{i_1}^\alpha  = 0 \,, \label{e=0} \\
 q^{(2)}_{i_1 i_2} &=&\frac{1}{N\!+\!1}  \sum_\alpha e_{i_1}^\alpha e_{i_2}^\alpha  = \delta_{i_1 i_2}\,. \label{ee=1}
\end{eqnarray}
When instead $p\geq 3$ we cannot generally simplify $q^{(p)}$ unless we specify the order of the permutation group.
Since our interest is to deal with a generic value for $q$ (and possibly analytically continue it)
we shall not require any further property, although it is possible to treat the cases $q=1$, $2$ and $3$, for which some simplification occurs, separately.

The most general local potential action for the $N$ fields $\phi_i$ is
\begin{eqnarray} \label{action-sq}
 S[\phi] &=& \int {\rm d}^d x \Bigl\{
 \frac{1}{2}(\partial\phi)^2  + V(\phi)\Bigr\}\,,
\end{eqnarray}
where the potential $V$ can be written as
\be
 V(\phi)
 = \sum_{p \geq 0} \frac{1}{p!} T^{(p)}_{i_1 \dots i_p} \phi_{i_1}\dots \phi_{i_p}
 \ee
in which the tensors up to the quintic order of interactions can be defined as 
\begin{eqnarray} 
 T^{(2)}_{i_1 i_2} &=& \zeta_2 \,\delta_{i_1 i_2}, \label{t2} \\
 T^{(3)}_{i_1 i_2 i_3} &=& \zeta_3 \,q^{(3)}_{i_1 i_2 i_3}, \label{t3} \\
 T^{(4)}_{i_1 i_2 i_3 i_4} &=& \zeta_{4,1} \,\delta_{(i_1 i_2} \delta_{i_3 i_4)} + \zeta_{4,2} \,q^{(4)}_{i_1 i_2 i_3 i_4}, \label{t4} \\
 T^{(5)}_{i_1 i_2 i_3 i_4 i_5} &=& \zeta_{5,1} \,\delta_{(i_1 i_2} q^{(3)}_{i_3 i_4 i_5)} + \zeta_{5,2} \,q^{(5)}_{i_1 i_2 i_3 i_4 i_5}. \label{t5}
\end{eqnarray}
The couplings from $\zeta_2$ to $\zeta_{5,2}$ have a rather straightforward meaning:
$\zeta_2$ plays the role of the mass for the multiplet $\phi_i$,
while all other couplings starting from $\zeta_3$ are genuine interactions with which one can construct a perturbative expansion.
Specifically: $\zeta_3$ is canonically marginal in $d=6$ so with it we can construct a perturbative expansion in $d=6$ and an $\epsilon$-expansion in $d=6-\epsilon$.
Likewise in $d=4$ and $d=4-\epsilon$ one has to consider a perturbative expansion in the couple 
$\left\{\zeta_{4,1},\zeta_{4,2}\right\}$.
In $d=\frac{10}{3}$ one has to consider a perturbative expansion in the couple $\left\{\zeta_{5,1},\zeta_{5,2}\right\}$.

For later convenience we want to introduce a basis of the $S_q$-invariant interactions
that is related to the above definition of the tensors $T^{(p)}$.
We denote the basis with ${\cal I}_{i,j}$: the index $i$ refers to the fact that
each element ${\cal I}_{i,j}$ is a fully $S_q$-invariant products of $i$ copies of the field components $\phi$,
while $j$ parametrizes the increasing size of the tensors $q^{(i)}$ in its construction
(the presence of the tensors $q^{(i)}$ instead of the Kronecker delta represents, to some extent,
the departure from an $O(N)$ invariant theory).
The first few invariants are
\be \label{eq:invariants-basis}
\ba{lll}
 {\cal I}_{2} = \phi_i\phi_i\,, & \qquad\qquad &
 {\cal I}_{3} = q^{(3)}_{i_1 i_2 i_3}\, \phi_{i_1}\phi_{i_2}\phi_{i_3}\,, \\[7pt]
 {\cal I}_{4,1} = \phi_i\phi_i \, \phi_j\phi_j\,, & \qquad\qquad &
 {\cal I}_{4,2} = q^{(4)}_{i_1 i_2 i_3 i_4}\phi_{i_1}\phi_{i_2}\phi_{i_3}\phi_{i_4}\,, \\[7pt]
 {\cal I}_{5,1} = \phi_i\phi_i \, q^{(3)}_{i_1 i_2 i_3}\phi_{i_1}\phi_{i_2}\phi_{i_3}\,, & \qquad\qquad &
 {\cal I}_{5,2} = q^{(5)}_{i_1 i_2 i_3 i_4 i_5}\phi_{i_1}\phi_{i_2}\phi_{i_3}\phi_{i_4}\phi_{i_5}\,.
\ea 
\ee
We have arranged their second index for increasing ``departure'' from $O(N)$ symmetry (in which the only allowed invariants are powers of $\phi_i\phi_i$):
notice that while the basis operators chosen in this paper are the same as the one of \cite{Zinati:2017hdy}, the two bases differ in the way the label $j$ is assigned.
By construction some invariants are algebraically related, for example
\begin{eqnarray} 
 {\cal I}_{4,1} = ({\cal I}_{2})^2\,, &\qquad&
 {\cal I}_{5,1} = {\cal I}_{2} {\cal I}_{3}\,.
\end{eqnarray}
For specific values of $q$ it is possible to find even more relations among the invariants.
In particular, given a natural value of $q$ there is a finite number of independent invariants that
we can build out of the field multiplet. We come back to this point later
when specializing some results to the first few low values of $q$.

One can write useful relations to simplify contractions of such $q^{(i)}$ tensor. We present some of them in the 
Appendix~B.\ref{ss:reduction}.

\subsection{Quadratic operators: Imposing $S_{N+1}$ invariance} \label{ss:qo-symm} 

In later sections, having specified the model and the explicit form of the potential, we will solve the eigenvalue equations \eqref{crit-cft} and \eqref{crit-cft-2} and determine the eigenvalues $\gamma^S_2$, which are the anomalous dimensions of the quadratic operators. However, considerable information can be extracted only from a knowledge of the symmetry, that is $S_{N+1}$ in our case, and without relying on the precise model. 
For this purpose we devote this section to understanding how much the symmetry alone can tell us about the quadratic scaling operators. As the first step, note that the $N$-dimensional space of fields $\phi_i$ or equivalently $\psi^\alpha$ carry the standard representation of $S_{N+1}$ which in the Young-Tableaux notation is nothing but the following diagram with $N+1$ boxes
{\setlength\arraycolsep{6pt}\def\arraystretch{0.7}
\be 
\ba{|c|c|c|c|c|}
\cline{1-4}
&&\multicolumn{1}{c|}{\dots}& \\ \cline{1-4}
& \multicolumn{4}{c}{}  \\ \cline{1-1}
\ea
\ee}%
From this, one can determine the decomposition of the symmetric product $\phi_{i}\phi_{j}$, or equivalently $\psi^{\alpha}\psi^{\beta}$, of two fields into irreducible representations. These irreducible representations are the representations carried by the quadratic scaling operators. Indeed the symmetric product of two standard representations is decomposed as
{\setlength\arraycolsep{5pt}\def\arraystretch{0.7}
\be 
Sym\!\left(\hspace{2pt}\ba{|c|c|c|c|c|}
\cline{1-4}
&&\multicolumn{1}{c|}{\dots}& \\ \cline{1-4}
& \multicolumn{4}{c}{}  \\ \cline{1-1}
\ea \hspace{38pt}
\otimes \hspace{2pt}
\ba{|c|c|c|c|c|}
\cline{1-4}
&&\multicolumn{1}{c|}{\dots}& \\ \cline{1-4}
& \multicolumn{4}{c}{}  \\ \cline{1-1}
\ea \hspace{38pt}\right)
=
\ba{|c|c|c|c|c|}
\cline{1-4}
&&\multicolumn{1}{c|}{\dots}& \\ \cline{1-4}
\ea \hspace{3pt}
\oplus \hspace{2pt}
\ba{|c|c|c|c|c|}
\cline{1-4}
&&\multicolumn{1}{c|}{\dots}& \\ \cline{1-4}
& \multicolumn{4}{c}{}  \\ \cline{1-1}
\ea \hspace{39pt}
\oplus \hspace{2pt}
\ba{|c|c|c|c|c|}
\cline{1-4}
&&\multicolumn{1}{c|}{\dots}& \\ \cline{1-4}
&& \multicolumn{3}{c}{}  \\ \cline{1-2}
\ea \\
\ee}%
where all the diagrams here include $N+1$ boxes. This shows that in $S_{N+1}$ invariant theories there are only three distinct anomalous dimensions, corresponding to three, possibly degenerate, set of scaling operators. In terms of dimensions, the above relation corresponds to the following decomposition
\be  \label{dim-decomp}
\frac{N(N+1)}{2} = 1 \oplus N \oplus \frac{(N+1)(N-2)}{2},
\ee
which can be obtained from the Hook length formula and specifies the degeneracy of each subspace of scaling operators with the same scaling dimension. More explicitly, one may introduce three projectors in the space of quadratic operators 
\be 
\phi_i\phi_j = (P_1)_{ij,kl}\,\phi_k\phi_l + (P_2)_{ij,kl}\,\phi_k\phi_l + (P_3)_{ij,kl}\,\phi_k\phi_l.
\ee
The explicit form of these projectors is given as follows
{\setlength\arraycolsep{2pt}
\bea \label{p1}
(P_1)_{ij,kl} &=& \frac{1}{N} \delta_{ij} \delta_{kl}, \\
(P_2)_{ij,kl} &=& \frac{1}{N-1} \left( q^{(4)}_{ijkl} - \delta_{ij} \delta_{kl} \right), \label{p2} \\
(P_3)_{ij,kl} &=& \delta_{i(k} \delta_{l)j} - \frac{1}{N-1}  q^{(4)}_{ijkl} + \frac{1}{N(N-1)}\delta_{ij}\delta_{kl}. \label{p3}
\eea}%
These are projection operators in the sense that
\be 
(P_a P_b)_{ij,kl} = \delta_{ab} (P_a)_{ij,kl}\quad , \qquad (P_1 + P_2 +P_3)_{ij,kl} = \delta_{i(k} \delta_{l)j}.
\ee
One may find the equivalent split of the product $\psi^\alpha\psi^\beta$ and the corresponding projectors simply by transforming the Latin indices to Greek indices by contracting the above results with the vectors $e^\alpha_i$ and including appropriate powers of $N+1$. 
So far we have gained general insight on the space of scaling operators. What is missing is information about the actual values of the anomalous dimensions. These may be obtained resorting to the eigenvalue equation 
\be  \label{gammaS2-general} 
(\gamma^{S}_2 - \eta) \; S_{ij} = \mathcal{M}_{ij,ab}\,S_{ab},
\ee
from which the explicit form of the above split also emerges, as will be shown shortly. Based on general grounds, for an $S_{N+1}$ invariant theory the stability matrix takes the general form which is a linear combination of four-index tensors constructed with the $q^{(n)}$ tensors and is symmetrized in its first and second pair of indices, that is
\be \label{M}
\mathcal{M}_{ij,ab} = \tau\, q^{(4)}_{ijab} + \rho\, \delta_{ij}\delta_{ab} + \kappa\, \delta_{i(a}\delta_{b)j}.
\ee
This leaves us with only three undetermined parameters in terms of which the anomalous dimensions can be expressed. When contracted with $S_{ab}$ this gives\footnote{We keep the summation on $a,b$ implicit, while summations on $\alpha,\beta$ are made explicit.}
\be 
\mathcal{M}_{ij,ab} \,S_{ab} = \tau' \sum_\alpha (e^\alpha_a S_{ab} e^\alpha_b)\, e^\alpha_ie^\alpha_j + \rho\, \delta_{ij} S_{aa}  + \kappa\, S_{ij}.  
\ee
where $(N+1)\tau' =\tau$. If the r.h.s is to be proportional to $S_{ij}$ itself, then either\footnote{Note that summing on $\alpha$ in \eqref{ese=0} gives $S_{aa}=0$.} 
\be \label{ese=0}
e^\alpha_a S_{ab}\, e^\alpha_b=0,
\ee 
for any $\alpha$, or $S_{ij}$ must have the following general structure as in the first term on the r.h.s
\be \label{s=aee}
S_{ij} =\sum_\alpha a_\alpha\, e^\alpha_ie^\alpha_j.
\ee
In the first case the eigenvalue is simply equal to $\kappa$, while in the second case one can use the relation
\be 
e^\alpha_a S_{ab}\, e^\alpha_b = (N^2-1)a_\alpha + \sum_\beta a_\beta,
\ee
to obtain
\be
M_{ij, ab} \,S_{ab} = \tau(N-1)S_{ij} + (\tau+\rho N)\, \delta_{ij}\sum_\beta a^\beta +\kappa S_{ij}.
\ee
This equation shows that either $\sum_\beta a^\beta=0$ in which case $S_{ij}$ is an eigenvector of the matrix $M_{ij, pq}$ with eigenvalue $\tau(N-1)+\kappa$, or $\sum_\beta a^\beta\neq 0$ in which case $S_{ij}$ must be proportional to the identity and therefore all $a^\beta$ must be equal (in particular $S_{ij}=\delta_{ij}$ if $a^\beta=1/(N\!+\!1)$). In this case the corresponding eigenvalue of $M_{ij, pq}$ is $(\tau+\rho) N + \kappa$. So in summary the three eigenvalues for $\gamma^S_2$ and their corresponding eigenvectors are respectively
{\setlength\arraycolsep{2pt}
\bea \label{es1} 
\gamma_2^1-\eta  &=& (\tau+\rho) N + \kappa, \hspace{22.1mm} S_{ij} = \delta_{ij}, 
\\[6pt]
\gamma_2^2-\eta  &=& \tau(N-1)+\kappa, \hspace{22.4mm} S_{ij} =\sum_\alpha a^\alpha\, e^\alpha_ie^\alpha_j, \quad \sum_\alpha a^\alpha=0, \label{es2} 
\\[-2pt]
\gamma_2^3-\eta  &=& \kappa, \hspace{44.5mm} S_{ij} =\sum_{\alpha,\beta} a^{\alpha\beta}\, e^\alpha_ie^\beta_j, \quad e^\alpha_p S_{pq} e^\alpha_q=0. \label{es3}
\eea}%
where in the last case the most general form of $S_{ij}$ has been written and the condition on the symmetric matrix $a^{\alpha\beta}$ is left implicit and will be determined shortly. Notice that the first eigenvalue is also consistent with \eqref{gamma20-eta}. As we will show below, these eigenvectors correspond to the three irreducible representations mentions earlier. The first eigenvector \eqref{es1} clearly corresponds to the one-dimensional representation \eqref{p1}. We will now solve and bring the other two eigenvectors into a more transparent form. In the second case \eqref{es2}, the vector $a^\alpha$ is an $(N+1)$-component vector which is restricted to an $N$-dimensional subspace characterized by $\sum_\alpha a^\alpha=0$. A convenient basis that spans this subspace is $e^\alpha_i$, $i=1,\cdots,N$, the elements of which are ensured by \eqref{e=0} to lie on the $N$-dimensional subspace. This choice makes the eigenvectors transform covariantly under rotations. Therefore there are $N$ eigenvectors $u_p$ with eigenvalue $\gamma^2_2$, which can most conveniently be written as
\be 
u_{p,ij} =\frac{1}{N\!+\!1}  \sum_\alpha e^\alpha_p e^\alpha_ie^\alpha_j = q^{(3)}_{pij}, \qquad p=1,\cdots N. \label{up}
\ee
Notice that $u_{p,ij}$ is symmetric in all three indices. Let us also set $u_{0,ij} = \delta_{ij}$ to make the notation uniform. This $N$-dimensional representation labelled by a single index is equivalent to the double-index redundant description \eqref{p2}. One can also verify that $q^{(3)}_{ijk}(P_2)_{jk,lm}=q^{(3)}_{ilm}$ and $\delta_{jk}(P_2)_{jk,lm}=0$.  

One may similarly find a convenient basis for the set of $a^{\alpha\beta}$ in \eqref{es3} that satisfy the condition \eqref{ese=0}. Let us first notice that the correspondence between the symmetric matrices $S_{ij}$ and $a^{\alpha\beta}$ which have respectively $N(N+1)/2$ and $(N+1)(N+2)/2$ independent components is certainly not one-to-one. However one notices that the expression of $S_{ij}$ in terms of $a^{\alpha\beta}$ is redundant under the transformation
\be 
a^{\alpha\beta} \rightarrow a^{\alpha\beta} - \sigma^\alpha,
\ee
which, by choosing $\sigma^\alpha$ appropriately, allows us to set $\sum_\beta a^{\alpha\beta} =0$ for any $\alpha$. This reduces the number of independent components to $N(N+1)/2$ and makes the correspondence to $S_{ij}$ one-to-one. Now, the condition \eqref{ese=0} on the general $a^{\alpha\beta}$ is
\be 
(N+1)^2a^{\gamma\gamma} - 2(N+1)\sum_\alpha a^{\gamma\alpha} + \sum_{\alpha,\beta} a^{\alpha\beta} =0  
\ee
for fixed $\gamma$, which upon imposing $\sum_\beta a^{\alpha\beta} =0$, i.e. fixing the redundancy, gives the extra $N+1$ conditions $a^{\gamma\gamma} =0$. By making a proper ansatz in terms of $e^\alpha_i$ one may find a convenient (i.e. covariant) basis spanning the matrices $a^{\alpha\beta}$ subject to the constraints
\be 
\sum_\alpha a^{\gamma\alpha} =0, \quad a^{\gamma\gamma}=0,
\ee
for any $\gamma$. Plugging these basis elements (labeled by two indices $p,q$) back into \eqref{es3} leads to the set of $(N+1)(N-2)/2$ independent eigenvectors $u_{pq,ij}$ labeled by $p,q$ and given explicitly by
\be \label{upq}
u_{pq,ij} = q^{(4)}_{pqij} - (N-1) \delta_{i(p} \delta_{q)j} - \frac{1}{N} \delta_{pq} \delta_{ij},
\ee
which spans a subspace orthogonal to the one generated by $\delta_{ij}$ and $q^{(3)}_{pij}$. This is clearly proportional to the projection operator \eqref{p3}. This analysis shows that the space of quadratic operators $\phi^i\phi^j$ that has a degenerate spectrum at the classical level is split by the leading quantum corrections into the three eigenspaces 
{\setlength\arraycolsep{2pt}
\bea 
u_{0,ij}\,\phi^i\phi^j  &=& \phi\!\cdot\!\phi, \label{U0} \\[5pt]
u_{p,ij}\,\phi^i\phi^j  &=& q^{(3)}_{pij}\phi^i\phi^j, \label{U1} \\
u_{pq,ij}\,\phi^i\phi^j &=& q^{(4)}_{pqij}\phi^i\phi^j - (N-1) \phi^p\phi^q  - \frac{1}{N} \delta_{pq} \phi\!\cdot\!\phi. \label{U2}
\eea}%
It is important to note that for $S_q$ invariant theories in $d_c=6$, which have a cubic interaction of the form \eqref{t3}, the second operator \eqref{U1} is a descendant.  Therefore, as pointed out in Section \ref{ss:qo}, even though \eqref{U1} is always an operator with definite scaling, in the particular case of models with a cubic critical interaction \eqref{t3}, the anomalous dimension of \eqref{U1} is not given by \eqref{es2} but is obtained from $\gamma^2_2 = \gamma+ \epsilon/2$. 

While the three set of eigenoperators \eqref{U0}-\eqref{U2} are orthogonal to each other, the bases chosen in \eqref{up} and \eqref{upq} also make the set of scaling operators in \eqref{U1} as well as those in \eqref{U2} separately mutually orthogonal in the free theory, which further motivates such a choice. This orthogonality is demonstrated by showing that the two point functions of different operators vanish. In fact the values of the two point functions of operators with the same anomalous dimension are
{\setlength\arraycolsep{2pt}
\bea  
\left\langle [\phi\!\cdot\!\phi](x) [\phi\!\cdot\!\phi](y)\right\rangle &=&  \frac{2N}{|x-y|^{8}} c^2, \label{es0-6} \\
\left\langle [u_{p,ij}\,\phi_i\phi_j](x) [u_{q,kl}\,\phi_k\phi_l](y)\right\rangle &=&  \frac{2u_{p,ij}u_{q,ij}}{|x-y|^{8}} c^2= \frac{2(N-1)}{|x-y|^{8}}c^2\delta_{pq}, \label{es1-6} \\
\left\langle [u_{pq,ij}\,\phi_i\phi_j](x) [u_{rs,kl}\,\phi_k\phi_l](y)\right\rangle &=&  \frac{2u_{pq,ij}u_{rs,ij}}{|x-y|^{8}} c^2= \frac{2(1-N)u_{pq,rs}}{|x-y|^{8}}c^2, \label{es2-6}
\eea}%
while the two point function of operators with different (anomalous) dimensions clearly vanish in the free theory. In the second equation, which shows the orthogonality of the operators \eqref{U2}, we have used the contraction formula \eqref{qq}. Repeated use of \eqref{fusion} and \eqref{trace} also leads to \eqref{es2-6}. The tensor $u_{pq,rs}$ on the r.h.s of Eq.~\eqref{es2-6} is the identity matrix in the $(N+1)(N-2)/2$ subspace $u_{pq,rs}$, while it vanishes in the $N+1$ dimensional subspace spanned by $\delta_{ij}$ and $u_{p,ij}$ (i.e. it is proportional to a projection operator). Notice also that from Eq.~\eqref{es2-6} the norm of $u_{rs,kl}\,\phi_k\phi_l$ is negative for $N>1$. 

The projection operators \eqref{p1}-\eqref{p3} allow for a more natural, though redundant, description of the three scaling operators \eqref{U0}-\eqref{U2} which treats them on the same footing, labeling them all with two indices. In this representation the two-point functions are
\be 
\left\langle [(P_a)_{ij,pq}\,\phi_p\phi_q](x) \; [(P_b)_{kl,rs}\,\phi_r\phi_s](y)\right\rangle =  
\frac{2(P_a P_b)_{ij,kl}}{|x-y|^{4\delta_c}} c^2= \frac{2\delta_{ab}(P_a)_{ij,kl}}{|x-y|^{4\delta_c}}c^2 ,
\ee
which leads us to define the normalized scaling operators $\mathcal{O}^{(2)}_a$ as
\be \label{o2}
\mathcal{O}^{(2)}_{a,ij} = \frac{1}{\sqrt{2}} (P_a)_{ij,kl}\,\phi_k\phi_l, \qquad a=1,2,3.
\ee
In the first two operators $a=1,2$ the redundancy is removed by contracting them with $\delta_{ij}$ and $q^{(3)}_{ijk}$ respectively, which brings us back to the definitions \eqref{U0} and \eqref{U1} but with a different normalization. For later use let us then define
\be  \label{o0}
\mathcal{O}^{(2)}_0 = \mathcal{O}^{(2)}_{1,ij}\,\delta_{ij} = \frac{1}{\sqrt{2}}\,\phi\!\cdot\!\phi,
\ee
\be \label{ok}
\mathcal{O}^{(2)}_{k} = \mathcal{O}^{(2)}_{2,ij}\,q^{(3)}_{ijk} = \frac{1}{\sqrt{2}}\,q^{(3)}_{ijk}\phi_i\phi_j,
\ee
which leads to an operator with no free index and an operator with one free index respectively. Notice that here \eqref{o0} and \eqref{ok} have not been normalized to unity. The description of the operator $\mathcal{O}^{(2)}_{3,ij}$ remains redundant and all we can say is that $\mathcal{O}^{(2)}_{3,ij}\,\delta_{ij} =0$ and $\mathcal{O}^{(2)}_{3,ij}\,q^{(3)}_{ijk}=0$, which put $N+1$ constraints on it.
The general analysis of this section can in principle be extended to higher order operators, using group theory arguments and perhaps also the equation \eqref{rr}. This will be beyond the scope of the present article. 

\section{Potts models with $d_c=6,4,\frac{10}{3}$}\label{sect:potts-cft}
\subsection{Cubic Potts model}\label{sect:cubic-potts-cft}
We will study in this section the $S_{N+1}$ invariant scalar theory in $d=6-\epsilon$. In this case the critical interaction is cubic and the action takes the form 
\begin{eqnarray} 
 S[\phi] 
 &=& \int {\rm d}^d x \Bigl\{
 \frac{1}{2}(\partial\phi)^2
 + \frac{1}{3!}\zeta_{3}  \,q^{(3)}_{i_1 i_2 i_3} \phi_{i_1}\phi_{i_2}\phi_{i_3}
 \Bigr\} \label{cubic-potts}\,,
\end{eqnarray}
which has thus a single critical coupling. In the following sections we obtain the leading order critical data of this model including anomalous dimensions and some structure constants, and determine the $\epsilon$-dependence of the critical coupling.
\subsubsection{Anomalous dimension}

For the cubic model where $n=3/2$ the general formula \eqref{eta-cft} for the field anomalous dimension, written in terms of $\eta=2\gamma$, reduces to 
\be  
\eta \,\delta_{ab} = \frac{c}{96}\;V_{aij}\, V_{bij}. 
\ee 
Inserting into this equation the critical couplings given by $V_{aij} = \zeta_3 \,q^{(3)}_{aij}$ and using Eq.~\eqref{qq} to contract the indices, 
one easily finds the expression for the anomalous dimension of the cubic Potts model in terms of the coupling
\be \label{eta-cubic}
\eta  = \frac{c}{96}\, \zeta^2_3\, (N-1),
\ee
in agreement with the leading order results of ref~\cite{CSVZ4} obtained from RG.

\subsubsection{Quadratic operators} \label{ss:fo-cubic}

For the cubic Potts model the general equation \eqref{crit-cft} which gives the critical exponents of the mass operators reduces to the following equation in which the field anomalous dimensions are equal because of symmetries
\be 
(\gamma^S_2 - \eta) \; S_{ij} = -\frac{c}{16}\, V_{i\,l\,(p}V_{q)j\,l} \,S_{pq}.
\ee
This equation is consistent with the RG flow equation \cite{CSVZ4} which governs the running of the coupling $J_{ab}$ in the operator $J_{ab}\phi_a\phi_b$. This becomes evident by setting $\beta_{J_{ab}} = -\gamma^S_2 J_{ab}$ in such an equation. It remains to diagonalize the stability matrix. In fact we need to find the eigenvectors and eigenvalues of the matrix 
\be 
\mathcal M_{ij, pq} \equiv -\frac{c}{16} V_{ai(p}V_{q)aj} = -\frac{c}{16}\zeta_3^2\, q^{(3)}_{a\,i(p} \,q^{(3)}{}^{a}{}_{q)j} = -\frac{c}{16} \zeta_3^2\left(q^{(4)}_{ipjq} -  \delta_{i(p}\delta_{q)j}\right).
\ee
This is done for a general stability matrix in Section~\ref{ss:qo-symm}. For specific models, all we need to know is the coefficients of the three terms in the stability matrix, i.e. the parameters $\tau,\rho,\kappa$ defined in \eqref{M}. In the present example these are
\be  
\tau = -\frac{c}{16}\zeta_3^2, \qquad 
\rho = 0, \qquad
\kappa = \frac{c}{16}\zeta_3^2.
\ee 
Using the relations \eqref{es1}-\eqref{es3} and the value of $\eta$ obtained in the previous section, one can immediately write down the two eigenvalues $\gamma^1_2$ and $\gamma^3_2$ corresponding to the scaling operators $(P_1)_{ij,kl}\,\phi_k\phi_l$ and $(P_2)_{ij,kl}\,\phi_k\phi_l$. The operator $(P_2)_{ij,kl}\,\phi_k\phi_l$ instead is the exception that we discussed in Section \ref{ss:qo}, the anomalous dimension of which does not satisfy \eqref{gammaS2}. This anomalous dimensions is given instead by the formula $\gamma^2_2 = \gamma + \epsilon/2$. In summary the anomalous dimensions and their corresponding eigenoperators are
{\setlength\arraycolsep{2pt}
\bea \label{es0-cubic} 
\gamma_2^1  &=& -\frac{5c}{96}\, (N-1) \zeta^2_3, \qquad\quad\;\;\;\, (P_1)_{ij,kl}\,\phi_k\phi_l, 
\\[2pt]
\gamma_2^2  &=& -\frac{c}{48}\, (2N-5)\zeta^2_3, \qquad\quad\;\, (P_2)_{ij,kl}\,\phi_k\phi_l, \label{es1-cubic} 
\\
\gamma_2^3  &=& +\frac{c}{96}\, (N+5)\zeta^2_3, \qquad\qquad (P_3)_{ij,kl}\,\phi_k\phi_l. \label{es2-cubic}
\eea}%
%
%

\subsubsection{Structure constants} \label{ss:ope-cubic}

Finally let us give some examples of structure constants in the case of the cubic model. 
The general structure constants presented in Sections~\ref{ssec:c12p2q} and~\ref{ssec:c112k} can be directly applied to the case $n=3/2$. In fact if for simplicity we avoid high order operators, the simplest examples that we can immediately find are the 
multi-field generalizations of $C_{122}$ and $C_{111}$, keeping in mind that the quadratic operators involved in $C_{122}$ must not be a descendant. The generalization of $C_{122}$ is obtained by \eqref{ci2p2q} upon setting $p=q=1$ and reads 
\be \label{ciuv-cft}
C_{\phi_i \cS_2 \tilde{\cS}_2} 
= -\zeta_3 \sqrt{c}\, q^{(3)}_{ilk}\,S_{jl} \, \tilde S_{kj},
\ee
where appropriate rescaling of the fields has been done to accord with the usual CFT normalization, as discussed at the end of Section \ref{ssec:c12p2q-1}. This can be compared with the OPE coefficient 
determined using the renormalization group~\cite{CSVZ4}, provided suitable rescalings are done in the beta function  (see \cite{Codello:2017epp}), i.e the replacement $\zeta_3 \rightarrow \zeta_3 /2(4\pi)^{3/2}$ is made, which in terms of $c$ is simply given as $\zeta_3 \rightarrow\zeta_3 \sqrt{c}/8$. Agreement between CFT and RG results is then verified immediately at this level. In order to obtain the explicit form of these OPE coefficients we choose the scaling operators $\cS_2$ and $\tilde{\cS}_2$ among \eqref{o2}, but excluding the descendant operator $\mathcal{O}^{(2)}_{2,ij}$. This gives
\be
C_{\phi_i\,\mathcal{O}_{a,pq}^{(2)} \mathcal{O}_{b,rs}^{(2)}} = -\frac{1}{2}\zeta_3 \sqrt{c}\, q^{(3)}_{ijk}\, (P_a)_{pq,jl}\,(P_b)_{rs,lk}.
\ee
with $a,b\neq 2$. These OPE coefficients vanish for the cases $(a,b)=(1,1),(1,3)$ and therefore the only nontrivial example is the following  
{\setlength\arraycolsep{2pt}
\bea 
C_{\phi_i\,\mathcal{O}_{3,pq}^{(2)} \mathcal{O}_{3,rs}^{(2)}} = \frac{1}{2}\zeta_3 \sqrt{c}\, &&\left[\frac{N}{(N-1)^2} q^{(5)}_{ipqrs}-\frac{1}{4}\left(\delta_{rp}q^{(3)}_{isq}+\delta_{rq}q^{(3)}_{isp}+\delta_{sp}q^{(3)}_{irq}+\delta_{sq}q^{(3)}_{irp}\right) \right. \nn\\
- && \left. \frac{1}{2(N-1)}\left(\delta_{ip}q^{(3)}_{qrs}+\delta_{iq}q^{(3)}_{prs}+\delta_{ir}q^{(3)}_{pqs}+\delta_{is}q^{(3)}_{pqr}\right) \right. \nn\\
- && \left. \frac{1}{(N-1)^2}\left(\delta_{pq}q^{(3)}_{irs}+\delta_{rs}q^{(3)}_{ipq}\right)\right].
\eea
In the next section we obtain the fixed point value of $\zeta_3$ from CFT considerations which can then be inserted into the above equations in order to get the physical $\epsilon$-dependent result.  

As the second example, Eq.~\eqref{ci2p-12q-1} gives for $p,q=1$ the generalization of the single field structure constant $C_{111}$. This is given by
\be \label{c111-cubic}
C_{\phi_i \phi_j \phi_k} = -\frac{1}{8} V_{ijk} C^{\mathrm{free}}_{211} = -\frac{c^2}{4} \zeta_3 \, q^{(3)}_{ijk} \rightarrow  -\frac{1}{4} \zeta_3 \sqrt{c} \, q^{(3)}_{ijk} = - \sqrt{\frac{2\epsilon}{7-3N}} \, q^{(3)}_{ijk},
\ee 
where the first two expressions are obtained from Eq.~\eqref{ci2p-12q-1}, while the third expression is given in the CFT normalization where the fields are rescaled to set their two point functions to unity. In the last equation $\zeta_3 \sqrt{c}/8$ has been set to the fixed point value computed in the next section in Eq.~\eqref{zeta3-fp}.

Apart from these structure constants that are proportional to the coupling constant it is straightforward to obtain by direct calculation some nontrivial structure constants in the free theory. For instance we have
\be 
\langle \mathcal{O}^{(2)}_{1,ij}(x) \mathcal{O}^{(2)}_{1,kl}(y) \mathcal{O}^{(2)}_{1,pq}(z)\rangle 
= \frac{2\sqrt{2}N^{-3}\delta_{ij}\delta_{kl}\delta_{pq}}{|x-y|^2|y-z|^2|x-z|^2} \label{O111},
\ee
\be 
\langle \mathcal{O}^{(2)}_{1,ij}(x) \mathcal{O}^{(2)}_{3,kl}(y) \mathcal{O}^{(2)}_{3,pq}(z)\rangle 
= \frac{2\sqrt{2}N^{-1}\delta_{ij}(P_3)_{kl,pq}}{|x-y|^2|y-z|^2|x-z|^2} \label{O133},
\ee
where in the second equation repeated use of the fusion and trace rules in Appendix B.\ref{ss:reduction} has been made. Notice that three-point functions involving the descendant operator $\mathcal{O}^{(2)}_{2,ij}$ do not define CFT data. The only nontrivial three-point function left that involves the primary quadratic operators is the one with three operators $\mathcal{O}^{(2)}_{3,ij}$ that we have avoided to write due to its long expression.


\subsubsection{Critical coupling $\zeta_3(\epsilon)$}\label{ss:cc-cubic}
One can also fix at leading order the relation linking the coupling $\zeta_3$ to $\epsilon$. This has been done in the single-field case in \cite{Codello:2017qek},
and extended to a general multi-field model in Section \ref{ss:fp-cubic}. However, it may still be instructive to repeat the argument directly for the permutation invariant case. 
Let us start from the relation
\be 
\langle \phi^i(x) \phi^j(y) \phi^k(z)\rangle = \frac{C_{ijk}}{|x-y|^2|y-z|^2|x-z|^2}  , \quad C_{ijk} = -\frac{1}{4}\zeta_3 c^2 q^{(3)}_{ijk} ,
\ee
where for the structure constant $C_{ijk}$ we have used the second expression in Eq.~\eqref{c111-cubic}. This is simply because here we are working with the original scalar field and not the rescaled one $\hat \phi$. 
Acting with three Laplacians on the general scaling form one obtains
{\setlength\arraycolsep{2pt}
\bea
\Box_x\Box_y\Box_z\langle \phi^i(x) \phi^j(y) \phi^k(z)\rangle &\stackrel{\mathrm{LO}}{=}& 32(\epsilon-6\gamma)\frac{C_{ijk}}{|x-y|^4|y-z|^4|x-z|^4}\nn\\
&=& 8(6\gamma-\epsilon)\zeta_3 c^2\frac{q^{(3)}_{ijk}}{|x-y|^4|y-z|^4|x-z|^4} ,
\label{eqcub3box1-2}
\eea}%
while using the SDE one gets
{\setlength\arraycolsep{2pt}
\bea
\langle \Box_x\phi^i(x) \Box_y\phi^j(y) \Box_z\phi^k(z)\rangle &=& \frac{\zeta_3^3}{2!^3}\langle q^{(3)}_{ii_1i_2}\phi^{i_1}(x)\phi^{i_2}(x)\; q^{(3)}_{jj_1j_2}\phi^{j_1}(y)\phi^{j_2}(y)\;q^{(3)}_{kk_1k_2}\phi^{k_1}(z)\phi^{k_2}(z)\rangle \nn\\
&\stackrel{\mathrm{LO}}{=}& \frac{\zeta_3^3}{2!^3}\,8(N-2)c^3\frac{q^{(3)}_{ijk}}{|x-y|^4|y-z|^4|x-z|^4}.
\label{eqcub3box2-2}
\eea}%
Equating the expressions on the right hand side of Eqs.~\eqref{eqcub3box1-2} and~\eqref{eqcub3box2-2} 
and recalling the value of the field anomalous dimension \eqref{eta-cubic}, this
gives
\be 
\frac{1}{4}\zeta_3^3\,(N-1)c^3-8\epsilon\zeta_3 c^2 = \zeta_3^3\,(N-2)c^3,
\ee
which can be trivially solved
\be \label{zeta3-fp}
\frac{\sqrt{c}}{8}\zeta_3(\epsilon) \stackrel{\mathrm{LO}}{=} \frac{1}{2}\sqrt{\frac{2\epsilon}{7-3N}}.
\ee
This is in agreement with the RG result~ \cite{CSVZ4} after making the replacement $\zeta_3 \rightarrow 8\zeta_3 /\sqrt{c}$ to accord with the RG conventions. One may now use this $\epsilon$ dependence of the coupling to express all the CFT data found in previous sections in terms of $\epsilon$. As an instance the field anomalous dimension given in \eqref{eta-cubic} and the anomalous dimension of the descendant operator $q^{(3)}_{ijk}\phi^j\phi^k$ reported in \eqref{es1-cubic} become 
\be 
\gamma  = \frac{1}{6}\, \frac{N-1}{7-3N}\, \epsilon, \qquad
\gamma^2_2 = \frac{2}{3}\,\frac{2N-5}{3N-7}\,\epsilon.
\ee
It is interesting to note that in the large-$N$ limit these critical data tend to those of the Lee-Yang model \cite{Rong:2017cow}
\be 
\gamma  = -\frac{1}{18}\, \epsilon, \qquad
\gamma^2_2 = \frac{4}{9}\,\epsilon.
\ee

\subsection{Quartic (restricted) Potts model}\label{sect:quartic-potts-cft}

Let us now move to the quartic theory. In this case we impose on top of permutation symmetry also a $\mathbb{Z}_2$ symmetry and hence refer to it as the restricted quartic Potts model \cite{Rong:2017cow}.
The action of the $S_{N+1} \times \mathbb{Z}_2$ invariant theory in $d=4-\epsilon$ is as follows
\begin{eqnarray}
 S[\phi] 
 &=&
 \int {\rm d}^d x \Bigl\{
 \frac{1}{2}(\partial\phi)^2 + \frac{1}{4}\zeta_{4,1} (\phi^2)^2+ \frac{1}{4!} \zeta_{4,2} \,q^{(4)}_{i_1 i_2 i_3 i_4} \phi_{i_1}\phi_{i_2}\phi_{i_3}\phi_{i_4}\Bigr\}.
\end{eqnarray}
As already mentioned in general the equation of motion
\be
\Box \phi_i =\zeta_{4,1} \phi^2 \phi_i +\frac{1}{3!} \zeta_{4,2} \,q^{(4)}_{i i_1 i_2 i_3} \phi_{i_1}\phi_{i_2}\phi_{i_3},
\ee
shows that turning on the interactions there is a recombination of the conformal multiplets such that the composite operator on the r.h.s has a scaling dimension $2+\Delta$, where $\Delta$ is the scaling dimension of the fields $\phi_i$ which in this case are all the same because of symmetries. 
\subsubsection{Anomalous dimension}

To compute the anomalous dimension for the quartic model where $n=2$ we need the eigenvalue of the quadratic tensor in $T^{(4)}_{abcd}$ defined in \eqref{t4}. This is given by 
\be 
T^{(4)}_{abc\, p} T^{(4)}{}^{abc}{}_{q} =  \left[\zeta^2_{4,1}\frac{1}{3}(N+2) +\zeta^2_{4,2}(N^2-N+1) + 2\,\zeta_{4,1}\zeta_{4,2}N\right]\delta_{pq}\,,
\ee
from which the anomalous dimension follows directly upon setting $n=2$ in \eqref{eta-cft} 
\be 
\eta  = \frac{c^2}{96} \left[\zeta^2_{4,1}\frac{1}{3}(N+2) +\zeta^2_{4,2}(N^2-N+1) + 2\,\zeta_{4,1}\zeta_{4,2}N\right]\,.
\ee
This expression agrees with the results obtained from RG analysis \cite{CSVZ4}.

\subsubsection{Quadratic operators}\label{sect:quad_op_4}

The computation of the critical exponents of the quadratic operators for the case of (restricted) quartic Potts model, which corresponds to $n=2$, is easier compared to the other cases as the eigenvalue equation \eqref{crit-cft-2} is linear in the potential. 
Indeed following closely the same strategy used for the single field $\phi^4$ theory~\cite{Codello:2017qek} and discussed in general in Section.~\ref{ss:qo}, 
one can start from the correlator 
\be
\langle \phi_i(x) \phi_j(y) [S_{pq} \phi_p \phi_q](z)\rangle,
\ee
where $[S_{pq} \phi_p \phi_q]$ must be a scaling operator with anomalous dimension $\gamma^S_2$.
Making use of the SDE on one hand we have
\begin{align}
\Box_x \braket{  \phi_i(x) \phi_j(y) [S_{pq} \phi_p \phi_q](z) }&\overset{{\rm LO}}{=}   \frac{{8c^2} \gamma\, S_{ij}}{|y-z|^{2 } |z-x|^{4}} 
- \frac{4c^2(2\gamma\!-\!\gamma^S_2) S_{ij}}{|x-y|^{2 }|z-x|^{4}} \,,
\label{box3pf1gamma2}
\end{align}
expression which should match at leading order
\begin{align}
 \braket{\Box_x \phi_i(x) \phi_j(y) [S_{pq} \phi_p \phi_q](z) } &= 
 \frac{T^{(4)}_{iabc} S_{pq}}{3!} \braket{[\phi_a\phi_b\phi_c](x) \phi_j(y)  [\phi_p \phi_q](z)  }\nn \\
 &\overset{{\rm LO}}{=}  T^{(4)}_{ijpq} S_{pq} \frac{c^3}{|x-y|^{2 }|z-x|^{4}} \,.
\end{align}
We can therefore deduce that at leading order $\gamma/\gamma^S_2 \to 0$, i.e. $\gamma^S_2$ depends linearly on the marginal couplings and write the eigenvalue equation
\be
\gamma^S_2 S_{ij}=\frac{c}{4} T^{(4)}_{ijpq} S_{pq}.
\label{gamma2eq}
\ee
To obtain the critical exponents one has to diagonalize the matrix 
\be \label{v=t4} 
\mathcal{M}_{i_1i_2i_3i_4} = \frac{c}{4} T^{(4)}_{i_1i_2i_3i_4} = \frac{c}{4}\zeta_{4,1}\delta_{(i_1i_2} \delta_{i_3i_4)}+\frac{c}{4}\zeta_{4,2}\,q^{(4)}_{i_1i_2i_3i_4}.
\ee
From this, the parameters defined in \eqref{M} are immediately read off
\be 
\tau = \frac{c}{4}\zeta_{4,2}, \qquad
\rho = \frac{c}{12}\zeta_{4,1}, \qquad
\kappa = \frac{c}{6}\zeta_{4,1}.
\ee
Given the values of these parameters and the fact that the anomalous dimension is of higher order, the eigensolutions of the stability matrix can then be summarized using \eqref{es1}-\eqref{es3} as follows
{\setlength\arraycolsep{2pt}
\bea \label{es0-quartic} 
\gamma_2^1  &=& \frac{c}{12} \zeta_{4,1} (N+2) +\frac{c}{4}\zeta_{4,2}N, \hspace{47pt} (P_1)_{ij,kl}\,\phi_k\phi_l, \label{es1-quartic} 
\\[2pt]
\gamma_2^2  &=& \frac{c}{6} \zeta_{4,1}+\frac{c}{4}\zeta_{4,2}(N-1),\hspace{64pt} (P_2)_{ij,kl}\,\phi_k\phi_l,
\\
\gamma_2^3  &=& \frac{c}{6}\zeta_{4,1}, \hspace{146pt} (P_3)_{ij,kl}\,\phi_k\phi_l. \label{es2-quartic}
\eea}%

\subsubsection{Critical couplings $\zeta_{4,1}(\epsilon)$ and $\zeta_{4,2}(\epsilon)$}\label{ss:cc-quartic} 

As in the cubic model, one can fix the $\epsilon$ dependence of the critical couplings $\zeta_{4,1}(\epsilon)$ and $\zeta_{4,2}(\epsilon)$ also in this case. This is of course a special case of the general analysis presented in Section \ref{ss:rr}, but let us take a slightly different root which can serve also as a crosscheck in this particular case. Before getting into the actual calculation let us first review the single-field case. For a single scalar field we have at leading order
\be  
\square_x\square_y \langle \phi(x) \,\phi(y)\, \phi(z)^{2}  \rangle
\stackrel{\mathrm{LO}}{=} C^{\mathrm{free}}_{112} \,  \frac{4(\eta-\gamma_2)(\eta +\gamma_2-\epsilon)}{|x-y|^{4}|x-z|^{2}|y-z|^{2}}, 
\ee
which we can compare, using the SDE, to
\be 
\langle \square_x\phi(x) \,\square_y\phi(y)\, \phi(z)^{2} \rangle 
\stackrel{\mathrm{LO}}{=} \frac{g^2}{(2n-1)!^2} \frac{C^{\mathrm{free}}_{2n-1,2n-1,2}}{|x-y|^{4}|x-z|^{2}|y-z|^{2}}, 
\ee 
where $g$ is the coefficient of $\frac{1}{4!}\phi^4$ in the Lagrangian. Equating the two and using the known result $\gamma_2 = cg/4$, which may be found by applying only one $\Box_x$ to the same correlator as the above, and the fact that $\eta=\mathcal{O}(\epsilon^2)$, one finds $g = 4c\epsilon/3$.   

The generalization of this calculation to the multi-field case is slightly more subtle because of the fact that there are more than one quadratic operators and one has to pick those with a definite scaling obtained in Section~\ref{sect:quad_op_4}. Using the projectors in the space of the scaling quadratic operators introduced in Section.~\ref{ss:qo-symm} we write on one hand
\be  
\square_x\square_y \langle \phi_i(x) \,\phi_j(y)\, [(P_a)_{kl,rs}\phi_r\phi_s](z) \rangle
\stackrel{\mathrm{LO}}{=} C^{\mathrm{free}}_{112} \,  \frac{4(\eta-\gamma^a_2)(\eta +\gamma^a_2-\epsilon)}{|x-y|^{4}|x-z|^{2}|y-z|^{2}} (P_a)_{kl,ij},
\ee
while on the other hand, using the SDE, we obtain
{\setlength\arraycolsep{2pt}
\bea 
&& \langle \square_x\phi_i(x) \,\square_y\phi_j(y)\, [(P_a)_{kl,rs}\phi_r\phi_s](z) \rangle
\nn\\
&=& \frac{1}{3!^2}\langle [T^{(4)}_{ii_1i_2i_3}\phi_{i_1}\phi_{i_2}\phi_{i_3}](x) \,[T^{(4)}_{jj_1j_2j_3}\phi_{j_1}\phi_{j_2}\phi_{j_3}](y)\, [(P_a)_{kl,rs}\phi_r\phi_s](z) \rangle \nn\\
&\stackrel{\mathrm{LO}}{=}& \frac{C^{\mathrm{free}}_{332}}{3!^2} \,  \frac{T^{(4)}_{pqir}T^{(4)}_{pqjs}(P_a)_{kl,rs}}{|x-y|^{4}|x-z|^{2}|y-z|^{2}}.
\eea}%
Equating the two expressions we get
\be 
T^{(4)}_{pqir}T^{(4)}_{pqjs}(P_a)_{kl,rs} c^4 = 8c^2 (\eta-\gamma^a_2)(\eta +\gamma^a_2-\epsilon)(P_a)_{kl,ij}
\ee
and noting that $\gamma^a_2=O(\epsilon)$ while $\eta=O(\epsilon^2)$ one can finally write
\be 
c^2T^{(4)}_{pqir}T^{(4)}_{pqjs}(P_a)_{kl,rs}  = 8\gamma^a_2(\epsilon-\gamma^a_2)(P_a)_{kl,ij}.
\ee
This equation can be specialized in the three different subspaces for the quadratic scaling operators. For $a=1$ we have
{\setlength\arraycolsep{2pt}
\bea
&& \frac{1}{3} \left[(N+2)\zeta^2_{4,1} +3(N^2-N+1)\zeta^2_{4,2} + 6N\zeta_{4,1}\zeta_{4,2}\right]c^2 \nn\\
&=& 8 \left(\frac{c}{12} \zeta_{4,1} (N+2) +\frac{c}{4}\zeta_{4,2}N\right)\left(\epsilon - \frac{c}{12} 
\zeta_{4,1} (N+2) -\frac{c}{4}\zeta_{4,2}N\right).
\eea}%
while for $a=2$ we get
{\setlength\arraycolsep{2pt}
\bea
&& \frac{1}{9} \left[(N+6)\zeta^2_{4,1} +9(N^2-2N+2)\zeta^2_{4,2} + 6(3N-2)\zeta_{4,1}\zeta_{4,2}\right]c^2 \nn\\
&=& 8 \left(\frac{c}{6} \zeta_{4,1}+\frac{c}{4}\zeta_{4,2}(N-1)\right)
\left(\epsilon - \frac{c}{6} \zeta_{4,1}-\frac{c}{4}\zeta_{4,2}(N-1)\right).
\eea}%
and finally $a=3$ gives 
\be 
\frac{1}{9} \left[(N+6)\zeta^2_{4,1} +9\zeta^2_{4,2} + 6N\zeta_{4,1}\zeta_{4,2}\right]c^2 = 8 \frac{c}{6}\zeta_{4,1}\left(\epsilon - \frac{c}{6}\zeta_{4,1}\right).
\ee
Despite the fact that there are three equations and two unknown variables, one can check that picking any pair of equations gives rise to the following four solutions for the critical couplings 
\be
\frac{c}{4}\left( \zeta_{4,1}, \zeta_{4,2}\right) = \!\Biggl\{\left(0, 0 \right), \left(\frac{3 \epsilon}{N\!+\!8}, 0 \right),  \frac{\epsilon}{N\!+\!3} \left(1 \,,\, \frac{1}{3} \right), 
\frac{\epsilon}{N^2\!-\!5N\!+\!8} \left(1 \,,\, \frac{N\!-\!4}{3} \right)\Biggr\} 
\label{FP4}
\ee
which are equal to those found from RG at leading order~\cite{CSVZ4}, after making the replacement $\zeta_{4,1}\rightarrow 4\zeta_{4,1}/c$.
Therefore with this method one can determine at leading order the three different nontrivial CFTs which correspond to the scale invariant theories (RG fixed points). 

Now that we have obtained the leading order $\epsilon$-dependence of the couplings, let us express some of the critical data that we have found in terms of $\epsilon$. These universal results are given here, as an example, at the third critical point in \eqref{FP4}. For this model the field anomalous dimension is
\be  
\eta  = \frac{(N+1)(N+7)}{54(N+3)^2} \epsilon\,.
\ee
Also, the anomalous dimensions of the three scaling operators are
\be \label{es-quartic-eps} 
\gamma_2^1  = \frac{2(N+1)}{3(N+3)} \epsilon, 
\qquad
\gamma_2^2  = \frac{N+1}{3(N+3)} \epsilon,
\qquad
\gamma_2^3  = \frac{2}{3(N+3)} \epsilon. 
\ee

\subsubsection{Quartic scaling operators}

Having at our disposal the general eigenvalue equation \eqref{gamma-k} for arbitrarily high order operators for even models, in the presence of permutation symmetry which significantly constrains the form of the potential we can take advantage of it to obtain some scaling operators and their corresponding anomalous dimensions for higher order operators. 
let us consider as an 
example the quartic operators, which will be used also in the next section. For simplicity we restrict ourselves to the space of invariant quartic operators, that is linear combinations of the form 
\be
\xi_{4,1} \,\delta_{(i_1 i_2} \delta_{i_3 i_4)} + \xi_{4,2} \,q^{(4)}_{i_1 i_2 i_3 i_4}.
\ee
Inserting this into \eqref{gamma-k} using the fact that $V_{i_1i_2i_3i_4} = T^{(4)}_{i_1i_2i_3i_4}$ leads to a two dimensional eigenvalue problem 
\be 
\gamma^S_4
\left(
\ba{c} \xi_{4,1} \\ \xi_{4,2} \ea
\right)
= 
\left(
\ba{cc} 
\frac{2}{3}(N+8)\zeta_{4,1}+2N\zeta_{4,2} & 2N\zeta_{4,1}+6\zeta_{4,2} \\
4\zeta_{4,2} & 4\zeta_{4,1}+6(N-1)\zeta_{4,2} 
\ea
\right)
\left(
\ba{c} \xi_{4,1} \\ \xi_{4,2} \ea
\right)
\ee
which can be solved easily.  Below, for each of the nontrivial fixed points of \eqref{FP4} we report the two scaling operators together with their corresponding anomalous dimensions, which are obtained by solving the above eigenvalue equation. For the first nontrivial fixed point in \eqref{FP4} these are
{\setlength\arraycolsep{2pt}
\bea \label{o1-quart1}
\mathcal{O}^{(4)}_1 &=& \frac{1}{\sqrt{8N(N+2)}} (\phi\!\cdot\!\phi)^2, 
\\ \label{o2-quart1}
\mathcal{O}^{(4)}_2 &=& \sqrt{\frac{N+2}{24N(N-2)(N^2-1)}} \left(q^{(4)}_{ijkl}\,\phi_i\phi_j\phi_k\phi_l - \frac{3N}{N+2}(\phi\!\cdot\!\phi)^2\right). 
\eea}%
\be 
\gamma^1_4 = 2\epsilon, \qquad \gamma^2_4 = \frac{12\epsilon}{N+8}
\ee
Notice that both expressions diverge when $N$ vanishes and also the last expression blows up in the limit $N\rightarrow -1$ which is a sign that the norm of the operators inside parenthesis vanishes in these limits. At the second nontrivial fixed point the scaling operators and their anomalous dimensions are
{\setlength\arraycolsep{2pt}
\bea  \label{o1-quart2}
\mathcal{O}^{(4)}_1 &=& \frac{1}{\sqrt{24N(N+1)(N+7)}} \left(q^{(4)}_{ijkl}\,\phi_i\phi_j\phi_k\phi_l +3(\phi\!\cdot\!\phi)^2\right). 
\\ \label{o2-quart2}
\mathcal{O}^{(4)}_2 &=& \frac{1}{\sqrt{2N(N-1)(N-2)(N+7)}} \left(q^{(4)}_{ijkl}\,\phi_i\phi_j\phi_k\phi_l -\frac{N+1}{2}(\phi\!\cdot\!\phi)^2\right). 
\eea}%
\be 
\gamma^1_4 = 2\epsilon, \qquad \gamma^2_4 = \frac{4(N+1)\epsilon}{3(N+3)}
\ee
Again, at $N=0$ both operators and at $N=-1$ the first operator blows up. Finally at the third nontrivial fixed point we have
{\setlength\arraycolsep{2pt}
\bea \label{o1-quart3}
\mathcal{O}^{(4)}_1 &=& \frac{N-4}{\sqrt{24N(N-1)(N-2)(N^2-6N+11)}} \left(q^{(4)}_{ijkl}\,\phi_i\phi_j\phi_k\phi_l +\frac{3}{N-4}(\phi\!\cdot\!\phi)^2\right). 
\\  \label{o2-quart3}
\mathcal{O}^{(4)}_2 &=& \frac{1}{\sqrt{8N(N+1)(N^2-6N+11)}} \left(q^{(4)}_{ijkl}\,\phi_i\phi_j\phi_k\phi_l -(N-2)(\phi\!\cdot\!\phi)^2\right). 
\eea}%
\be 
\gamma^1_4 = 2\epsilon, \qquad \gamma^2_4 = \frac{2(N-1)(N-2)\epsilon}{3(N^2-5N+8)}
\ee
As the reader might have already noticed, the first operators at each fixed point, i.e. operators \eqref{o1-quart1}, \eqref{o1-quart2}, \eqref{o1-quart3} all have anomalous dimension $\gamma^1_4 = 2\epsilon$. In fact, apart from a normalization factor, these operators are nothing but the potential itself evaluated at the corresponding fixed point, and according to the discussion at the end of Section \ref{ss:gem}, regardless of the fixed point, the critical operator \eqref{V} is always a scaling operator with anomalous dimension \eqref{gamma-V-2n} which in the present case of $n=2$ is equal to $2\epsilon$.

\subsubsection{Structure constants: some examples}
As an example of a structure constant for the quartic Potts model one can consider for instance the expression \eqref{ciuv-even} evaluated for $n=2$. The simplest case corresponds to $p=2$ and $q=1$, which gives a generalization of $C_{114}$. The general expression for this structure constant, when the rescaling $\phi_i \rightarrow \phi_i/\sqrt{c}$ has been done to bring the two-point functions to unity, is 
\be \label{C114}
C_{\phi_i \phi_j \cS_ 4} = \frac{V_{iklm}\,S_{j klm}}{3!} \frac{c}{8}C^{\mathrm{free}}_{1,3,4} = \frac{c}{2}T^{(4)}_{iklm}\,S_{j klm},
\ee 
In order to evaluate this explicitly one needs to choose the operator $\cS_4$ appropriately, i.e. such that it is a scaling operator satisfying \eqref{gamma-k}. As pointed out earlier in Section \ref{ss:gem} and shown explicitly in the previous section, at order $\epsilon$ one of the scaling operators is always the potential itself which corresponds to taking $S_{i_1i_2i_3i_4} = T^{(4)}_{i_1i_2i_3i_4}$. Such operators have been normalized to unity and reported in equations \eqref{o1-quart1}, \eqref{o1-quart2} and \eqref{o1-quart3} respectively for the three nontrivial critical theories given in \eqref{FP4}. Using the explicit form of these scaling operators which all have eigenvalue $2\epsilon$, and inserting them into the general equation \eqref{C114} we obtain the structure constants, respectively for the three nontrivial critical points of \eqref{FP4}
\be
C_{\phi_i \phi_j {\cal O}^{(4)}_1} 
= \sqrt{\frac{N+2}{2N}} \,\frac{\epsilon \delta_{ij}}{N+8},
\ee
\be
C_{\phi_i \phi_j {\cal O}^{(4)}_1} 
= \sqrt{\frac{N^2+8N+7}{6N}} \,\frac{\epsilon \delta_{ij}}{3(N+3)},
\ee
\be
C_{\phi_i \phi_j {\cal O}^{(4)}_1}  = \sqrt{\frac{N^4-9N^3+31N^2-45N+22}{6N}} \,\frac{\epsilon \delta_{ij}}{3(N^2-5N+8)}.
\ee
These results are in complete agreement with RG analysis~\cite{CSVZ4}. As the next step one may be tempted to calculate the structure constants involving the operators ${\cal O}^{(4)}_2$. However, one can argue that, regardless of the critical theory, replacing the operator $\cS_4$ in \eqref{C114} with the operators ${\cal O}^{(4)}_2$ or any other quartic operator with an eigenvalue different from that of ${\cal O}^{(4)}_1$, the resulting structure constants vanish. This can be seen as follows. By symmetry arguments, the structure constant \eqref{C114} which has two free indices can only be proportional to $\delta_{ij}$. This means that \eqref{C114} is proportional to its trace, i.e. setting $i=j$ and summing over the index. But the trace is proportional to the two-point function of the operators $V$ and $\cS_4$ which vanishes if they have different eigenvalues. This can also be checked explicitly for the three ${\cal O}^{(4)}_2$ operators given in the previous section.

\subsection{Quintic Potts model}\label{sect:quintic-potts-cft}

The last model that we consider is the critical quintic model which is defined in $d_c=\frac{10}{3}$. The $S_{N+1}$-invariant action takes the form
\be 
 S[\phi] = \int {\rm d}^d x \Bigl\{
 \frac{1}{2}(\partial\phi)^2
 + \frac{1}{5!}\Bigl(\zeta_{5,1} \delta_{(i_1 i_2} q^{(3)}_{i_3 i_4 i_5)} +\zeta_{5,2} \,q^{(5)}_{i_1 i_2 i_3 i_4 i_5}\Bigr)\phi_{i_1}\dots \phi_{i_5}
 \Bigr\}\,,
\ee
with two marginal couplings. In the following we give several results for the critical data in the $\epsilon$ expansion at leading order. Notice that one finally has to set $\epsilon=1/3$ in order to get the results in three dimensions.

\subsubsection{Anomalous dimension}

The quintic model corresponds to the multicriticality label  $n=5/2$. In this case the computation of the anomalous dimension requires the following quadratic tensor appearing in \eqref{t5}
{\setlength\arraycolsep{2pt}
\bea
T^{(5)}_{abcd\, p} T^{(5)}{}^{abcd}{}_{q} &=& \left[\zeta^2_{5,1}\frac{3}{100}(N+18) + \zeta^2_{5,2}(N^2-N+1)+ \zeta_{5,1}\zeta_{5,2}\frac{3}{5}(3N-2)\right] N q^{(2)}_{pq} \nn\\
&+& \left[\zeta^2_{5,1}\frac{1}{100}(N^2+8N-42)- \zeta^2_{5,2}+ \zeta_{5,1}\zeta_{5,2}\frac{1}{5} N(N-4)\right] q^{(2)}_{pq} \nn\\
&+& \left[\zeta^2_{5,1}\frac{3}{50}(N-3)\right]N q^{(2)}_{pq} + \left[\zeta^2_{5,1}\frac{3}{50}(N-3)\right]q^{(2)}_{pq} \nn\\[3mm]
&=& \left[\zeta^2_{5,1}\frac{1}{10}(N+6) +\zeta^2_{5,2}(N^2+1) + 2\,\zeta_{5,1}\zeta_{5,2}N\right](N-1)q^{(2)}_{pq}\,.
\eea}%
This can then be used to obtain the anomalous dimension in terms of the two critical couplings
\be \label{eta-quintic} 
\eta  = \frac{3c^3}{640}\left[\zeta^2_{5,1}\frac{1}{10}(N+6) +\zeta^2_{5,2}(N^2+1) + 2\,\zeta_{5,1}\zeta_{5,2}N\right](N-1)\,.
\ee
This expression agrees with the findings of RG analysis~\cite{CSVZ4}.

\subsubsection{Quadratic operators} \label{ss:fo-103} 
Let us now turn to the critical exponents of operators of the form $S_{ab}\phi_a\phi_b$. For these to be eigenoperators in the quintic model they must satisfy 
\be 
(\gamma^S_2 - \eta) \; S_{ij} =\frac{3c^3}{32}\, V_{i\,p\,i_2i_3 i_4}V_{j\,q\,i_2 i_3 i_4} \,S_{pq},
\ee
which is a special case of \eqref{crit-cft} for $n=5/2$. This eigenvalue equation is consistent with the linear flow of the couplings of quadratic operators $\phi_a\phi_b$ obtained directly with RG methods~\cite{CSVZ4}.
In order to diagonalize this equation one needs to find the eigenvectors and eigenvalues of the matrix
\be \label{M=T5T5}
\mathcal M_{ij, pq} = \frac{3c^3}{32} T^{(5)}_{abc\, ip} T^{(5)}{}^{abc}{}_{jq}, \qquad 
T^{(5)}_{i_1i_2i_3i_4i_5} = \zeta_{5,1}\delta_{(i_1i_2} q^{(3)}_{i_3i_4i_5)}+\zeta_{5,2}\,q^{(5)}_{i_1i_2i_3i_4i_5}.
\ee
The details of this computation are collected in Appendix~B.\ref{ss:details}. Here we report the final result, i.e. the eigenvalues 
{\setlength\arraycolsep{2pt}
\bea 
\label{gamma21-quintinc} 
\gamma_2^1  &=& \frac{63c^3}{640}\!\!\left[\frac{1}{10}(N+6)\zeta^2_{5,1} +(N^2+1)\zeta^2_{5,2} +  2N\zeta_{5,1}\zeta_{5,2}\right]\!(N-1) = 21\eta,
\\[1mm]
\label{gamma22-quintinc}
\gamma_2^2  &=& \frac{3c^3}{640}\bigg[\frac{9\zeta_{5,1}^2}{10}(N^2+15N-26)  + (21N^3-41N^2+41N-41)\zeta_{5,2}^2 
\nn\\[-1mm]
&& \hspace{10mm} + 6(7N^2-13N+4)\zeta_{5,1}\zeta_{5,2}\bigg], 
\\[1mm]
\label{gamma23-quintinc}
\gamma_2^3  &=&\! \frac{3c^3}{640}\left[\frac{3}{10}\zeta^2_{5,1}(N+14) +\zeta^2_{5,2}(N^2+2N+7) + 6\zeta_{5,1}\zeta_{5,2}N\right]\!(N-3),
\eea}%
respectively corresponding to the scaling operators $(P_a)_{ij,kl}\,\phi_k\phi_l$  for $a=1,2,3$. The value of $\gamma_2^1$ is again seen to be consistent with the general result \eqref{gamma20-eta}.   

\subsubsection{Structure constants}

Among the general classes of structure constants that we have obtained in this work, the ones in Sections~\ref{ssec:c12p2q},~\ref{ssec:c111} and~\ref{ssec:c112k} can be applied to the case $n=5/2$. The simplest examples, that do not involve high order operators are the 
multi-field generalizations of $C_{122}$ and $C_{111}$. The former is given by \eqref{ci2p2q} upon setting $p=q=1$ and reads 
\be \label{c122-quintic}
C_{\phi_i \cS_2 \tilde{\cS}_2} 
= -\frac{9}{4}\; c^{\scriptscriptstyle 3/2}\, T^{(5)}_{i\,a_1a_2\,b_1b_2}\,S_{a_1a_2} \, \tilde S_{b_1b_2},
\ee
where appropriate rescaling of the fields has been done to accord with the usual CFT normalization, as discussed at the end of Section \ref{ssec:c12p2q-1}. This can be compared with the OPE coefficient 
determined using the renormalization group~\cite{CSVZ4}, provided suitable rescalings are done in the beta function.
In order to obtain the explicit form of these structure constants (OPE coefficients) we choose the scaling operators $\cS_2$ and $\tilde{\cS}_2$ among \eqref{o2}. This gives  
\be
C_{\phi_i\,\mathcal{O}_{a,pq}^{(2)} \mathcal{O}_{b,rs}^{(2)}} = -\frac{9}{8}\; c^{\scriptscriptstyle 3/2}\, T^{(5)}_{i\,ab\,ef}\, (P_a)_{pq,ab}\,(P_b)_{rs,ef}.
\ee
These structure constants vanish for the cases $(a,b)=(1,1),(1,3)$ and for the nontrivial cases they are given as follows 
\be  
C_{\phi_i\,\mathcal{O}_{1,pq}^{(2)} \mathcal{O}_{2,rs}^{(2)}} = -\frac{9((N+6)\zeta_{5,1}+10N\zeta_{5,2})}{80N}\, c^{\scriptscriptstyle 3/2} \,\delta_{pq} q^{(3)}_{irs},
\ee
\be  
C_{\phi_i\,\mathcal{O}_{2,pq}^{(2)} \mathcal{O}_{2,rs}^{(2)}} = -\frac{9((4N-6)\zeta_{5,1}+5(N-1)^2\zeta_{5,2})}{40(N-1)^2}\, c^{\scriptscriptstyle 3/2}\left(q^{(5)}_{ipqrs}-\delta_{pq}q^{(3)}_{irs}-\delta_{rs}q^{(3)}_{ipq}\right),
\ee
{\setlength\arraycolsep{2pt}
\bea 
C_{\phi_i\,\mathcal{O}_{2,pq}^{(2)} \mathcal{O}_{3,rs}^{(2)}} = \frac{9}{40}(N-3)\zeta_{5,1} \,c^{\scriptscriptstyle 3/2} &&\left[\frac{1}{(N-1)^2} q^{(5)}_{ipqrs}-\frac{1}{N-1}\delta_{i(r}q^{(3)}_{s)pq} \right. \nn\\
- && \left. \frac{1}{(N-1)^2}\delta_{pq}q^{(3)}_{irs}-\frac{1}{N(N-1)^2}\delta_{rs}q^{(3)}_{ipq}\right],
\eea
{\setlength\arraycolsep{2pt}
\bea 
C_{\phi_i\,\mathcal{O}_{3,pq}^{(2)} \mathcal{O}_{3,rs}^{(2)}} = \frac{9}{20}\,\zeta_{5,1} \,c^{\scriptscriptstyle 3/2} &&\left[\frac{N}{(N-1)^2} q^{(5)}_{ipqrs}-\frac{1}{4}\left(\delta_{rp}q^{(3)}_{isq}+\delta_{rq}q^{(3)}_{isp}+\delta_{sp}q^{(3)}_{irq}+\delta_{sq}q^{(3)}_{irp}\right) \right. \nn\\
- && \left. \frac{1}{2(N-1)}\left(\delta_{ip}q^{(3)}_{qrs}+\delta_{iq}q^{(3)}_{prs}+\delta_{ir}q^{(3)}_{pqs}+\delta_{is}q^{(3)}_{pqr}\right) \right. \nn\\
- && \left. \frac{1}{(N-1)^2}\left(\delta_{pq}q^{(3)}_{irs}+\delta_{rs}q^{(3)}_{ipq}\right)\right].
\eea
As discussed earlier, the indices $ij$ provide a redundant description of the set of operators.  
Using the nonredundant descriptions given by \eqref{o0} and \eqref{ok}  one may re-express the above structure constants apart from the last one. These are   
\be  
C_{\phi_i\,\mathcal{O}_0^{(2)} \mathcal{O}_j^{(2)}} = -\frac{9}{80}\, c^{\scriptscriptstyle 3/2} (N-1)((N+6)\zeta_{5,1}+10N\zeta_{5,2})\,\delta_{ij},
\ee
\be  
C_{\phi_i\,\mathcal{O}_j^{(2)} \mathcal{O}_k^{(2)}} = -\frac{9}{40}\, c^{\scriptscriptstyle 3/2}\,((4N-6)\zeta_{5,1}+5(N-1)^2\zeta_{5,2}) \, q^{(3)}_{ijk},
\ee
\be 
C_{\phi_i\,\mathcal{O}_j^{(2)} \mathcal{O}_{3,kl}^{(2)}} = -\frac{9}{40}\,c^{\scriptscriptstyle 3/2}\,\zeta_{5,1} (P_3)_{ij,kl}.
\ee
The structure constant $C_{111}$ can also be generalized in this case. Contrary to the cubic model for which $C_{ijk}$ was obtained from \eqref{ci2p-12q-1}, for the quintic model this is obtained from Eq.~\eqref{cijk}. Setting $\ell=2$ we get
\be 
C_{\phi_i \phi_j \phi_k} = \frac{729}{4096}\, c^{\scriptscriptstyle 9/2} \, V_{ia_1a_2b_1b_2} V_{jb_1b_2c_1c_2} V_{kc_1c_2a_1a_2}
\ee
Replacing the potential derivatives with the expression \eqref{t4} and contracting the indices we get
{\setlength\arraycolsep{2pt}
\bea
C_{\phi_i \phi_j \phi_k} &=& \frac{729}{4096}\, c^{\scriptscriptstyle 9/2} \left[\frac{1}{250}(11N^2+146N-240) \zeta_{5,1}^3  + (N-1)^2(N^2+2)\zeta_{5,2}^3 \right. \nn\\ 
&& \left. \hspace{15mm}+ \frac{3}{100}(N^3+88N^2-159N+54) \zeta_{5,1}^2\zeta_{5,2} \right. \nn\\  
&& \left.  \hspace{15mm}+ \frac{3}{5}(5N^3-10N^2+9N-6) \zeta_{5,1}\zeta_{5,2}^2 \right] q^{(3)}_{ijk}  \,.
\eea}%
Let us stress that contrary to the cubic and the quartic models, in our CFT+SDE approach, and without exploiting the RG analysis, we cannot fix the relation among the critical couplings and $\epsilon$ for the same reasons as that in the single-field case. Therefore in this case we can use our RG results~\cite{CSVZ4}, which have been reported in Appendix 
B.\ref{sec:quintic_RG} in order to write explicit expressions in $\epsilon$.

\subsubsection{Some universal results}

We collect in this section a few examples of critical quantities expressed in terms of $\epsilon$. The results are reported for two critical points that are infrared, one for the case $N=0$ corresponding to percolation theory and one for $N=-1$ corresponding to spanning forest. For $N=0$ we take the second critical point \eqref{n0-2} and report as a few examples the following critical quantities which include the field anomalous dimension, the three anomalous dimensions of the quadratic operators as well as two structure constants
\be 
\eta = -0.00431599 \epsilon, \quad
\ba{l}
\gamma^1_2 = -0.0906358 \epsilon, \\[1mm]
\gamma^2_2 = -0.118809 \epsilon, \\[1mm]
\gamma^3_2 = -0.0906358 \epsilon, 
\ea
\quad
\ba{c}
C_{\phi_i\,\mathcal{O}_0^{(2)} \mathcal{O}_j^{(2)}} = 0.39996\, \epsilon^{\scriptscriptstyle 1/2} \,\delta_{ij}, 
\\
C_{\phi_i \phi_j \phi_k} = - 0.08266  \,\epsilon^{\scriptscriptstyle 3/2} \, q^{(3)}_{ijk}  \,.
\ea
\ee
For the case $N=-1$ we pick the second critical point \eqref{n1-2}. For this model the anomalous dimensions of the quadratic operators and the two structure constants are 
\be 
\eta = -0.000509232 \epsilon, \quad
\ba{l}
\displaystyle \gamma^1_2 = -0.0106939 \epsilon, \\[1mm]
\gamma^2_2 = -0.0183324 \epsilon, \\[1mm]
\gamma^3_2 = -0.0816536 \epsilon, 
\ea
\quad
\ba{c}
C_{\phi_i\,\mathcal{O}_0^{(2)} \mathcal{O}_j^{(2)}} = - 0.3708\, \epsilon^{\scriptscriptstyle 1/2} \,\delta_{ij}, 
\\
C_{\phi_i \phi_j \phi_k} = - 2.04362  \,\epsilon^{\scriptscriptstyle 3/2}  \, q^{(3)}_{ijk}  \,.
\ea
\ee
The $\epsilon$-dependence of the critical points presented in Appendix B.\ref{sec:quintic_RG} can be calculated also analytically, from which analytic results could have been given for the above quantities. However, these are complicated expressions involving square roots. We therefore avoid such expressions and present here only the approximate numerical values.

\section{Conclusions}\label{sect:conclusions}

We have employed a general method based on the use of conformal symmetry and Schwinger-Dyson equations to investigate multicritical multi-field scalar QFTs,
characterized by a critical potential of order $m$. At the leading order in the perturbative $\epsilon$-expansion,
this method gives access, in a simple way, to some universal data which includes both nontrivial anomalous dimensions and structure constants.
These results generalize the method applied to generic multicritical models of a single scalar field presented in~\cite{Codello:2017qek}. 
Even without assuming any symmetry and considering only two- and three-point functions,
one can already find a considerable amount of information which includes the anomalous dimensions of the fields, the scaling dimension of the quadratic composite operators,
a tower of all-order composite operators for the "even" ($m$=$2n$) unitary multicritical theories,
and the explicit form of several structure constants. For $m=2n$ and $m=3$ one can also find the equations that constrain the interaction potential at criticality.
In particular we show how these constraints can be cast in exactly the same form as fixed-point conditions which could be obtained from a functional perturbative RG analysis.
Most of the general results and computational strategies presented in this work are new: while part of the results could in principle be obtained from more standard perturbative RG methods,
one of the objectives of this paper has been to show how conformal symmetry alone can give access to such critical information. 

We remark here some interesting general features of our investigation. The results obtained in the first part of the paper for a general class of multi-field scalar QFTs
are derived without the use of any internal symmetry, but of course can be specialized to cases characterized by any (continuous or discrete) symmetry.
We have focused on the derivation of many universal quantities, but we stress that
also
the criticality conditions that
we have obtained on the set of possible couplings of the potential in the Landau-Ginzburg description are important by themselves.
In fact these conditions, which coincide with fixed point of functional beta functions determined with standard perturbative RG, are in many cases expected
to lead to the emergence of some symmetries at criticality, which is a fact that has already been observed in the literature (see~\cite{Osborn:2017ucf} and references therein). 
Clearly, with a growing number of fields the pattern of solutions is increasingly complex, but in principle
any multicritical multi-field scalar theory can be analyzed specializing our general framework.
Pursuing such an approach one can access -- at the perturbative level -- all the possible internal continuous
or discrete symmetries of a theory with given critical dimension $d_c$ and number of fields $N$ that is a CFT at criticality.
In other words one can expect constrained emergent symmetries at a critical point.

We have then specialized the analysis to potentials characterized by $S_q$ invariance,
which encompass the Potts model and some of its multicritical generalizations.
We have used the standard representation theory of $S_q$ to construct all the symmetric interaction terms in the multicritical potential.
Even before embarking on explicit calculations for particular models, we have explored how far we could get from a knowledge of the symmetry group alone.
We have given explicit expressions for the decomposition of the quadratic operators into scaling operators,
which carry irreducible representations of $S_q$, and we have presented model independent formulas for anomalous dimensions of such operators.  

The formalism gives the possibility to perform an analytic continuations of $q$ to some specific values,
e.g.\ the ones of special interest in statistical mechanics: $q=1$ (percolation) and $q=0$ (spanning forest).
Therefore, as an application we have analyzed in detail all the theories which have an upper critical dimension $d_c>3$,
which are nontrivial critical models in any integer dimensions $d$ in the range $3 \le d<d_c$, and specialized $q$ to the values of interest.
The results we have found match with the ones that will be presented in a companion paper~\cite{CSVZ4} which is devoted to the study of
Potts-like field theories with functional perturbative renormalization group methods
and which puts an emphasis on those with quintic interactions ($d_c=10/3$).
The results found there are confirmed by the present investigation for all critical and multicritical Potts-like models.
In the cubic (standard Potts) and quartic (restricted Potts)
critical models one can fix, with the aid of relations based on conformal invariance,
the critical values of the couplings as a function of $\epsilon$.
Unfortunately this is not possible in the quintic case, therefore we completed the analysis of the quintic model
using the RG results of the companion paper which for convenience we have included and briefly discussed in Appendix B.\ref{sec:quintic_RG}.
Likewise in the main text, we have also focused on the generalization of percolation and spanning forests universality classes,
for which we show that critical solutions associated to second order phase transition do exist, depending on the kind of multicritical theory considered.

There are several directions and extensions of this work that one can take for future investigations.
One such direction would be the inclusion of large-$N$ types of analyses.
It is not immediately clear if in this very general framework the large-$N$ expansion can be of help because we lack the constraints offered by a symmetry such as $O(N)$,
but we expect that there can be intermediate semi-general situations in which it could be useful. 
We also expect that extending the analysis to higher order correlation functions one can have access to further relations
and informations on the conformal data. In particular one could wish to extend the results to the next-to-leading nontrivial order in the $\epsilon$ expansion.
Furthermore, one could change the structure and the degrees of freedom of the theories: e.g.\ considering general tensor models or theories with fermion \cite{Torres:2018jij} or vector fields.
Again interesting constraints on the symmetries should be investigated both in this framework
and in the functional perturbative RG to get access to
their universal data.
Among these possible extensions we would like to include
the study of nonunitary higher-derivative theories, which have recently been investigated in the single-field case in \cite{Safari:2017irw,Safari:2017tgs}
and which are important to extend critical theories to higher dimensions.

A final future direction that we would like to point out is inspired by the works \cite{Gorbenko:2018ncu,Gorbenko:2018dtm}
and involves the study of ``walking transitions''. In some
systems scale invariance can be approximately realized because the renormalization group runs ``close'' to complex fixed points.
These complex CFTs are nonunitary, but otherwise fully consistent conformal theories with complex conformal data.
For these models too we expect that the general conditions of criticality derived in Section~\ref{sec:FPcond} select
the allowed internal symmetries compatible with the number of fields and the upper critical dimensions.
Furthermore, the $q$-states Potts model considered in this paper is a prototypical example for the investigation of complex CFTs because,
as a function of $q$, pairs of fixed points that are related by the reflection $\phi^i \to -\phi^i$
annihilate and morph into pairs of purely imaginary complex CFTs (structurally similar to the PT-invariance of the Lee-Yang model's potential \cite{Zambelli:2016cbw}).
Therefore, by opportunely tuning $q$ and $d$ it is realistically possible to encounter scenarios in
which walking transitions are realized.

\bigskip

\noindent \emph{Acknowledgements.}\\
OZ acknowledges support from the DFG under Grant No.~Za~958/2-1.

\appendix

\section{Free theory}
\label{free}

In this appendix we collect the expressions for the two and three-point functions in the free theory which will be used throughout the text. The two-point function is simply given as 
\be \label{phiphi-free}
\braket{\phi_i(x)  \phi_j(y)} \overset{\mathrm{free}}=\frac{c\, \delta_{ij}}{|x-y|^{2 \delta_c}}\,,
\ee
where  $\delta_c=\frac{1}{2}d_c-1$ is the dimension of the field $\phi_i$ in the free theory and
\be
c=\frac{1}{4 \pi} \frac{\Gamma(\delta_c)}{\pi^{\delta_c}}.
\label{c}
\ee
To bring the two-point function into the canonical form with a normalized coefficient one can work instead with the rescaled field $\tilde{\phi}_i$ defined through $\phi_i=\sqrt{c}\tilde\phi_i$. This can be used to obtain the two-point function 
\be
\braket{[\phi^{i_1}\cdots \phi^{i_k}](x)\,  [\phi_{j_1}\cdots \phi_{j_k}](y)} \overset{\mathrm{free}}{=}\delta^{i_1}_{(j_1}\cdots \delta^{i_k}_{j_k)} \,   \frac{k!\,c^k}{|x-y|^{2 k \delta_c}}\,,
\label{2pf-free}
\ee
where on the r.h.s the $j$ indices are symmetrized, and the symmetrization includes the inverse factor of $k!$. Consider now a generic three-point function 
\bea
&& \braket{[\phi_{i_1}\cdots \phi_{i_{n_1}}](x_1)\; [\phi_{j_1}\cdots \phi_{j_{n_2}}](x_2)\;[\phi_{k_1}\cdots \phi_{k_{n_3}}](x_3)}  \nn\\
&\overset{\mathrm{free}}{=}& 
  \frac{C_{i_1\cdots i_{n_1},j_1\cdots j_{n_2},k_1\cdots k_{n_3}}^{\mathrm{free}}}{|x_1\!-\!x_2|^{\delta_c (n_1+n_2-n_3)}|x_2\!-\!x_3|^{\delta_c (n_2+n_3-n_1)}
  |x_3\!-\!x_1|^{\delta_c (n_3+n_1-n_2)}} \,.
\label{3pf_free}
\eea
The coefficients on the r.h.s are nonvanishing only when the number of propagators in each edge of the diagram in Fig.~\ref{3pfcounting} 

\begin{figure}[h]
\begin{center}
\includegraphics[width=6cm]{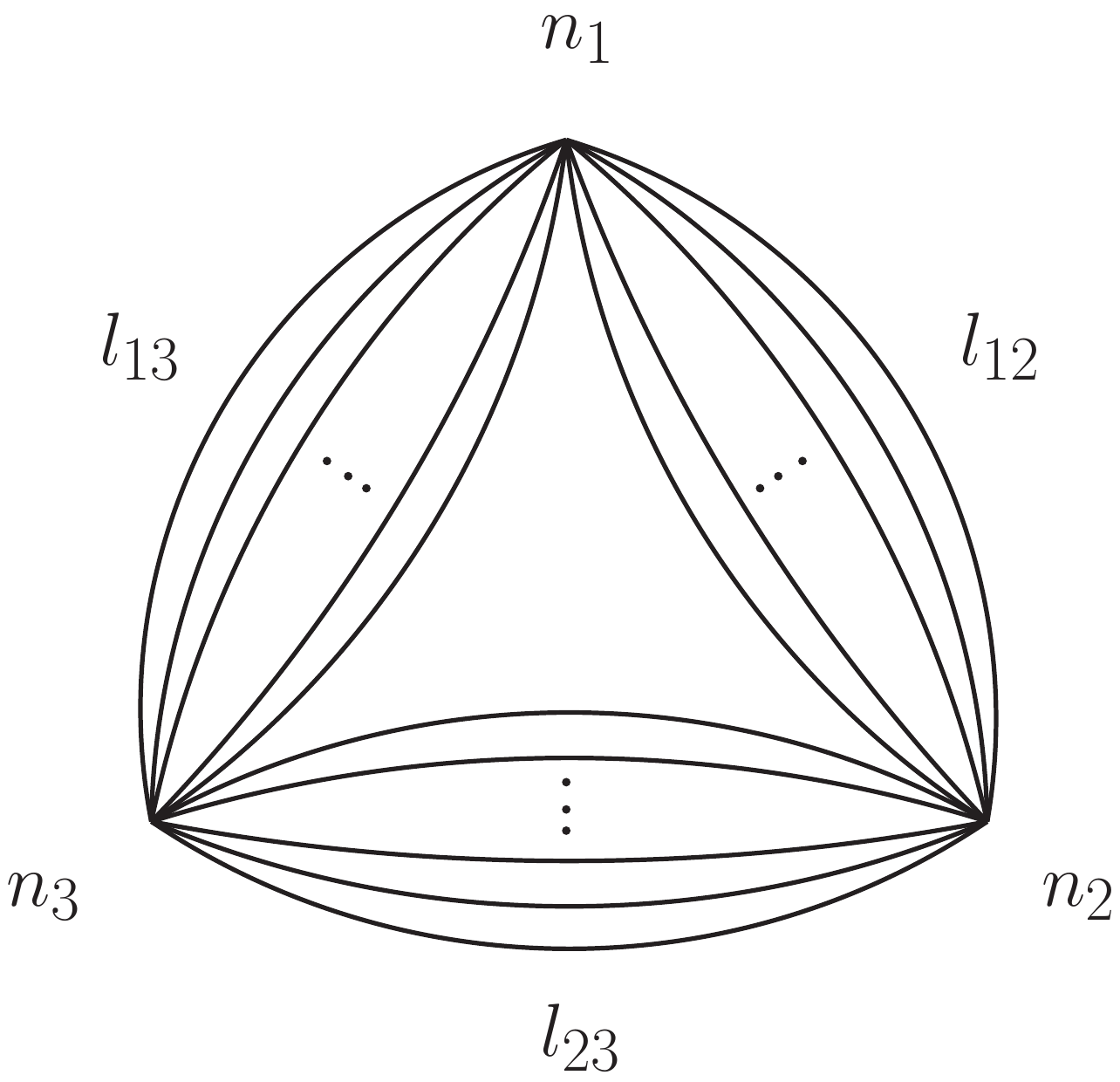}
\end{center}
\caption{Wick contraction counting of a three point correlator.
The vertices are labelled by $i=1,2,3$, the order of the $i$-th vertex is $n_i$, and there are $l_{ij}$ lines connecting two distinct vertices $i$ and $j$.}
\label{3pfcounting}
\end{figure}

\be 
l_{ij} = \frac{1}{2}(n_i+n_j-n_k), \qquad i\neq j \neq k
\ee
turns out to be nonnegative. They are obtained from the condition $n_i=l_{ij}+l_{ki}$ for $i\neq j \neq k$. In this case the coefficients are related to the their single-field counterpart 
\be 
C_{n_1,n_2,n_3}^{\mathrm{free}} = 
 \frac{n_1! \ n_2! \ n_3!}{\left(\frac{n_1+n_2-n_3}{2}\right)!\left(\frac{n_2+n_3-n_1}{2}\right)!\left(\frac{n_3+n_1-n_2}{2}\right)!} c^{\frac{n_1+n_2+n_3}{2}} \,,
\label{c3_free}
\ee
through the following equation
\be 
C^{\mathrm{free}}_{i_1\cdots i_{n_1},j_1\cdots j_{n_2},k_1\cdots k_{n_3}} = C_{n_1,n_2,n_3}^{\mathrm{free}} \, (\delta_{i_1j_1} \cdots \delta_{i_{l_{12}}j_{l_{12}}}\,\delta_{i_{l_{12}+1}k_1} \cdots \delta_{i_{n_1}k_{l_{13}}}\,\delta_{k_{l_{13}+1}j_{l_{12}+1}} \cdots \delta_{k_{n_3}j_{n_2}}),
\label{3pfree}
\ee
where the parenthesis enclosing the Kronecker deltas indicates that the $i$s the $j$s and the $k$s are separately symmetrized (including an inverse factor of $l_{12}!l_{13}!l_{23}!$).

\section{Some details on the Potts models}\label{sect:pottsdetails}

\subsection{Reduction algorithms}\label{ss:reduction}

The tensors $q^{(p)}=Q^{(p)}/(N+1)$ appear contracted as vertices in the diagrammatic expansion of cubic quartic and quintic theories.
They allow us to bypass the use of the Greek indices such as $\alpha=1,\dots,N+1$ altogether,
which is quite useful because they do not constitute a basis of a vector space.
The simplification of Feynman diagrams thus involves the contraction of several vertices built with $q^{(p)}$ and
is quite long, but it can be implemented recursively with an algorithm for reduction of the tensors $q^{(p)}$.

The algorithm uses the consecutive iteration of the fusion rule of two $Q$-tensors
\begin{eqnarray} \label{fusion}
 q^{(p)}_{i_1,\dots,i_{p-1},j}\,  q^{(r)}_{i_{p},\dots,i_{p+r-2}}{}^{j} &=& q^{(p+r-2)}_{i_1,\dots , i_{p+r-2}} - q^{(p-1)}_{i_1,\dots,i_{p-1}}\,  q^{(r-1)}_{i_{p},\dots,i_{p+r-2}},
\end{eqnarray}
and the trace rule
\begin{eqnarray} \label{trace}
 q^{(p)}_{i_1,\dots,i_{p-2},j}{}^j &=& N\, q^{(p-2)}_{i_1,\dots,i_{p-2}}.
\end{eqnarray}
While the fusion rule generally increases the index of the fused $Q$-tensor as $p \times r \to p+r-2$,
the traces of the indices of the Feynman diagrams ensure that the trace rule reduces the order as $p\to p-2$.
Each manipulation terminates whenever the cases $q^{(1)}_i=0$ and $q^{(2)}_{ij}=\delta_{ij}$ are encountered.
This is always ensured by the fact that the cubic and quintic models are renormalized
by considering diagrams with at most three and five external lines respectively.

As an instructive example, let us reduce the following product of two copies of $q^{(3)}$
\be \label{qq}
q^{(3)}_{i_1,j_1,j_2} \,q^{(3)}_{i_2}{}^{j_1,j_2} = q^{(4)}_{i_1,j_1,i_2}{}^{j_1} - q^{(2)}_{i_1,j_1}q^{(2)}_{i_2}{}^{j_1} 
= (N-1)\,\delta_{i_1,i_2}.
\ee
In the first equation we have fused $3\times 3 \to 4$, and in the second equation we have traced $4 \to 2$ and fused $2\times 2 \to 2$. As a second example consider the following contraction of three copies of $q^{(3)}$
\begin{eqnarray}
 q^{(3)}_{i_1,j_1}{}^{j_2} \,q^{(3)}_{i_2,j_2}{}^{j_3} \, q^{(3)}_{i_3,j_3}{}^{j_1}
 &=&
  q^{(3)}_{i_1,j_1}{}^{j_2} \, q^{(4)}_{i_2,j_2,i_3,}{}^{j_1} - q^{(3)}_{i_1,i_2,i_3} \nn\\
 &=&
  q^{(5)}_{i_1,j_1,i_2,i_3}{}^{j_1} - 2 q^{(3)}_{i_1,i_2,i_3}  \nn\\
 &=&
 (N-2)q^{(3)}_{i_1,i_2,i_3}. \label{qqq}
\end{eqnarray}
In the first line we fused $3\times 3 \to 4$, in the second line we fused $3\times 4 \to 5$, and in the third line we traced $5\to 3$.
At each step we substituted the predetermined values for $q^{(1)}$ and $q^{(2)}$.

\subsection{Some computational details for the quintic model} \label{ss:details}

In this appendix we give the details of the computation of the scaling operators quadratic in the fields along with their anomalous scaling dimensions, for the quintic model. For this purpose we first need to compute the quantity $T^{(5)}_{abc\, ip} T^{(5)}{}^{abc}{}_{jq}$ and then find the eigensystem of the stability matrix which is proportional to it. Let us start with the (one-index) contraction of two $q^{(5)}$s which is nothing but a special case of the fusion rule \eqref{fusion}
\be 
q^{(5)}_{a\,i_1i_2i_3i_4}q^{(5)}{}^{a}{}_{j_1j_2j_3j_4} = q^{(8)}_{i_1i_2 i_3i_4\,j_1j_2j_3j_4} - q^{(4)}_{i_1i_2i_3i_4} \,q^{(4)}_{j_1j_2 j_3j_4}.
\ee
Contracting another index, this becomes
\be
q^{(5)}_{ab\,i_2 i_3i_4}q^{(5)}{}^{ab}{}_{j_2j_3j_4} = (N-1) q^{(6)}_{i_2i_3i_4\,j_2j_3j_4} + q^{(3)}{}_{i_2i_3i_4}\,q^{(3)}_{j_2j_3j_4}.
\ee
Finally we contract the third index to get
\be
q^{(5)}_{abc\,i_3i_4}q^{(5)}{}^{abc}{}_{j_3j_4} = (N^2-N+1) q^{(4)}_{i_3i_4j_3j_4} - q^{(2)}_{i_3i_4}\,q^{(2)}_{j_3j_4}.
\ee
Recall that $q^{(2)}_{ij}$ is nothing but the Kronecker delta $\delta_{ij}$. The same procedure is carried out for the contraction of $q^{(5)}$ with the other symmetric five-index tensor $\delta_{(ij} q^{(3)}_{klm)}$. In this case contraction of one index gives
\be 
\delta_{(ai_1} q^{(3)}_{i_2i_3i_4)}q^{(5)}{}^{a}{}_{j_1j_2j_3j_4} = \frac{3}{5} \delta_{(i_1i_2}q^{(6)}_{i_3i_4)j_1j_2j_3j_4} -\frac{3}{5} \delta_{(i_1i_2}q^{(2)}_{i_3i_4)} \,q^{(4)}_{j_1j_2 j_3j_4} + \frac{2}{5} q^{(3)}_{(i_1i_2i_3} \,q^{(5)}_{i_4)j_1j_2 j_3j_4}.
\ee
Contracting two indices leads to 
\be
\delta_{(ab} q^{(3)}_{i_2i_3i_4)}q^{(5)}{}^{ab}{}_{j_2j_3j_4} = \frac{3}{5} q^{(6)}_{i_2i_3i_4j_2j_3j_4} + \frac{3}{10}(N-3) q^{(2)}_{(i_2i_3} \,q^{(4)}_{i_4)j_2 j_3j_4} + \frac{1}{10} N q^{(3)}_{i_2i_3i_4} \,q^{(3)}_{j_2 j_3j_4}.
\ee
and finally contracting the third index we find 
\be
\delta_{(ab} q^{(3)}_{ci_3i_4)}q^{(5)}{}^{abc}{}_{j_3j_4} = \frac{3}{10}(3N-2) q^{(4)}_{i_3i_4j_3j_4} + \frac{1}{10} N(N-4) q^{(2)}_{i_3i_4} \,q^{(2)}_{j_3j_4}.
\ee
We also need to contract two tensors of the type $\delta_{(ij} q^{(3)}_{klm)}$ for which we directly give the final result with all three indices contracted 
{\setlength\arraycolsep{2pt}
\bea
\delta_{(ab} q^{(3)}_{ci_3i_4)} \delta^{(ab} q^{(3)}{}^c{}_{j_3j_4)} &=& \frac{3}{100}(N+18) q^{(4)}_{i_3i_4j_3j_4} + \frac{1}{100}(N^2+8N-42)q^{(2)}_{i_3i_4}q^{(2)}_{j_3j_4} \nn\\
&+& \frac{3}{50}(N-3)q^{(2)}_{i_3j_3}q^{(2)}_{i_4j_4} + \frac{3}{50}(N-3)q^{(2)}_{i_3j_4}q^{(2)}_{i_4j_3}.
\eea}%
Using these results we can now compute the contraction of two tensors of the form 
\be 
T^{(5)}_{i_1i_2i_3i_4i_5} = \zeta_{5,1}\delta_{(i_1i_2} q^{(3)}_{i_3i_4i_5)}+\zeta_{5,2}q^{(5)}_{i_1i_2i_3i_4i_5}.
\ee
This is given by
{\setlength\arraycolsep{2pt}
\bea
T^{(5)}_{abc\, ip} T^{(5)}{}^{abc}{}_{jq} &=& \left[\zeta^2_{5,1}\frac{3}{100}(N+18) + \zeta^2_{5,2}(N^2-N+1)+ \zeta_{5,1}\zeta_{5,2}\frac{3}{5}(3N-2)\right] q^{(4)}_{ijpq} \nn\\
&+& \left[\zeta^2_{5,1}\frac{1}{100}(N^2+8N-42)- \zeta^2_{5,2}+ \zeta_{5,1}\zeta_{5,2}\frac{1}{5} N(N-4)\right]q^{(2)}_{ip} q^{(2)}_{jq} \nn\\
&+& \left[\zeta^2_{5,1}\frac{3}{50}(N-3)\right]q^{(2)}_{ij}q^{(2)}_{pq} + \left[\zeta^2_{5,1}\frac{3}{50}(N-3)\right]q^{(2)}_{iq}q^{(2)}_{jp}.
\eea}%
As we can see here, this is not necessarily symmetric in the $ij$ or $pq$ indices, but the stability matrix which is always contracted with $S_{pq}$ can be chosen to be symmetric without loss of generality. The symmetrized version of the above tensor is 
{\setlength\arraycolsep{2pt}
\bea
T^{(5)}_{abc\, i(p} T^{(5)}{}^{abc}{}_{q)j} &=& \left[\zeta^2_{5,1}\frac{3}{100}(N+18) + \zeta^2_{5,2}(N^2-N+1)+ \zeta_{5,1}\zeta_{5,2}\frac{3}{5}(3N-2)\right] q^{(4)}_{ijpq} \nn\\
&+& \left[\zeta^2_{5,1}\frac{1}{100}(N^2+14N-60)- \zeta^2_{5,2}+ \zeta_{5,1}\zeta_{5,2}\frac{1}{5} N(N-4)\right]q^{(2)}_{i(p} q^{(2)}_{q)j} \nn\\
&+& \left[\zeta^2_{5,1}\frac{3}{50}(N-3)\right]q^{(2)}_{ij}q^{(2)}_{pq}.
\eea}%
Given the proportionality factor in \eqref{M=T5T5} between the above tensor and the stability matrix, the three parameters defined in \eqref{M} are obtained as follows  
{\setlength\arraycolsep{2pt}
\bea
\tau &=& \frac{3c^3}{32}\left[\zeta^2_{5,1}\frac{3}{100}(N+18) + \zeta^2_{5,2}(N^2-N+1)+ \zeta_{5,1}\zeta_{5,2}\frac{3}{5}(3N-2)\right], \nn\\
\kappa &=& \frac{3c^3}{32}\left[\zeta^2_{5,1}\frac{1}{100}(N^2+14N-60)- \zeta^2_{5,2}+ \zeta_{5,1}\zeta_{5,2}\frac{1}{5} N(N-4)\right], \nn\\
\rho &=& \frac{3c^3}{32}\left[\zeta^2_{5,1}\frac{3}{50}(N-3)\right].
\eea}%
These are then inserted into the relations \eqref{es1}-\eqref{es3} to obtain the anomalous dimensions $\gamma^S_2$, using also the value of $\eta$ given in \eqref{eta-quintic}. The final results for $\gamma_2^i$ are reported in \eqref{gamma21-quintinc}-\eqref{gamma23-quintinc}.

\subsection{Fixed point results from RG for the quintic model}\label{sec:quintic_RG}

In Sections \ref{ss:cc-cubic} and \ref{ss:cc-quartic} we have obtained the critical values of the couplings of the cubic and quartic model respectively relying only on CFT methods.
Unfortunately, similar methods cannot be generalized to the quintic model, a situation which is reminiscent of the single-field case studied in \cite{Codello:2017qek}.
It comes to the rescue the fact that the critical values of the couplings as a function of $\epsilon$ can be understood as fixed point values of the models' RG equations.
Luckily, the leading fixed points can be obtained from standard RG practice.
In this appendix we report the necessary results of a soon-to-appear paper \cite{CSVZ4} in which we determine the $\epsilon$-dependence of the fixed point couplings of the quintic model with RG methods.
These fixed points are used throughout Section.~\ref{sect:quintic-potts-cft}.

The quintic model generalizes structurally the single-field model given in \cite{Codello:2017epp} to the multi-field case.
Specifically, the multi-field model has two critical couplings $\zeta_{5,1}$ and $\zeta_{5,2}$,
and consequently two beta functions $\beta_{5,1}$ and $\beta_{5,2}$.
The beta functions are rather long because they are cubic in the couplings themselves and depend parametrically on $N$;
they will be presented in the companion RG paper \cite{CSVZ4}
which will also discuss in far greater detail the content of this appendix.

The zeroes of these beta functions are also very complicate parametric functions of $N$ being the intersection of two cubics.
For brevity we omit showing the general fixed points too.\footnote{The analytic continuation in $N$ of the results
will also play a pivotal role in the discussion of \cite{CSVZ4}.} We just state the result that nontrivial real fixed points
are found for the two natural values $N=0$ and $N=-1$, while there are no further fixed points for $N\geq 1$.
This is rather interesting considering that also the cubic model
is nontrivial for those two cases which are generally associated to the notions of {\tt Percolation} and {\tt Spanning Forest}
universality classes.

\subsubsection{Fixed Points of the quintic model for $N=0$: Percolation}

For $N=0$, which corresponds to the single-state limit $q=1$, there are two nontrivial independent fixed points (factoring out the isospectral reflection $\zeta_{5,i} \to - \zeta_{5,i}$
in which the potential undergoes a parity transformation and neglecting complex solutions).
We give them conveniently in terms of the normalization used in this paper
\begin{eqnarray}
  {\rm FP}_1: &\qquad& \frac{3c^{\scriptscriptstyle 3/2}}{8} (\zeta_{5,1},\zeta_{5,2})=\left(0.0697,-0.0605\right)\sqrt{\epsilon}\,, \label{n0-1} \\
  {\rm FP}_2: &\qquad& \frac{3c^{\scriptscriptstyle 3/2}}{8} (\zeta_{5,1},\zeta_{5,2}) = \left(0.2222,0.3160\right)\sqrt{\epsilon}\,. \label{n0-2}
\end{eqnarray}
For completeness we provide also the spectra of critical exponents of the relevant and almost marginal operators \eqref{eq:invariants-basis}
at these fixed points
\begin{eqnarray}
{\rm Spectrum}({\rm FP}_1) &=&
 \left\{
 2 + 0.0046 \epsilon, \frac{4}{3} + 0.4183 \epsilon, \frac{2}{3} - 0.0655 \epsilon, \frac{2}{3} + 1.0008 \epsilon, -3 \epsilon, 1.2925 \epsilon
 \right\},\nonumber
 \\
{\rm Spectrum}({\rm FP}_2) &=&
 \left\{
 2 + 0.0907 \epsilon, \frac{4}{3} + 0.3096 \epsilon, \frac{2}{3} - 3.4239 \epsilon, \frac{2}{3} - 0.1216 \epsilon, -13.34 \epsilon, -3 \epsilon
 \right\}.\nonumber
\end{eqnarray}
The second fixed point ${\rm FP}_2$ is a realistic candidate for controlling the large scale limit of a system with the opportune degrees of freedom
being the true IR point of the pair. On symmetry grounds, one could argue that the universality class described by its scaling exponents is a multicritical
generalization of the standard percolation at criticality which is nontrivial even in $d=3$ for $\epsilon=\frac{1}{3}$ \cite{CSVZ4}.

\subsubsection{Fixed Points of the quintic model for $N = -1$: Spanning Forest}\label{section:quintic_Nm1}

For $N=-1$, which corresponds to the no-states limit $q=0$, there are four nontrivial independent fixed points.
Like the previous case, we factor out the isospectral reflection $\zeta_{5,i} \to - \zeta_{5,i}$
and neglect all complex solutions. We label them in such a way that the first two can actually be connected with those of the $N=0$ case by analytically continuing $N$.
The result is
\begin{eqnarray}
  {\rm FP}_1: &\qquad& \frac{3c^{\scriptscriptstyle 3/2}}{8} (\zeta_{5,1},\zeta_{5,2})= \left(0.0309,-0.0463\right)\sqrt{\epsilon}\,, \label{n1-1} \\
  {\rm FP}_2: &\qquad& \frac{3c^{\scriptscriptstyle 3/2}}{8} (\zeta_{5,1},\zeta_{5,2})= \left(0.4956,0.3096\right)\sqrt{\epsilon}\,, \label{n1-2} \\
  {\rm FP}_3: &\qquad& \frac{3c^{\scriptscriptstyle 3/2}}{8} (\zeta_{5,1},\zeta_{5,2})= \left(0.2229,0.0497\right)\sqrt{\epsilon}\,, \\
  {\rm FP}_4: &\qquad& \frac{3c^{\scriptscriptstyle 3/2}}{8} (\zeta_{5,1},\zeta_{5,2})= \left(\frac{5}{2\sqrt{46}},\frac{5}{4\sqrt{46}}\right) \sqrt{\epsilon}\,.
\end{eqnarray}
We omit giving here their full spectra, but we mention the interesting property that
the first three fixed points share the same anomalous dimension
\begin{eqnarray}
 \eta_{1,2,3} = -\frac{1}{1965}\epsilon \simeq -0.00051 \epsilon\,,
\end{eqnarray}
while the fourth has a Gaussian dimension $\eta_{4}=0$.
Like the previous section, the second fixed point ${\rm FP}_2$ might be controlling the large scale limit of a system with the opportune degrees of freedom
being the true IR point of the model. By analogy, we think that the universality class described by its scaling exponents could be a multicritical
generalization of the spanning forest's model at criticality that should be nontrivial even in $d=3$ \cite{CSVZ4}.


\end{document}